\documentclass[twocolumn,showpacs,preprintnumbers,amsmath,amssymb, rmp, showkeys, reprint]{revtex4-1}
\usepackage{amsmath}
\usepackage{stmaryrd}
\usepackage{txfonts}
\usepackage{amssymb}
\usepackage{mathrsfs}
\usepackage{graphicx}
\usepackage{dcolumn}
\usepackage{bm}
\usepackage{epsfig}
\usepackage{color}
\usepackage[colorlinks=true, linkcolor=blue, citecolor=blue, urlcolor=blue]{hyperref} 

\begin{document}
\preprint{\href{http://dx.doi.org/10.1088/0953-8984/26/49/493202}{S.-Z. Lin,  J. Phys.: Condens. Matter {\bf 26}, 493202 (2014) [Topical Reviews, 34 pages]. }}

\title{Ground state, collective mode, phase soliton and vortex in multiband superconductors}
\author{Shi-Zeng Lin}
\email{szl@lanl.gov}
\affiliation{Theoretical Division, Los Alamos National Laboratory, Los Alamos, New Mexico
87545, USA}

\begin{abstract}
This article reviews theoretical and experimental work on the novel physics in multiband superconductors. Multiband superconductors are characterized by multiple superconducting energy gaps in different bands with interaction between Cooper pairs in these bands. The discovery of prominent multiband superconductors $\mathrm{MgB_2}$ and later iron-based superconductors has triggered enormous interests in multiband superconductors. Most recently discovered superconductors exhibit multiband features. The multiband superconductors possess novel properties that are not shared by their single-band counterpart. Examples include the time-reversal symmetry broken state in multiband superconductors with frustrated interband couplings, the collective oscillation of number of Cooper pairs between different bands, known as the Leggett mode, the phase soliton and fractional vortex, which are the main focus of this review. This review presents a survey of a wide range of theoretical exploration and experimental investigations of novel physics in multiband superconductors. Vast information derived from these studies is shown to highlight unusual and unique properties of multiband superconductors, and to reveal the challenges and opportunities in the research on the multiband superconductivity.
\end{abstract}
\date{\today}
\maketitle

\tableofcontents

\section{Introduction}

\begin{figure}[t]
	\psfig{figure=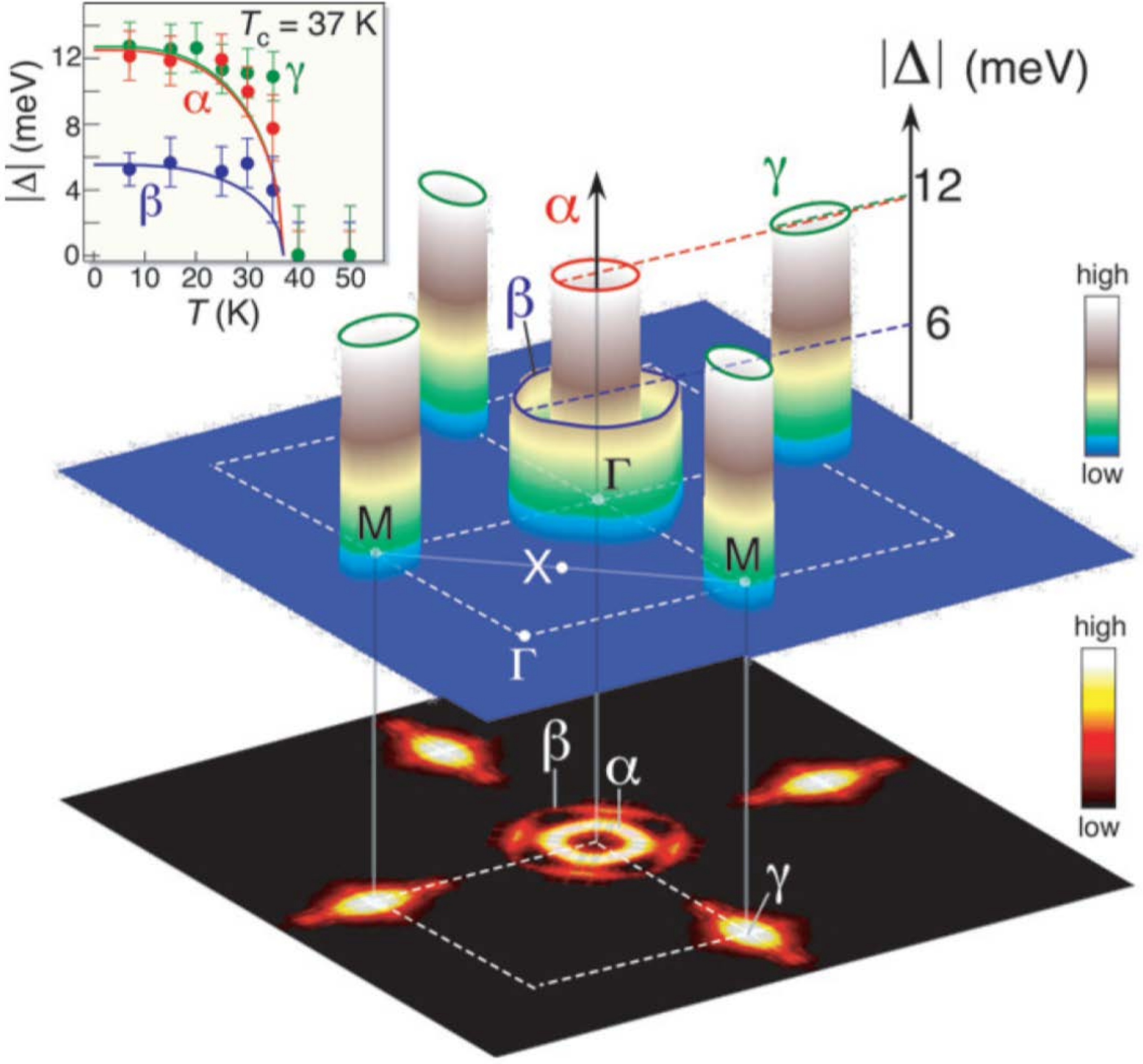,width=\columnwidth} \caption{\label{multigap_fig}
		(color online). Three-dimensional plot of the superconducting energy gaps $|\Psi_j|$ for the iron-based superconductor $\mathrm{Ba_{0.6}K_{0.4}Fe_2As_2}$ measured by ARPES at 15 K on the three observed Fermi surface sheets (shown at the bottom as an intensity plot) and their temperature evolutions (inset). From Ref. \cite{Ding08}.}
\end{figure}

The advent of the BCS theory \cite{BCSTheory57} provides a solid theoretical framework to understand various physical properties of superconductors. According to this theory, electrons near Fermi surface form Cooper pairs and condense into a macroscopic quantum state. Real superconducting materials usually involve multiple Fermi surfaces. It is possible that electrons/holes in each Fermi surface form superconducting condensate with interaction between them as a result of electron/hole hopping between different bands, see Fig. \ref{multigap_fig} for an example. These multiband superconductors, mostly found in transition metals in 20th century, can be described by a multiband BCS theory \cite{PhysRevLett.3.552,Moskalenko1959}, which was proposed shortly after the BCS theory. The research on multiband superconductors was refueled by the discovery of $\mathrm{MgB_2}$ with pronounced multiband characteristics in 2001. \cite{Nagamatsu01} The discovery of multiband iron-based superconductors in 2008 added momentum to the study on multiband superconductors. \cite{Kamihara08} With advances in crystal growth, experimental measurements and theoretical modeling, many superconductors originally labeled as single-band superconductors are rediscovered as multiband superconductors. The discovery of these multiband superconductors adds a new dimension to the superconductivity research.

The physical properties of multiband superconductors deviate significantly from their single-band counterpart. This deviation is usually a signature of multiband behavior in experiments. One line of research is to calculate physical properties with a more realistic multiband model by taking details of band structure, interactions and crystal structure into account. Moreover multiband superconductors possess novel physics that is not shared by single-band superconductors. One famous example is the collective oscillation of number of Cooper pairs between different bands, known as the Leggett mode. \cite{Leggett66} Caution must be taken for temperatures close to superconducting transition temperature $T_c$. According to the Landau argument, it is sufficient to describe a symmetry broken phase with one order parameter near $T_c$ where only one symmetry is broken. This means that multiband superconductors with interband couplings behave as single-band superconductors for temperatures sufficiently close to $T_c$. \cite{Geilikman1967,Kogan2011} Nevertheless there is pronounced multiband characteristics at low temperatures as revealed by various measurements. In short, multiband superconductivity at low temperatures is not a straightforward extension of single-band superconductivity. Instead new physics appears due to the multiband nature, which makes multiband superconductors interesting and promising for applications 
 
The present review is intended to give an overview about the novel physics in multiband superconductors. For purpose of demonstration, we adopt a minimal model by focusing on isotropic $s$-wave multiband superconductors. Such an approach is legitimate as we mainly focus on the qualitative new features in multiband superconductors. It is also very interesting when superconducting condensates in different bands have different pairing symmetries. \cite{PhysRevLett.102.217002,PhysRevLett.84.4445} In the followings, we will review the ground state, collective excitation, phase soliton and vortex in multiband superconductors. The phase soliton and vortex as topological objects, their overall properties should not depend on the microscopic details of the Hamiltonian. A review emphasizing the thermodynamic properties and materials realizations can be found in Ref. \cite{zehetmayer_review_2013}.

Superconductivity in each band can be described by a complex gap function $\Psi_j=\Delta_j\exp(i\phi_j)$. The phase differences between different bands are determined by the interband coupling. The interband coupling can be either attractive which favors the same superconducting phase, or repulsive which favors a $\pi$ phase shift. Frustration may arise in three or more bands superconductors. Without frustration, the phase difference is either $0$ or $\pi$. With a strong frustration, it is possible for superconductors to break the $Z_2$ time reversal symmetry in addition to the $U(1)$ symmetry. In this case the phase differences can take a value neither $0$ nor $\pi$. This state without time-reversal symmetry was first considered by Agterberg \emph{et al.} \cite{Agterberg99}, and later by Stanev and Te\u{s}anovi\'{c} \cite{Stanev10} in the context of iron-based superconductors. As a consequence of the time-reversal symmetry breaking, new phenomena such as the appearance of spontaneous magnetic fields in the presence of non-magnetic defects \cite{LinNJP2012,PhysRevLett.112.017003}, existence of a gapless Leggett mode \cite{Lin2012PRL} and phase solitons between two distinct time-reversal symmetry broken systems emerge. \cite{LinNJP2012, PhysRevLett.107.197001} 

Superconductivity as a consequence of symmetry breaking allows for the existences of several collective modes. \cite{PhysRevB.26.4883,kulik_pair_1981} One is the Goldstone mode associated with the breaking of $U(1)$ continuous symmetry, and in the context of superconductors is known as the Bogoliubov-Anderson-Goldstone boson. \cite{Anderson58,Bogoliubov59} This mode becomes a gapped plasma mode when couples to electromagnetic fields due to the Anderson-Higgs mechanism. \cite{PhysRev.130.439,PhysRevLett.13.508} Near $T_c$, the Bogoliubov-Anderson-Goldstone mode with electromagnetic fields is not pushed up to the plasma frequency because the conversion rate between normal current and supercurrent is slow. In the two-fluid model picture, the normal current and supercurrent oscillate in order to maintain charge neutrality. This mode is called the Carlson-Goldman mode \cite{PhysRevLett.34.941,Artemenko1975} and has been observed in Al thin film near $T_c$. \cite{PhysRevLett.34.11}  The collective oscillation of the amplitude of the order parameter $\Delta$ is known as the Schmid mode. \cite{Schmid1968} It has a gap of $2\Delta$ and can be regarded as a Higgs boson. The amplitude mode has been observed in the $2H$-$\mathrm{NbSe_2}$ superconductor. \cite{PhysRevLett.45.660,PhysRevLett.47.811} For multiband superconductors, besides the modes mentioned above, it hosts another collective excitation associated with oscillation of number of Cooper pairs between different bands due to the interband coupling, known as the Leggett mode. This mode corresponds to a small out of phase oscillation of the phase mode in different bands and was first pointed by Leggett in 1966. \cite{Leggett66} The Leggett mode is gapped. The Leggett mode has been observed experimentally in $\mathrm{MgB_2}$. \cite{Blumberg07} The multiband superconductors without time-reversal symmetry have significant effects on the collective modes. In the time-reversal symmetry broken state, the phase mode hybridizes with the amplitude mode, and form a phase-amplitude composite mode. \cite{Stanev11} At the time-reversal symmetry breaking transition, one of the Leggett modes becomes gapless. \cite{Lin2012PRL}

Besides the small out of phase oscillation in multiband superconductors, there exist phase soliton excitations in the superconducting phase difference between different condensates. For a Josephson-like interband coupling, there are energy degenerate ground states for the phase difference and it allows for the phase soliton between any pair of the degenerate ground states. Such a phase soliton unique to multiband superconductors was first discussed by Tanaka in 2001, \cite{Tanaka2001} which has been observed experimentally in an artificial multiband superconductor. \cite{Bluhm06} In multiband superconductors without time-reversal symmetry, one can also have phase solitons between two time-reversal symmetry broken states, similar to the domain walls in ferromagnets. The phase soliton can only be stable in one dimensional systems but it could be stabilized by defects or vortices in higher dimensions. \cite{PhysRevLett.107.197001,PhysRevLett.112.017003} Inside the phase soliton, the superconducting phase differences are neither $0$ or $\pi$ and therefore the time-reversal symmetry is broken locally. Spontaneous magnetic fields can exist in the phase soliton region under proper conditions. \cite{LinNJP2012,PhysRevLett.112.017003} The same as the Leggett mode, the phase soliton is neutral and does not couple to magnetic fields. However the phase soliton can be excited dynamically by an electric field in nonequilibrium region. \cite{Gurevich03}

Another hallmark of the $U(1)$ symmetry breaking in superconductors is the existence of vortex carrying quantized magnetic flux $n\Phi_0=nhc/(2e)$ with the integer $n$ being the phase winding number. In multiband superconductors the phase winding number for superconducting condensates in different bands may be different, which results in vortex carrying fractional quantum flux. This fractional vortex was first studied by Babaev in 2002. \cite{Babaev02} Fractional vortices with the same polarization in the same band interact repulsively due to the magnetic interaction. The fractional vortices in different condensates also repel each other due to the exchange of massive photon. Meanwhile they attract each other due to the coupling to the same gauge field. \cite{szlin13PRL} They also attract because of the interband coupling in superconducting channel. The attraction outweighs the repulsion and the net interaction of fractional vortices in different condensates is attractive.  Therefore fractional vortices in different condensates in the ground state bind together with their normal cores locked together to form a composite vortex with the standard integer quantum flux $\Phi_0$. In the flux flow region with a high current drive, the composite vortex lattice can dissociate into fractional vortex lattices with different velocities because of the disparity in the vortex viscosity and magnetic flux of the fractional vortices in different bands. \cite{szlin13PRL} After turning off the current, the fractional vortices can be trapped when pinning centers are present, therefore results in metastable fractional vortices. \cite{LinPRB13} The fractional vortices can also be stabilized in a mesoscopic multiband superconductor. \cite{chibotaru_thermodynamically_2007,Chibotaru10,Geurts10,PhysRevB.84.144504,Pina12,PhysRevB.89.024512} Because of the possible existence of distinct length scales for condensates in different bands at lower temperatures, vortex may interact repulsively at short distance, attractively at intermediate distance and repulsively at large distance due to the demagnetization effect. The existence of nonmonotonic inter-vortex interaction in multiband superconductors was first discussed by Babaev and Speight in 2005.\cite{Babaev05} This kind of vortex interaction leads to unusual magnetic response in multiband superconductors, which differs from the conventional type I and type II superconductors.  

The remainder of this review is organized as follows. In Sec. \ref{Sec2} we will introduce the models, discuss the ground state properties, behavior near $T_c$ and material realizations. Section \ref{Sec3} is devoted to the Leggett modes. In Sec. \ref{Sec4} the phase solitons are discussed and in Sec. \ref{Sec5} we will review vortices in multiband superconductors. The paper is concluded by discussions in Sec. \ref{Sec6}.

\section{Model and ground state}\label{Sec2}
In this section, we will introduce the isotropic Ginzburg-Landau free energy functional and BCS Hamiltonian. Their relation will be discussed. We then will show that multiband superconductors undergoing a single $U(1)$ symmetry breaking at $T_c$ behave as a single-band superconductor near $T_c$. We will present a zero magnetic field phase diagram for multiband superconductors, focusing on three-band superconductors with frustrated interband couplings where time-reversal symmetry may be broken inside the superconducting phase. Finally material realizations of multiband superconductivity will be reviewed.

\subsection{Model}\label{Sec2A}
Here we introduce models to describe multiband superconductors.  A phenomenological description can be obtained by generalizing the single-band Ginzburg-Landau free energy functional to multiband case. Such as a phenomenological theory can be derived from a microscopic theory with material specified Hamiltonian for temperatures close to $T_c$. The coefficients in the Ginzburg-Landau free energy functional are functions of microscopic coupling constants responsible for superconductivity and temperature. For $\mathrm{MgB_2}$, the Ginzburg-Landau theory was derived in Ref. \cite{PhysRevB.69.054508}.  We neglect microscopic details such as the complicated band structure and crystal anisotropy, and focus on the isotropic Ginzburg-Landau free energy density functional
\begin{equation}\label{meq1}
\begin{array}{l}
{\cal F} = \sum\limits_j {\left[ {{\alpha _j}{{\left| {{\Psi _j}} \right|}^2} + \frac{{{\beta _j}}}{2}{{\left| {{\Psi _j}} \right|}^4} + \frac{1}{{2{m_j}}}{{\left| {\left( { - i \hbar\nabla  - \frac{2e}{c}{\bf{A}}} \right){\Psi _j}} \right|}^2}} \right]} \\
+ \frac{1}{{8\pi }}{(\nabla  \times {\bf{A}})^2} + \sum\limits_{l < j} {{\gamma _{lj}}} \left( {{\Psi _l}\Psi _j^* + {\Psi _l}^*\Psi _j} \right).
\end{array}
\end{equation}
Here $\alpha_j$ depends on temperature, while $\beta_j$ and the interband Josephson-like coupling $\gamma_{lj}$ are temperature-independent. Here $m_j$ is the electron effective mass. The supercurrent density is
\begin{align}
\mathbf{J}_s=\sum _j \left[\frac{-i \hbar e  }{m_j}\left(\Psi _j^* \nabla \Psi _j-\Psi _j \nabla\Psi _j^*\right)-\frac{4 e^2 }{m_j c}|\Psi _j|^2 \mathbf{A}\right].
\end{align}
We have assumed a Josephson-like interband coupling. Other forms of coupling such as coupling between superfluid density, $|\Psi_j\Psi_l|^2$, can also exist. \cite{gurevich_limits_2007} In additional to the interband Josephson coupling, the superconducting phases in different bands couple to the same gauge field $\mathbf{A}$. When interband coupling is absent $\gamma_{ij}=0$, we can define a coherence length $\xi_i$ for each band, 
\begin{equation}
\xi_i=\sqrt{\frac{\hbar^2}{2m_i|\alpha_i|}}.
\end{equation} 
Since only one gauge field is involved in Eq. \eqref{meq1}, there is only one London penetration depth
\begin{equation}
\lambda^{-2}=\sum_i\lambda_i^{-2},
\end{equation}
with the parameter $\lambda_i=\sqrt{m_i c^2/(16\pi e^2 \Psi_{i0}^2)}$ and the uniform amplitude of the order parameter $\Psi_{i0}=\sqrt{|\alpha_i|/\beta_i}$. The interband coupling mixes different condensates and $\xi_i$, $\Psi_{i0}$ need to be redefined. In the strong coupling limit, $|\gamma_{lj}|\gg |\alpha_i|$, $\xi_i$ for different bands becomes the same.

The dynamics of superconductivity can be described by the time-dependent Ginzburg-Landau theory
\begin{equation}\label{meq2}
\frac{{{\hbar ^2}}}{{2{m_j}D_j}}\left({\partial _t} + i\frac{{{2e}}}{\hbar }\varphi \right)\Psi_j   =   - \frac{{\delta \mathcal{F}}}{{\delta {\Psi_j ^*}}},
\end{equation}
\begin{equation}\label{meq3}
\frac{\sigma }{c}\left(\frac{1}{c}{\partial _t}{\bf{A}} + \nabla \varphi \right)  =   - \frac{{\delta \mathcal{F}}}{{\delta {\bf{A}}}},
\end{equation}    
with $D_j$ the diffusion constant, $\sigma$ the normal conductivity, and $\varphi$ the electric potential. The time-dependent Ginzburg-Landau equation can be derived from a microscopic theory near $T_c$.

The generalized BCS model for multiband superconductors has the form \cite{PhysRevLett.3.552,Moskalenko1959}
\begin{equation}\label{meq4}
\begin{array}{l}
\mathcal{H} = \sum\limits_{l,\sigma } {\int {{d^3}r\psi _{l\sigma }^\dag (\mathbf{r}){(\varepsilon_l-\mu)} {\psi _{l\sigma }}(\mathbf{r})} } \\
- \sum\limits_{j,l} {\int } {d^3}r\psi _{j\sigma }^\dag (\mathbf{r})\psi _{j\bar{\sigma} }^\dag (\mathbf{r}){V_{jl}}{\psi _{l \bar{\sigma} }}(\mathbf{r}){\psi _{l\sigma }}(\mathbf{r}),
\end{array}
\end{equation}
where $\psi _{l\sigma }^\dag$ (${\psi _{l\sigma }}$) is the electron creation (annihilation) operator in the $l$-th band with the dispersion $\varepsilon_l(\mathbf{k})$ and the chemical potential $\mu$ and spin index $\sigma$. We consider a parabolic dispersion for electrons $\varepsilon_l(\mathbf{k})=\hbar \mathbf{k}^2/2m_l$ with an electron mass $m_l$. $V_{jl}$ is the intraband for $l=j$ and interband for $l\neq j$ scattering respectively, which can be either repulsive or attractive depending, for instance, on the strength of the Coulomb and electron-phonon interaction. Here we have assumed a contact interaction for $V_{jl}$. Equation \eqref{meq4} reduces to Eq. \eqref{meq1} at temperature close to $T_c$ in the clean limit, i.e. $(T_c-T)/T_c\ll 1$. \cite{0370-1328-84-4-313,PhysRevB.69.054508} In the dirty limit, the interband impurity scattering induces additional coupling between different bands, other than the Josephson coupling in Eq. \eqref{meq1}. \cite{gurevich_limits_2007} The applicability region of Eq. \eqref{meq1} near $T_c$ depends on materials. It was argued in Refs. \cite{PhysRevLett.92.107008,PhysRevB.72.064523} that the applicability region of Eq. \eqref{meq1} shrinks practically to zero for $\mathrm{MgB_2}$.

\subsection{Behavior of multiband superconductors at temperatures close to $T_c$}\label{Sec2B}
Here we discuss the behavior of multiband superconductors near $T_c$ when interband Josephson couplings are present. We restrict to the case that there is only a single continuous phase transition associated with the breaking of $U(1)$ symmetry at $T_c$. We will show that in this region the multiband superconductors behave as single-band superconductors, i.e. there exists only one coherence length for all superconducting condensates. This was realized long time ago by Geilikman \emph{et al.} \cite{Geilikman1967}, and later independently by Kogan \emph{et al.} \cite{Kogan2011}.  We follow the derivation in Ref. \cite{Kogan2011,PhysRevB.82.104521}. As an example we consider a two-band isotropic Ginzburg-Landau free energy functional in Eq. \eqref{meq1}.

The critical temperature $T_c$ for Eq. \eqref{meq1} is given by the condition that the determinant of the coefficient matrix for the quadratic terms $\Psi_i^*\Psi_j$ is zero. For a two-band superconductor, it is given by $\alpha_1(T_c)\alpha_2(T_c)-\gamma_{12}^2=0$. We denote $\alpha_{1c}\equiv \alpha_1(T_c)$ and $\alpha_{2c}\equiv \alpha_2(T_c)$. The interband Josephson coupling enhances $T_c$, \cite{kondo_superconductivity_1963} i.e. superconductivity can exist even when $\alpha_i(T)>0$. In this sense superconductivity near $T_c$ is induced by interband coupling. This is the reason why condensates in different bands are strongly locked with each other, hence effectively become a single-band superconductor. Using $\delta\mathcal{F}/\delta\Psi_i^*=0$, we obtain
\begin{equation}\label{Tceq1}
\alpha_1 \Psi_1  + \beta_1 {|\Psi_1| ^2}\Psi_1  + \frac{1}{{2{m_1}}}{\left( - {i}\hbar \nabla  - \frac{{2e}}{c}\mathbf{A}\right)^2}\Psi_1+\gamma_{12}\Psi_2  = 0,
\end{equation}
\begin{equation}\label{Tceq2}
\alpha_2 \Psi_2  + \beta_2 {|\Psi_2| ^2}\Psi_2  + \frac{1}{{2{m_2}}}{\left( - {i}\hbar \nabla  - \frac{{2e}}{c}\mathbf{A}\right)^2}\Psi_2+\gamma_{12}\Psi_1  = 0.
\end{equation}
In the Ginzburg-Landau approximation, Eqs. \eqref{Tceq1} and \eqref{Tceq2} are valid up to the order $\tau^{3/2}$ with $\tau\equiv (T_c-T)/T_c\ll 1$. Keeping terms up to $\tau^{3/2}$, Eqs. \eqref{Tceq1} and \eqref{Tceq2} can be rewritten as \cite{Kogan2011}
\begin{align}\label{Tceq3}
(\alpha_1\alpha_2-\gamma_{12}^2)\Psi_1+(\beta_1\alpha_2+\beta_2\alpha_1^3/\gamma_{12}^2){|\Psi_1| ^2}\Psi_1 \nonumber\\
-\left(\frac{\alpha_1\hbar^2}{2m_2}+\frac{\alpha_2\hbar^2}{2m_1}\right)\left(\nabla-i\frac{2\pi}{\Phi_0}\mathbf{A} \right)^2\Psi_1=0,
\end{align}
\begin{align}\label{Tceq4}
(\alpha_1\alpha_2-\gamma_{12}^2)\Psi_2+(\beta_2\alpha_1+\beta_1\alpha_2^3/\gamma_{12}^2){|\Psi_2| ^2}\Psi_2 \nonumber\\
-\left(\frac{\alpha_1\hbar^2}{2m_2}+\frac{\alpha_2\hbar^2}{2m_1}\right)\left(\nabla-i\frac{2\pi}{\Phi_0}\mathbf{A} \right)^2\Psi_2=0.
\end{align}
Close to $T_c$, we expand $\alpha_i(T)=\alpha_{ic}-\tilde{\alpha}_i\tau$. Equations \eqref{Tceq3} and \eqref{Tceq4} become
\begin{equation}\label{Tceq5}
-\bar{\alpha}\tau\Psi_1+\bar{\beta}_1|\Psi_1| ^2\Psi_1-\bar{K}\left(\nabla-i\frac{2\pi}{\Phi_0}\mathbf{A} \right)^2\Psi_1=0,
\end{equation}
\begin{equation}\label{Tceq6}
-\bar{\alpha}\tau\Psi_2+\bar{\beta}_2|\Psi_2| ^2\Psi_2-\bar{K}\left(\nabla-i\frac{2\pi}{\Phi_0}\mathbf{A} \right)^2\Psi_2=0.
\end{equation}
with 
\begin{eqnarray}
\bar{\alpha}=\tilde{\alpha}_1\alpha_{2c}+\tilde{\alpha}_2\alpha_{1c},\ \ \ \bar{K}=\frac{\hbar^2\alpha_{1c}}{2m_2}+\frac{\hbar^2\alpha_{2c}}{2m_1}, \nonumber\\
\bar{\beta}_1=\beta_1\alpha_{2c}+\beta_2\alpha_{1c}^3/\gamma_{12}^2, \ \ \ \bar{\beta}_2=\beta_2\alpha_{1c}+\beta_1\alpha_{2c}^3/\gamma_{12}^2. \nonumber
\end{eqnarray}
Equation \eqref{Tceq6} reduces to Eq. \eqref{Tceq5} if we replace $\Psi_2(\mathbf{r}, T)$ by $\Psi_1(\mathbf{r}, T)\sqrt{\bar{\beta}_1/\bar{\beta}_2}$, i.e. $[\Psi_1(\mathbf{r}, T),\ \Psi_2(\mathbf{r}, T)]=[\Psi_1(\mathbf{r}, T),\ \Psi_1(\mathbf{r}, T)\sqrt{\bar{\beta}_1/\bar{\beta}_2}]$ is always a solution to Eqs. \eqref{Tceq1} and \eqref{Tceq2} near $T_c$. The superconducting order parameters for different bands vary in space with the same length scale and a two-band superconductor is equivalent to a single-band superconductor. This is in accordance with the Landau theory for a continuous phase transition: one order parameter is sufficient to describe physical properties of a symmetry broken state near $T_c$.

The above statement is only valid when the interband coupling is present, no matter how weak it is. Without the interband coupling such as for the proposed superconductivity in liquid hydrogen under high pressure \cite{Ashcroft68,Jaffe81}, the electron and proton superconducting condensates can have two distinct length scales even close to $T_c$.

If another symmetry, such as time-reversal symmetry, is broken simultaneous with $U(1)$ symmetry at $T_c$, (see Fig. ~\ref{gs_fig2}), the superconductor can have two diverging length scales associate with the breaking of two symmetries. This was demonstrated in Ref. \cite{Hu11} for three-band superconductors where the time-reversal symmetry and $U(1)$ symmetry break at the same time at $T_c$.

We cannot rule out the possible existence of many length scales for different condensates at low temperatures since the Ginzburg-Landau theory cannot apply there. Also the above derivation is valid for $\tau\ll 1$. In fact, it was shown recently using a microscopic approach that there exists distinct length scales for different condensates at low temperatures, \cite{Silaev11} thus allows for the emergence of novel properties unique to multiband superconductors.

\subsection{Ground state and phase diagram}\label{Sec2C}
In this subsection we discuss the zero-field phase diagram for Eqs. \eqref{meq1} and \eqref{meq4}. For the Ginzburg-Landau theory in Eq. \eqref{meq1}, $\Psi_i$ can be obtained readily using the condition $\delta\mathcal{F}/\delta\Psi_i^*=0$. Here we determine the phase diagram using the BCS Hamiltonian in Eq. \eqref{meq4} because it is valid at all temperatures. Let us first introduce the general gap equations for multiband superconductors without external magnetic fields.

Introducing the energy gap $\Psi_j$ through the Hubbard-Stratonovich transform and the Nambu spinor operator $\hat{\Theta}_j=(\psi_{j\uparrow},\ \psi_{j\downarrow}^\dag)^T$, we obtain the following action for Eq. \eqref{meq4} in the imaginary time representation after integrating out the fermionic fields \cite{SimonsQFT}
\begin{equation}\label{gseq2}
{S} = \int d \tau d^3 r\sum_{j,l}^M{\Psi _j}{g_{jl}}\Psi _l^* - \sum\limits_j {{\rm{Tr}}} \ln {\cal G}_j^{ - 1},
\end{equation}
with $\hat{g}=\hat{V}^{-1}$, $M$ the number of bands and the Gor'kov green function
\begin{equation}\label{gseq3}
{\cal G}_j^{ - 1} =  - \left( {\begin{array}{*{20}{c}}
	{{\partial _\tau } + ( {{\varepsilon_j} - \mu } )}&{ - {\Psi _j}}\\
	{ - \Psi _j^*}&{{\partial _\tau } - ( {{\varepsilon_j} - \mu } )}
	\end{array}} \right).
\end{equation}
Here $\hat{V}$ is a matrix form of interaction in Eq. \eqref{meq4}. The ground state $\Psi_j=\Delta_j\exp(i\phi_j)$ is given by the condition $\delta S/\delta \Psi_j=0$, which yields
\begin{equation}\label{gseq4}
\sum_{l = 1}^M {\Psi _l}{g_{lj}} = {N_j}(0){\Psi _j} F_j(\Psi_j, T),
\end{equation}
with
\begin{equation}\label{gseq5}
F_j(\Psi_j, T)\equiv  \int_0^{{\omega _{cj}}}  \frac{{{d}}\xi}{{\sqrt {{\xi ^2} + |{\Psi _j}{|^2}} }}\tanh \left[\frac{{\sqrt {{\xi ^2} + |{\Psi _j}{|^2}} }}{{2{k_B}T}}\right].
\end{equation}
with $\omega _{cj}$ a cutoff frequency, which depends on the pairing mechanism. For phonon mediated superconductivity, $\omega_{cj}$ is the Debye frequency. Here $N_j(0)$ the density of states at the Fermi surface in normal state. 

\begin{figure*}[t]
	\psfig{figure=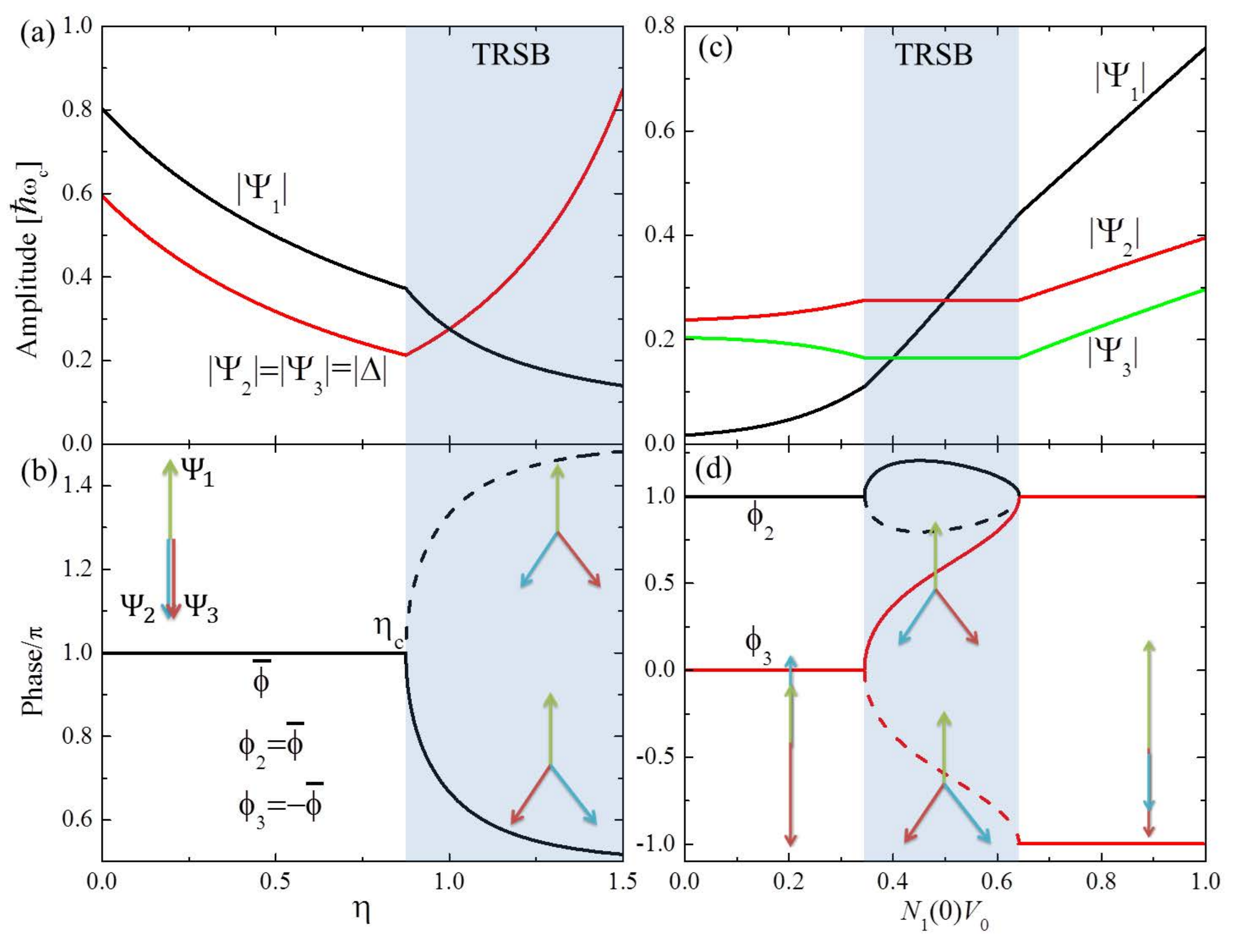,width=14cm} \caption{\label{gs_fig1}
		(color online). Amplitudes and phases of order parameters at time-reversal symmetry breaking phase transition, in (a) and (b) as a function of $\eta$, and in (c, d) as a function of density of state $N_1(0) V_0$ of the first component. $\Delta_1$ is taken as real and positive. In (a) and (b), an identical density of state $N(0)V_0=0.5$ is taken for the three bands and $\alpha=2$ in Eq.~\eqref{gseqg1}. In (c) and (d), $N_2(0) V_0=0.5$ and $N_3(0) V_0=0.4$, $\alpha=2$ and $\eta=1$. In the time-reversal symmetry breaking regime, there are two degenerate ground states $(\Psi_1, \Psi_2, \Psi_3)$ (solid lines) and $(\Psi_1^*, \Psi_2^*, \Psi_3^*)$ (dashed lines). The two solid lines for $N_1(0) V_0>0.64$ in (d) refer to the same state with time-reversal symmetry. Here $\Psi_j$ is in unit of $\omega_{c}$. Here TRSB refers to the state with time-reversal symmetry breaking. From Ref. \cite{Lin2012PRL}.}
\end{figure*}

It is particularly interesting when the interband interactions are frustrated and the system may break time-reversal symmetry in addition to the $U(1)$ symmetry. \cite{Agterberg99,Stanev10} We consider a three-band case since it is a minimal model to demonstrate the time-reversal symmetry breaking. We also focus on the case with identical density of state $N_j(0)=N(0)$ and cutoff frequency $\omega_{cj}=\omega_c$ at $T=0$. Here $F(\Psi_j, T=0)=\sinh^{-1}(\omega_c/|\Psi_j|)$. We also take a set of simplified interband couplings
\begin{equation}\label{gseqg1}
\hat g = \frac{1}{V_0}\left( {\begin{array}{*{20}{c}}
	\alpha &1&1\\
	1&\alpha &\eta \\
	1&\eta &\alpha
	\end{array}} \right).
\end{equation}
Here $\alpha>0$ and $\eta>0$ correspond to a repulsive interaction. We can always take $\Psi_1=\Delta_1$ as real by properly choosing the gauge. As $\hat{g}$ is symmetric, the solution for $\Psi_2$ and $\Psi_3$ can be written as $\Psi_2=\Delta\exp(i\bar{\phi} )=\Psi_3^*$. For a small $\eta\ll 1$, the repulsion between $\Psi_1$ and $\Psi_2$ (or $\Psi_3$) dominates over the repulsion between $\Psi_2$ and $\Psi_3$ and $\bar{\phi}=\pi$. As $\eta$ increases, the repulsion between $\Psi_2$ and $\Psi_3$ becomes more important and at a critical $\eta$, $\bar{\phi}$ starts to deviate from $\pi$, which breaks the time-reversal symmetry. In the state with time-reversal symmetry, $\bar{\phi}=\pi$ and $\Delta$, $\Delta_1$ are given by
\begin{equation}\label{gseqg2}
\frac{\alpha {\Delta _1} - 2\Delta }{{N(0)V_0}} = {\Delta _1}{\sinh ^{ - 1}}\left(\frac{{{\omega _c}}}{{{\Delta _1}}}\right)
\end{equation}
\begin{equation}\label{gseqg3}
\frac{{{\Delta _1} - (\alpha  + \eta )\Delta }}{{N(0)V_0}} =  - \Delta {\sinh ^{ - 1}}\left(\frac{{{\omega _c}}}{\Delta }\right).
\end{equation}
In the state without time-reversal symmetry, $\Delta$, $\Delta_1$ are given by
\begin{equation}\label{gseqg4}
\frac{\Delta }{{{\omega _c}}} = 1/\sinh \left(\frac{{\alpha  - \eta }}{{N(0)V_0}}\right), \ \ \frac{{{\Delta _1}}}{{{\omega _c}}} = 1/\sinh \left(\frac{{\alpha \eta  - 1}}{\eta }\frac{1}{{N(0)V_0}}\right).
\end{equation}
The time-reversal symmetry breaking occurs at 
\begin{equation}\label{gseqg5}
\cos\bar{\phi}  =  - {\Delta _1}/(2\eta \Delta ) =  - 1.
\end{equation}
The results for $\Psi_j$ as a function of $\eta$ are displayed in Fig. ~\ref{gs_fig1} (a) and (b) for $N(0) V_0=0.5$ and $\alpha=2$, where there is a continuous phase transition associated with the breaking of time-reversal symmetry.

The time-reversal symmetry breaking phase transition can also be driven by $N_j(0)$, which can be tuned in experiments by careful chemical doping. As an example, we calculate $\Psi_j$ as a function of $N_1(0) V_0$ for $N_2(0) V_0=0.5$, $N_3(0) V_0=0.4$, $\alpha=2$, $\eta=1$. As displayed in Fig. ~\ref{gs_fig1} (c) and (d), the time-reversal symmetry broken phase is stabilized in an intermediate region of $N_1(0) V_0$.

The associated free energy density at $T=0$ is
\begin{equation}\label{gseqg6}
\mathcal{F} = \sum_{lj} {\Psi _l}{g_{lj}}\Psi _j^* - \sum_j N_j(0)|{\Psi _j}|^2\left[ \frac{1}{2} + \ln \left(\frac{{{2\omega _{cj}}}}{{|{\Psi _j}|}}\right)\right], 
\end{equation}
and it can be verified that the time-reversal symmetry broken state indeed has lower free energy, thus is a thermodynamically stable phase.

\begin{figure}[t]
	\psfig{figure=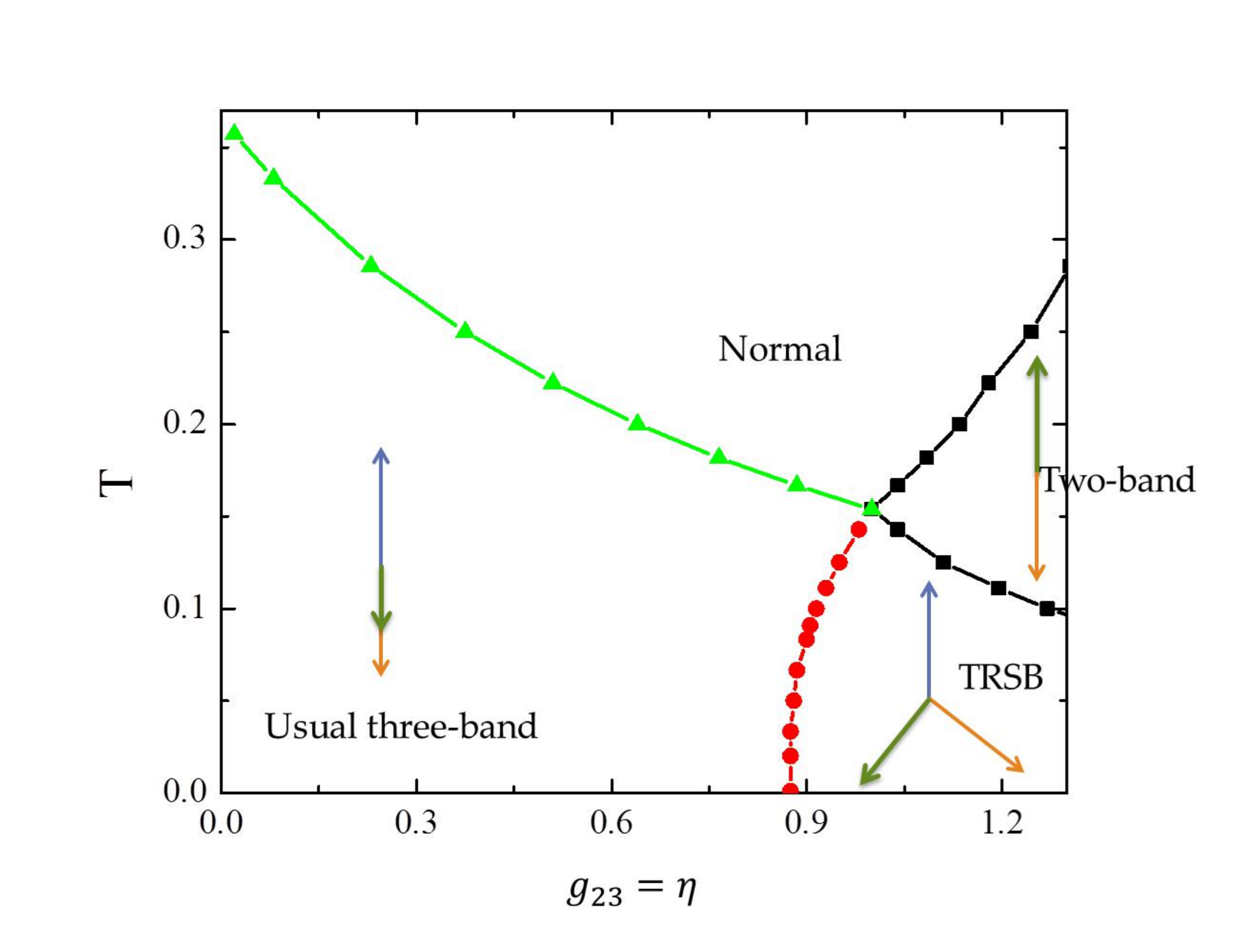,width=\columnwidth} \caption{\label{gs_fig2}
		(color online). $T$-$\eta$ phase diagram of a three-band superconductor with frustrated interband coupling. We take $N_j(0) V_0=0.5$ and $\alpha=2$. Here TRSB refers to the state with time-reversal symmetry breaking.}
\end{figure}

The phase diagram at $T>0$ can be obtained numerically. The $T$-$\eta$ phase diagram for a symmetric coupling $\hat{g}$ in Eq. \eqref{gseqg1} with $N_j(0) V_0=0.5$ and $\alpha=2$ is shown in Fig. ~\ref{gs_fig2}. For a large $g_{23}=\eta \gg 1$ at $T>0$, the superconductivity in the first band is strongly frustrated resulting in $\Psi_1=0$, and the system behaves as a two-band superconductor. The region without time-reversal symmetry shrinks when $T$ approaches to $T_c$, and contracts into a point at $\eta^*$ at $T_c$. At $T_c$ and $\eta^*$, the $U(1)$ and $Z_2$ symmetries are broken simultaneously. 

\begin{figure}[b]
	\psfig{figure=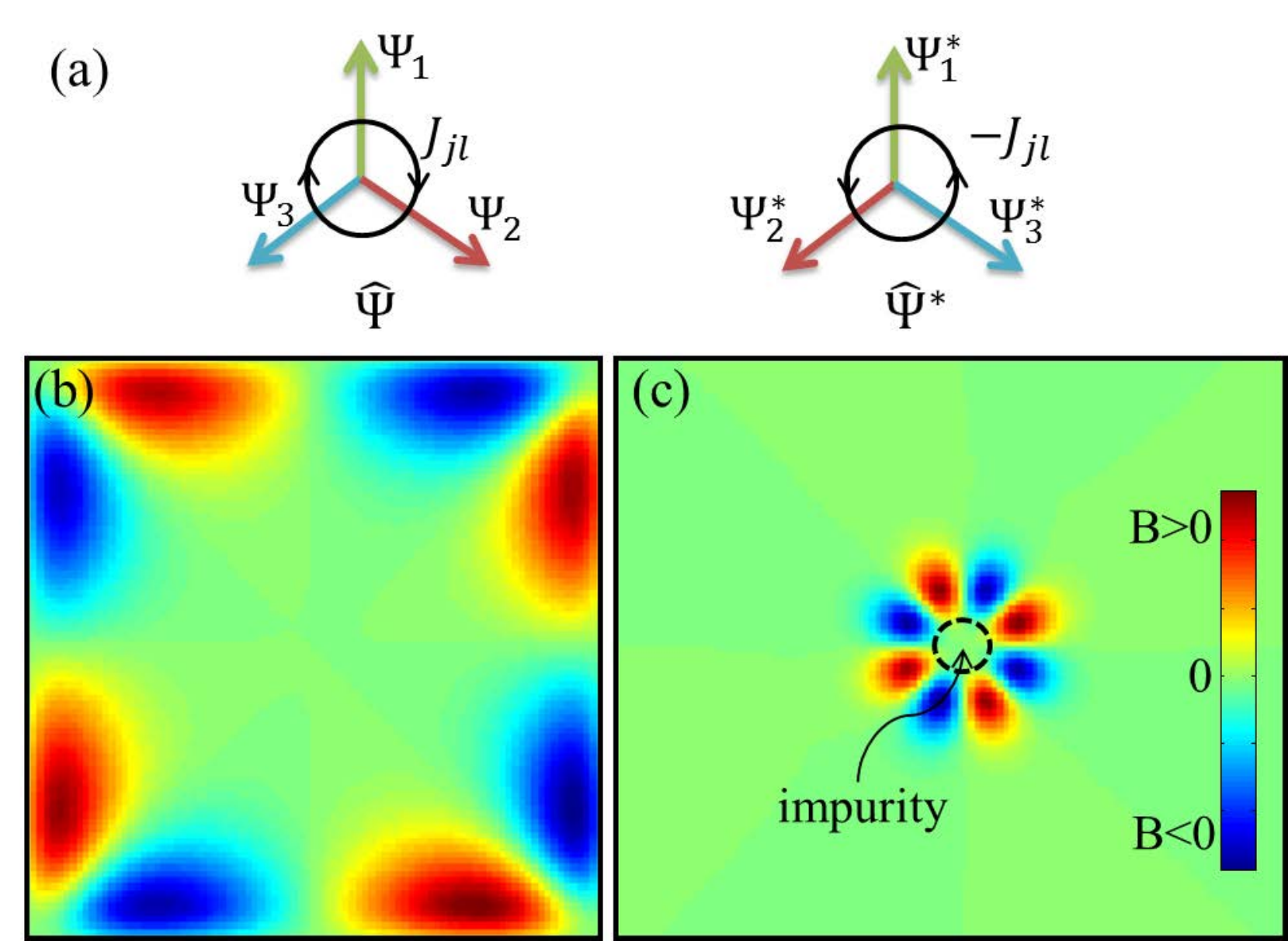,width=\columnwidth} \caption{\label{gs_fig3}
		(color online). (a) Schematic views of circulating interband Josephson current in the band space for a three-band superconductor without time-reversal symmetry. (b) and (c): Numerical results of the profile of spontaneous magnetic fields for (b) a three-band superconductor without time-reversal symmetry in contact with a normal metal, and for (c) a three-band superconductor without time-reversal symmetry in the presence of an non-magnetic impurity. The parameters for (b) are $\alpha_j=0$, $\beta_j=m_j=1$, $\gamma_{12}=1$, $\gamma_{13}=1.2$ and $\gamma_{23}=1.5$, $p_{jj}=1$ and $p_{j\neq l}=\infty$; for (c), the same as (b) except for $\alpha_j=0.5$ inside the impurity area, and $p_{jl}=\infty$. (b) and (c) are from Ref. \cite{LinNJP2012}.}
\end{figure}

One characteristic consequence of time-reversal symmetry breaking in most systems is the appearance of spontaneous magnetic fields. For homogeneous multiband superconductors without time-reversal symmetry, there is interband Josephson current flowing between different bands $J_{jl}\propto \gamma_{jl}\sin(\phi_j-\phi_l)$ for the state $\hat{\Psi}\equiv (\Psi_1,\ \Psi_2,\ \cdots,\ \Psi_M)$, while $J_{jl}\propto -\gamma_{jl}\sin(\phi_j-\phi_l)$ for the state $\hat{\Psi}^*\equiv (\Psi_1^*,\ \Psi_2^*,\ \cdots,\ \Psi_M^*)$, as sketched in Fig. \ref{gs_fig3} (a). In this sense, the ground is chiral given by the direction of the interband Josephson current. The interband Josephson current occurs in the band space and does not couple to gauge fields. Therefore the circulation of interband Josephson current does not generate spontaneous magnetic fields for homogeneous multiband superconductors without time-reversal symmetry. When inhomogeneities exist due to non-magnetic impurities, proximity effect at sample edges or local heating, spontaneous magnetic fields may appear. This is due to the fact in the time-reversal symmetry broken state, the spatial variation of the amplitude is coupled with the spatial variation of phase of superconducting order parameters. There is induced supercurrent near the inhomogeneities where the amplitudes of the superconducting order parameters are modified, and magnetic fields are generated. 

We consider the proximity effect between a three-band superconductor without time-reversal symmetry and a normal metal by numerical simulations of the time-dependent Ginzburg-Landau equations in Eqs. \eqref{meq2} and \eqref{meq3}. The boundary condition at the superconductor-normal metal interface is given by \cite{TinkhamBook,Brinkman04}
\begin{equation}\label{gseqg7}
\left(-i \nabla -\frac{2\pi}{\Phi_0}\bf{A}\right)\Psi _j=i\sum_k \frac{\Psi _k}{p_{jk}},
\end{equation}
where the diagonal coefficient $j=k$ accounts for the suppression of superconductivity due to the leakage of Cooper pairs at the interface, while the off-diagonal coefficient $p_{jk}$ with $j\neq k$ represents the interband coupling. As displayed in Fig. \ref{gs_fig3} (b), spontaneous magnetic fields with a total flux equal to zero are produced at the corners of the superconductors due to the proximity effect. \cite{LinNJP2012} We then investigate the effect of a single non-magnetic impurity. The impurity is introduced in simulations by modifying $\alpha_j$ locally. As shown in Fig. \ref{gs_fig3} (c), there are spontaneous magnetic fields alternating in space around the impurity. Therefore the appearance of spontaneous magnetic fields near non-magnetic impurities or surfaces of superconductors due to the proximity effect can be used to detect the time-reversal symmetry breaking in multiband superconductors with frustrated interband couplings. 

The ground state in two-band isotropic $s$-wave superconductors is much simple. The phase difference between two gap functions $\Psi_1$ and $\Psi_2$ can be either $0$ or $\pi$, depending on the sign of interband coupling $g_{12}$ or $\gamma_{12}$. For an attractive interband coupling $\gamma_{12}<0$, there is no phase difference between $\Psi_1$ and $\Psi_2$; while for an repulsive interband coupling $\gamma_{12}>0$, the system favors $s\pm$ pair symmetry with a $\pi$ phase shift between $\Psi_1$ and $\Psi_2$. A typical dependence of the amplitude of the energy gap for different bands on temperature is shown in the inset of Fig. \ref{multigap_fig}, where all gaps vanish at the same $T_c$ due to the interband couplings. 

Recently, a general classification of the ground states for phase-frustrated multiband superconductors using a graph-theoretical approach was reported by Weston and Babaev. \cite{PhysRevB.88.214507}

So far we have adopted the mean-field approximation. The phase diagram for $s$-wave three-band superconductors with frustrated interband couplings was calculated beyond the mean-field approximation by Monte Carlo simulations. \cite{PhysRevB.88.220511,PhysRevB.89.104509,bojesen_strong_2014}  A novel phase with $U(1)$ symmetry but without $Z_2$ (time-reversal symmetry) symmetry was found. In the $U(1)$ and $Z_2$ symmetry broken phase, the proliferation of vortex and antivortex restores the $U(1)$ symmetry and the proliferation of phase soliton recovers the $Z_2$ symmetry. The former transition belong the $XY$ universality class and the latter belongs to the Ising universality class. It was found in certain parameter space that the energy cost for the vortex proliferation is lower than that for phase soliton proliferation.  In this case, the $U(1)$ symmetry is restored prior to $Z_2$ symmetry upon increasing temperatures. Therefore a new dissipative metallic phase with $U(1)$ symmetry but without time-reversal symmetry appears.

The multiband nature also has profound effects on the magnetic field-temperature $H$-$T$ phase diagram. One example is in the case for superconductors with a nonmonotonic inter-vortex interaction as will be discussed in Sec. \ref{Sec5F}. The upper critical field $H_{c2}(T)$ as a function of $T$ is particularly interesting from the experimental point of view. The dependence $H_{c2}(T)$ for multiband superconductors differs from the single-band case. \cite{askerzade_ginzburg-landau_2002,Gurevich03b,PhysRevB.82.184504,gurevich_iron-based_2011} On the other hand, one can extract microscopic parameters by fitting the measured  $H_{c2}(T)$ to a theoretical model.

\subsection{Material realizations of multiband superconductivity}\label{Sec2D}
Most superconductors have multiple Fermi surfaces, where electrons/holes form superconducting condensate below $T_c$. Therefore multiband superconductors are ubiquitous and strictly speaking, most superconductors can be labeled as multiband superconductors. However in most cases, superconductivity in these superconductors is dominant by one band and the superconductor behaves as a single-band superconductor. In this subsection, we will present several typical examples of multiband superconductors. The list nevertheless is incomplete, see Ref. \cite{zehetmayer_review_2013} for more discussions. 

Some binary compounds were found to exhibit prominent multiband superconductivity long time ago, such as $\mathrm{NbSe_2}$ \cite{yokoya_fermi_2001}, $\mathrm{V_3Si}$ \cite{nefyodov_microwave_2005},  $\mathrm{ZrB_{12}}$ \cite{PhysRevB.73.094510}. It was found from the microwave surface impedance and complex conductivity measurements that interband coupling for $\mathrm{V_3Si}$ is extremely weak \cite{yokoya_fermi_2001}, and $\mathrm{V_3Si}$ could be served as a playground to observe the decoupling of phases of superconducting order parameters discussed below.  The revival of the research on multiband superconductivity more or less can be attributed to the discovery of $\mathrm{MgB_2}$ with $T_c\approx 39$ K.  \cite{Nagamatsu01} Superconductivity in $\mathrm{MgB_2}$ is mediated by phonons. $\mathrm{MgB_2}$  is well characterized after intensive studies in the past decade. \cite{Xi08, Xi09} Most of its superconducting properties can be described by a two-band $s$-wave superconductor model. The energy gap for the $\sigma$ band is about $\Delta_\sigma=5.5-6.5$ meV and for the $\pi$ band is $\Delta_\pi=1.5-2.2$ meV. \cite{PhysRevLett.87.137005, PhysRevLett.89.187002, PhysRevLett.87.177006, souma_origin_2003} The interband coupling matrix $\hat{V}$ has been obtained using the first-principle calculations. \cite{PhysRevLett.87.087005,choi_origin_2002} The reported interband coupling ranges from weak to intermediate coupling.  The phases of superconducting order parameters in the two bands are the same, which requires that $\gamma_{12}<0$ in Eq. \eqref{meq1} and $g_{12}<0$ in Eq. \eqref{gseq2}.

The discovery of iron-based superconductors \cite{Kamihara08} attracts growing interests in the study of multiband superconductivity. There are large families of iron-based superconductors and their Fermi surface topology and pairing symmetry vary, see Refs. \cite{Ishida09,Paglione10,Johnston10,Wen11,Wang11,Hirschfeld11,Stewart11,RevModPhys.85.849} for a review. Most of them have five bands that contribute to superconductivity \cite{Ding08}. A simplified two-band model has been proposed to account for superconductivity in these materials. \cite{PhysRevB.77.220503} Theories predict that the phases of superconducting order parameters change sign between bands with a full gap in each band, and the pairing symmetry is denoted $s\pm$. \cite{Mazin08,Kuroki08} Many experimental evidences supports the $s\pm$ pairing symmetry. This pair symmetry can be modeled by the simplified models in Eqs.  \eqref{meq1} and  \eqref{meq4} with $\gamma_{12}>0$ in Eq. \eqref{meq1} and $g_{12}>0$ in Eq. \eqref{gseq2}.

The compound $\mathrm{Ba_{1-x}K_xFe_2As_2}$ has attracted lots of attention recently. It was revealed by various measurements near the optimal doping $x=0.4$ that the superconducting gaps at the two $\Gamma$-centered hole pockets are fully gapped and have the same sign. \cite{Ding08,PhysRevB.83.020501,PhysRevLett.102.187005,PhysRevB.80.140503,christianson_unconventional_2008,reid_d-wave_2012} The gap at the electron pockets has a $\pi$ phase shift with respect to the gaps at hole pockets. This pairing symmetry is denoted as $s++$. At $x=1$, it was found from ARPES measurements that only hole pockets exist. \cite{PhysRevLett.103.047002,okazaki_octet-line_2012} Both the $d$-wave pairing and $s$-wave pairing were proposed for the $x=1$ case. \cite{PhysRevB.84.144514,PhysRevB.80.180505,PhysRevLett.107.117001,PhysRevB.84.224505,PhysRevB.85.014511} For the $s$-wave pairing, the energy gaps are largest at the two $\Gamma$-centered hole pockets and they have a $\pi$ phase difference. We denote this pairing symmetry as $s\pm$. The existing experimental data either favor the $d$-wave pairing or $s$-wave pairing. \cite{okazaki_octet-line_2012,PhysRevLett.109.087001,PhysRevB.89.064510,PhysRevB.87.180507} Maiti and Chubukov assumed the $s$-wave pairing for the $x=1$ case. \cite{Maiti2013} Then the pairing symmetry of $\mathrm{Ba_{1-x}K_xFe_2As_2}$ changes from $s++$ to $s\pm$ when $x$ is increased. This transition is possible through an intermediate $s+i s$ state which breaks the time-reversal symmetry. To describe this transition, a three-band Hamiltonian with frustrated interband coupling is needed. They solved the three-band Hamiltonian and found a phase diagram similar to that in Fig. ~\ref{gs_fig2}. Thus  $\mathrm{Ba_{1-x}K_xFe_2As_2}$ is a promising playground to test the time-reversal symmetry broken state and the related novel physics.

Several heavy fermion superconductors were shown to exhibit multiband superconductivity, such as $\mathrm{UNi_2Al_3}$ \cite{PhysRevLett.93.097001}, $\mathrm{PrOs_4Sb_{12}}$ \cite{PhysRevLett.95.107004}, $\mathrm{URu_2Si_2}$ \cite{PhysRevLett.99.116402}, $\mathrm{CePt_3Si}$ \cite{mukuda_multiband_2009}. The recently discovered $\mathrm{BiS_2}$ based superconductors, \cite{mizuguchi_superconductivity_2012} such as $\mathrm{LaO_{1-x}F_xBiS_2}$ were also revealed to exhibit multiband characteristics. 

Multiband superconductivity may also exist in the proposed liquid hydrogen under high pressure, where both protons and electrons contribute to superconductivity \cite{Ashcroft68,Jaffe81}. In this case the interband Josephson tunneling is absent, $\gamma_{12}=0$. Moreover there is a large disparity between the electron and proton superconducting condensate due to the huge mass difference. Methods to identify this hypothetical novel metallic superfluid phase are proposed in Ref. \cite{PhysRevLett.95.105301} and tested numerically in Ref. \cite{PhysRevLett.95.135301}.

\section{Collective mode: the Leggett mode}\label{Sec3}
Having determined the ground states of multiband superconductors, in this section we will investigate the collective excitations in the ground state. We will first present the basic concept of the Leggett mode and give a phenomenological description based on the phase of superconducting order parameters or the number of Cooper pairs. Then we will provide a microscopic derivation of the Leggett mode in a two-band superconductor. In multiband superconductors undergoing time reversal symmetry breaking phase transition, we will show the existence of a gapless Leggett mode at the transition point. At the end of this section, we will discuss the detection of the Leggett mode by Raman spectroscopy and review the experimental observations of the Leggett mode in $\mathrm{MgB_2}$. Possible observation of the Leggett mode by measurements of thermodynamical quantities will also be discussed.

\subsection{Basic concept}\label{Sec3A}
In multiband superconductors, electrons/holes in different bands form superconducting condensate, which can be described by a complex gap function $\Psi_j=\Delta_j\exp(i\phi_j)$. Because electron/hole can hop between different bands, the number of Cooper pairs in different bands fluctuates. The collective oscillation of the Cooper pairs between different bands was first discussed by Leggett in 1966, now known as the Leggett mode. \cite{Leggett66} The number of Cooper pairs $n_c$ and phase $\phi$ are conjugate variables satisfying the uncertain relation $\Delta n_c\Delta \phi\ge 1$. The collective oscillation of Cooper pairs between different bands therefore can be described in term of the superconducting phase difference between different bands $\phi_j-\phi_l$. The dispersion for the Leggett mode can be obtained using a phenomenological approach where each band is described by the Lagrangian \cite{SimonsQFT},
\begin{equation}\label{blmeq1}
\mathcal{L}_{\phi_j}={\hbar ^2}{N_j}(0){({\partial _t}{\phi _j}  )^2} -\frac{\hbar^2{\Psi_{sj}^2}}{{2{m_j}}}{(\nabla {\phi _j})^2},
\end{equation}
where $\Psi_{sj}^2$ is the superfluid density with dimension 1/volume. Here we consider a two-band superconductor without coupling to electromagnetic fields. The interband Josephson coupling is
\begin{equation}\label{blmeq2}
{{\mathcal L}_J} =  - 2{\gamma _{12}}{\Psi_{s1}}{\Psi_{s2}}\cos ({\phi _1} - {\phi _2}).
\end{equation}
The collective modes for Eqs. \eqref{blmeq1} and \eqref{blmeq2} in the long wavelength limit are
\begin{equation}\label{blmeq3}
\Omega _{\text{BAG}}^2 = \frac{1}{{{N_1}(0) + {N_2}(0)}}\left(\frac{{\Psi_{s1}^2}}{{2{m_1}}} + \frac{{\Psi_{s2}^2}}{{2{m_2}}}\right)q^2,
\end{equation}
\begin{align}\label{blmeq4}
\Omega _{\text{L}}^2 = \frac{{[{N_1}(0) + {N_2}(0)]|{\gamma _{12}}|{\Psi_{s1}}{\Psi_{s2}}}}{{{N_1}(0){N_2}(0){\hbar ^2}}} \nonumber \\
+ \frac{1}{{{N_1}(0) + {N_2}(0)}}\left(\frac{{\Psi_{s1}^2{N_2}(0)}}{{2{m_1}{N_1}(0)}} + \frac{{\Psi_{s2}^2{N_1}(0)}}{{2{m_2}{N_2}(0)}}\right) q^2.
\end{align}
The first mode $\Omega _{\text{BAG}}$ is the gapless Bogoliubov-Anderson-Goldstone mode associated with the in phase oscillations of phases $\phi_1$ and $\phi_2$. \cite{Anderson58,Bogoliubov59} The second mode is the Leggett mode corresponding to the out of phase oscillation of phase $\phi_1-\phi_2$. \cite{Leggett66} The Leggett mode $\Omega _{\text{L}}$ is gapped with a gap being proportional to the interband coupling. 

One can also adopt a hydrodynamic description based on the number of Cooper pairs assuming that the total number of Cooper pairs is conserved, i.e. $\sum_j n_{cj}=\mathrm{constant}$. \cite{Leggett66} Accounting for the tunneling of Cooper pairs between bands, we can write a set of equations to describe $n_{cj}$
\begin{equation}\label{blmeq5}
\partial_t^2 n_{c1}=\frac{v_1^2}{3}\nabla^2 n_{c1}+T_{12}[N_1(0) n_{c2}-N_2(0) n_{c1}],
\end{equation}
\begin{equation}\label{blmeq6}
\partial_t^2 n_{c2}=\frac{v_2^2}{3}\nabla^2 n_{c2}+T_{12}[N_2(0) n_{c1}-N_1(0) n_{c2}],
\end{equation}
where $T_{12}>0$ is the tunneling coefficient. The dispersion for the collective modes in Eqs. \eqref{blmeq5} and \eqref{blmeq6} in the long wavelength limit is
\begin{equation}\label{blmeq7}
\Omega _{\text{BAG}}^2 =\frac{1}{3} \frac{ v_1^2 N_1(0)+v_2^2 N_2(0)}{N_1(0)+N_2(0)}q^2,
\end{equation}
\begin{equation}\label{blmeq8}
\Omega _{\text{L}}^2 =[N_1(0)+N_2(0)]T_{12}+\frac{1}{3} \frac{v_1^2 N_2(0)+v_2^2 N_1(0)}{N_1(0)+N_2(0)} q^2
\end{equation}
Equations \eqref{blmeq7} and \eqref{blmeq8} have similar forms to those in Eqs. \eqref{blmeq3} and \eqref{blmeq4}. If we set $T_{12}=\mathcal{D}_1/[2\hbar^2N_1(0)N_2(0)]$,  Eqs. \eqref{blmeq7} and \eqref{blmeq8} coincide with these derived from a microscopic theory, see Eqs. \eqref{lmeq6} and \eqref{lmeq7} in Sec. \ref{Sec3B}.

The tunneling of Cooper pairs between different bands shares certain similarities to that in a Josephson junction. The dispersion of the Leggett mode has the same form as the collective excitation in a Josephson junction. In Josephson junctions, the collective mode couples directly to gauge fields and becomes a plasma mode, known as the Josephson plasma. \cite{BaroneBook} In contrast, the Leggett mode does not respond to gauge fields and is a neutral mode.

\subsection{Microscopic description of the Leggett mode in two-band superconductors}\label{Sec3B}
In this subsection, we present a microscopic description of the Leggett mode in a two-band superconductor using a field theoretical approach. \cite{Sharapov02} We use the two-band BCS Hamiltonian in Eq. \eqref{meq4} and consider the $T=0$ case. The ground state $\Psi_j=\Delta_j\exp(i\phi_j)$ is determined by the two-band version of Eqs. \eqref{gseq4} and \eqref{gseq5}. As we are interested in the low-energy phase fluctuations, we can treat the amplitude of the energy gaps as fixed. We expand the action in Eq. \eqref{gseq2} up to the second order in the phase fluctuations. This can be done by the following gauge transformation to separate the phase and amplitude of gaps \cite{Loktev01,Sharapov02}
\begin{equation}\label{lmeq1}
{\Psi _j} \rightarrow {\Delta _j}{e^{i{\vartheta _j}}}{\text{\ and\ }}{\hat{\Theta} _j}(\tau ,r) \rightarrow \left( {\begin{array}{*{20}{c}}
	{{e^{i{\vartheta _j}/2}}}&0\\
	0&{{e^{ - i{\vartheta _j}/2}}}
	\end{array}} \right){\hat{\Theta} _j}(\tau ,r).
\end{equation}
We then obtain the action for the phase fluctuations
\begin{equation}\label{lmeq2}
S = \int d \tau d^3r\sum_{j,l} {{\Delta _l}{g_{{{lj}}}}{\Delta _j}e^{i( {\vartheta _l - \vartheta _j})}} - \sum\limits_j {{\text{Tr}}} \left[ {\ln \left( {{\cal G}_j^{ - 1} - {\Sigma _j}} \right)} \right]
\end{equation}
where 
\[
\Sigma_j={ - \frac{{{\hbar
				^2}}}{{2{m_j}}}}\left({\frac{i}{2}{\nabla ^2}{\vartheta _j}  + i\nabla
	{\vartheta _j}\nabla }\right)\sigma_0+\left[{i\frac{{{\partial _\tau }{\vartheta
				_j}}}{2}}+{\frac{{{\hbar ^2}}}{{8{m_j}}}}{{{\left( {\nabla {\vartheta
					_j}} \right)}^2}}\right]\sigma_3,
\]
with $\sigma_j$ being the Pauli matrices and $\sigma_0$ the unit matrix \cite{Palo99,Benfatto04}. From this action, one can obtain the time-dependent nonlinear Schr\"{o}dinger Lagrangian for the phase fluctuations \cite{Aitchison95,Aitchison00}. In $S$ in Eq. \eqref{lmeq2} the most important term for the Leggett’s mode is the Josephson coupling, $\Delta_l\Delta_j\cos(\vartheta_l-\vartheta_j)$ which explicitly depends on the relative phase of two condensates $\vartheta_l-\vartheta_j$. Considering small phase fluctuations around the saddle point $\theta_j=\vartheta_j-\phi_j$ and expanding $S$ up to the second order in $\theta_j$, we have
\begin{equation}\label{lmeq3}
S_\theta\left[ {{\theta _j}} \right] = \frac{1}{8}\sum\limits_l \int{{d^3}}q \hat{\theta}({-\Omega _l},-q)^T {\bf{M}}
\hat{\theta}({\Omega _l},q)
\end{equation}
with $\hat{\theta}({\Omega _l},q)\equiv [\theta _1({\Omega _l},q),\ \theta _2({\Omega _l},q)]^T$ and
\begin{equation}\label{lmeq4}
{\bf{M}} = \left( {\begin{array}{*{20}{c}}
	{ P_1 - {\mathcal{D}_1}}&{{\mathcal{D}_1}}\\
	{{\mathcal{D}_1}}&{ P_2- {\mathcal{D}_1} }
	\end{array}} \right)
\end{equation}
with ${\mathcal{D}_1} = {8}|g_{12}|\Delta_1\Delta_2$. Here $\Omega_l=2l\pi k_B T $ with $k_B$ the Boltzmann constant and the excitations are bosons. In the hydrodynamic limit at $T=0$, the dissipation is absent and
\begin{equation}\label{lmeq5}
P_j={ 2N_j(0)\left(-\Omega^2 +  {\frac{1}{3}v_j^2} {q^2}\right)},
\end{equation}
after the analytical continuation $i\Omega_l\leftarrow\Omega+i 0^+$, where $v_j$ is the Fermi velocity. In the calculation of $P_j$, we have used the random phase approximation
\[
\text{Tr}\left[\ln  \left({\mathcal{G}_j^{-1}}-\Sigma_j \right)\right]={\text{Tr}}\ln{\mathcal{G}_j^{-1}}-\text{Tr} \sum _{n=1} \frac{(\mathcal{G}_j \Sigma_j )^n}{n},
\]
to the second order $n=2$. For details on the evaluation of $\text{Tr}(\mathcal{G}_j \Sigma_j )^n$, please refer to Ref. \cite{Sharapov02}.

From $\text{Det}[\mathbf{M}] =0$, we obtain the dispersion relations in the long wavelength limit $q\ll 1$
\begin{eqnarray}\label{lmeq6}
\Omega _{\text{BAG}}^2 &=& \frac{1}{3} \frac{ v_1^2 N_1(0)+v_2^2 N_2(0)}{N_1(0)+N_2(0)}q^2, \\
\Omega _{\text{L}}^2 &=& \frac{N_1(0)+N_2(0)}{2\hbar^2 N_1(0) N_2(0)}\mathcal{D}_1+\frac{1}{3} \frac{v_1^2 N_2(0)+v_2^2 N_1(0)}{N_1(0)+N_2(0)} q^2. \label{lmeq7}
\end{eqnarray}
The first mode $\Omega _{\text{BAG}}$ is the gapless Bogoliubov-Anderson-Goldstone boson \cite{Anderson58,Bogoliubov59}. The second  mode $\Omega _{\text{L}}$ is the neutral gapped Leggett mode. \cite{Leggett66} The Leggett mode in two-band superconductors does not depend on the sign of the interband Josephson coupling $g_{12}$. The modes in Eqs. \eqref{blmeq3} and \eqref{blmeq4} obtained from a phenomenological Lagrangian have similar forms as those in Eqs. \eqref{lmeq6} and \eqref{lmeq7} if one identifies $\Psi_{sj}^2\sim m_j N_j(0) v_j^2$. To compare quantitatively, one needs to relate the parameters in Eqs. \eqref{blmeq1} and \eqref{blmeq2} to those in microscopic theories.  When coupled to gauge field $\mathbf{A}$, the Bogoliubov-Anderson-Goldstone mode gains a mass according to the Anderson-Higgs mechanism. \cite{PhysRev.130.439,PhysRevLett.13.508} In contrast, the Leggett mode does not couple to the gauge field $\mathbf{A}$.

We have neglected the coulomb repulsion in the above derivation. The effects of Coulomb interaction was studied by Leggett \cite{Leggett66} and by Sharapov \emph{et al.} \cite{Sharapov02}. The coulomb interaction does not change the gap of the Leggett mode, but modifies its velocity. 

Recently the Leggett mode in iron-based superconductors was considered in Refs. \cite{Burnell10}. Using a strong-coupling two-orbital model that is relevant for iron-based superconductors, it was shown that the Leggett mode lies below the two-particle continuum in certain parameter space. This could facilitate the experimental observation of the Leggett mode because the damping is weak. Meanwhile it is possible to detect the pairing symmetry for the iron-based superconductors by utilizing the Leggett mode, because the dispersion of the Leggett mode depends on the pairing symmetry. \cite{Burnell10} Ota \emph{et al.} studied the Leggett mode in three-band superconductors with time reversal symmetry. \cite{Ota11} They found that the gap of the Leggett mode is reduced when the Josephson coupling between different bands cancels each other, but it is still larger than zero.

In the discussions as far we have neglected quasiparticles, which is valid when the gap of the Leggett mode lies below the superconducting gaps $\Delta_i$. When the gap of the Leggett mode is above one of the superconducting energy gap, the Leggett mode is damped by transferring energy into quasiparticles, a process called the Landau damping. When the damping is strong, the lifetime of the Leggett mode is short and the Leggett mode becomes ill-defined collective excitations. On the other hand, when the gap of the Leggett mode is below the superconducting energy gap, the damping due to quasiparticles is weak. The damping of the Leggett boson can also arise due to the interaction between the Leggett bosons when the amplitude of the Leggett mode is strong. The Leggett mode can lose energy to other bosonic degrees of freedom, such as phonons.

Here we have considered the Leggett modes in clean multiband superconductors. The collective modes in dirty multiband superconductors were investigated by Anishchanka \emph{et al.}, \cite{PhysRevB.76.104504} where the interplay between the Leggett mode and the Carlson-Goldman mode was studied. Finally an alternative derivation of the dispersion of the Leggett mode using the Ward-Takahashi identity was present in Ref. \cite{koyama_collective_2014}.

\subsection{The gapless Leggett mode in frustrated superconductors with time-reversal symmetry breaking}\label{Sec3C}

\begin{figure}[t]
	\psfig{figure=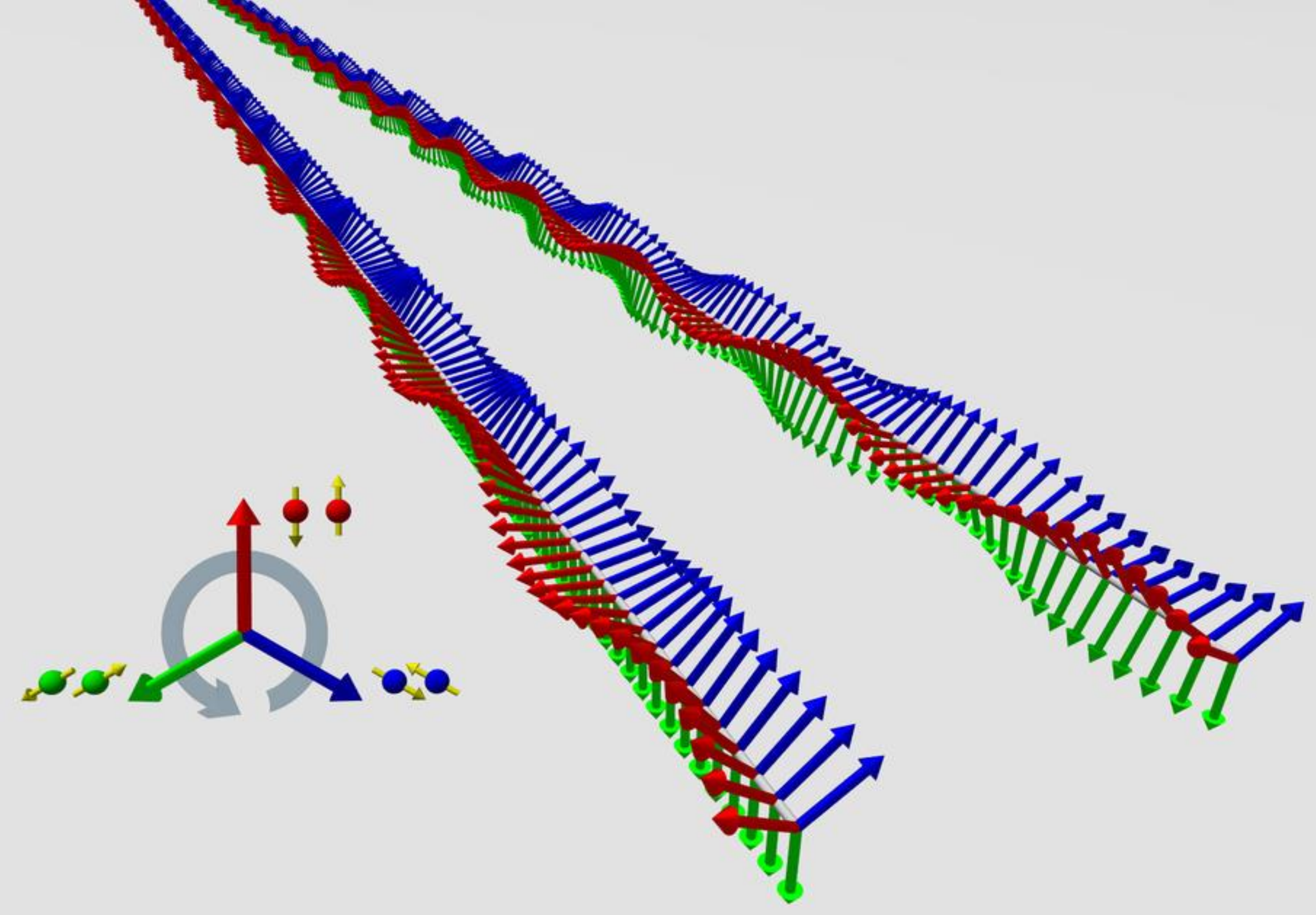,width=\columnwidth} \caption{\label{lm_fig1}
		(color online). Frustrated interband scatterings force Cooper pairs in different bands to carry different phases, which results in interband Josephson currents. There exist two dynamical modes associated with superconducting phases in three-band superconductors: the Leggett mode, where two phases oscillate out-of-phase while the third one stays unchanged, becoming gapless at the time-reversal symmetry breaking transition (left), and the Bogoliubov-Anderson-Goldstone mode, where all the three phases rotate in the same direction during the propagation of plasma wave in space (right). From Ref. \cite{Lin2012PRL}.}
\end{figure}

For two-band superconductors, the Leggett mode is always gapped with a gap value proportional to the interband Josephson coupling. This statement is valid for all $s$-wave multiband superconductor with time-reversal symmetry. As pointed out in Ref. \cite{Lin2012PRL}, one Leggett mode becomes gapless when a multiband superconductor undergoes time-reversal symmetry breaking transition because of the frustrated interband couplings. This is based on the observation that for any continuous phase transition, there always exists a soft mode at the transition point which restores the symmetry under consideration. This is can be illustrated with the following simple Lagrangian for a scalar field $\vartheta$
\begin{equation}\label{glmeq1}
\mathcal{L}_{\vartheta}=\frac{1}{2}(\partial_t\vartheta)^2-\frac{c_\vartheta^2}{2}(\nabla\vartheta)^2-\frac{\alpha}{2}\vartheta^2-\frac{\beta}{4}\vartheta^4,
\end{equation} 
with $\alpha=(T-T_t)/T_t$ with $T_t$ the transition temperature. In the symmetry broken phase $\vartheta_0=\sqrt{-\alpha/\beta}$, the dispersion of the collective excitation $\tilde{\vartheta}=\vartheta-\vartheta_0$ is
\begin{equation}\label{glmeq2}
\omega^2=c_\vartheta^2k^2-2\alpha.
\end{equation}
It is gapped below $T_t$ and the gap vanishes at $T_t$, where the symmetry is restored.

\begin{figure}[t]
	\psfig{figure=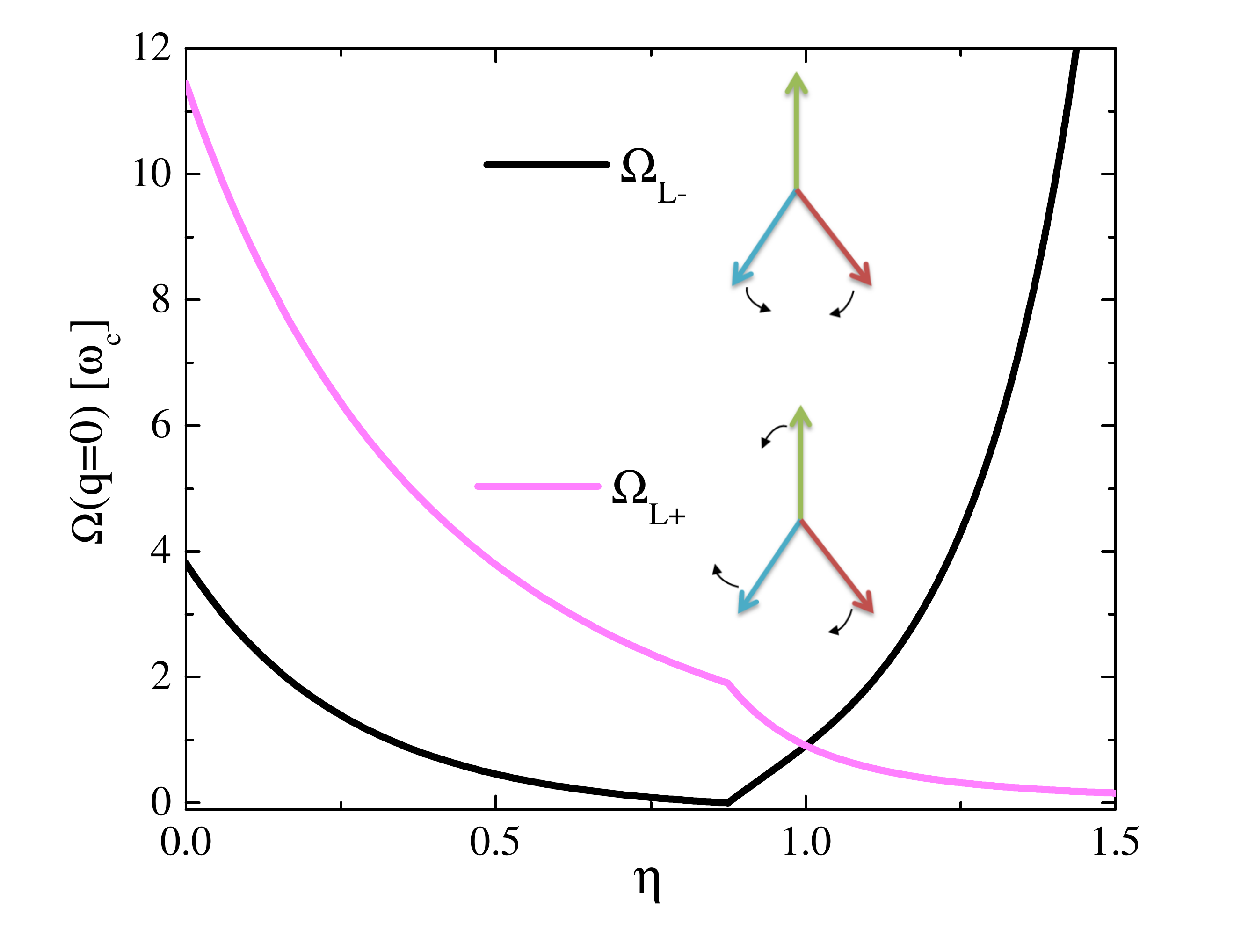,width=\columnwidth} \caption{\label{lm_fig2}
		(color online). Dependence of the gap of the Leggett modes on the interband coupling $\eta$. Here $N(0) V_0=0.5$ and $\alpha=2$, and the gaps are in units of $\omega_c$. From Ref. \cite{Lin2012PRL}.}
\end{figure}

In $s$-wave multiband superconductors experiencing time-reversal symmetry breaking transition, the corresponding soft mode is the Leggett mode. To demonstrate the existence of a gapless Leggett mode, we consider a three-band superconductor with a continuous time-reversal symmetry breaking phase transition. We study the Leggett mode in the state with time-reversal symmetry and calculate its gap as the system is tuned to the time-reversal symmetry breaking transition. In the state with time-reversal symmetry, the amplitude and phase fluctuations are decoupled; while in the state without time-reversal symmetry, the amplitude and phase fluctuations are coupled and one has to treat these fluctuations consistently, as done by Stanev \cite{Stanev11}. For simplicity, we again consider a symmetric interband coupling in Eq. \eqref{gseqg1}. Generalizing the calculations in Sec. \ref{Sec3B} to the three-band case, we obtain the action for the phase fluctuations
\begin{equation}\label{glmeq3}
S_\theta\left[ {{\theta _j}} \right] = \frac{1}{8}\sum\limits_l \int{{d^3}}q \hat{\theta}({-\Omega _l},-q)^T {\bf{M}}
\hat{\theta}({\Omega _l},q)
\end{equation}
with $\hat{\theta}({\Omega _l},q)\equiv [\theta _1({\Omega _l},q),\ \theta _2({\Omega _l},q),\ \theta _3({\Omega _l},q)]^T$ and  
\begin{equation}\label{glmeq4}
{\bf{M}} = \left( {\begin{array}{*{20}{c}}
	{ P_1 - 2{\mathcal{D}_1}}&{{\mathcal{D}_1}}&{{\mathcal{D}_1}}\\
	{{\mathcal{D}_1}}&{ P_2- {\mathcal{D}_1} - {\mathcal{D}_2}}&{{\mathcal{D}_2}}\\
	{{\mathcal{D}_1}}&{{\mathcal{D}_2}}&{ P_3 - {\mathcal{D}_1} - {\mathcal{D}_2}}
	\end{array}} \right)
\end{equation}
with ${\mathcal{D}_1} = {8}{\Delta _1}\Delta \cos{{\bar{\phi}}}/V_0$ and ${\mathcal{D}_2} = {{8\eta }}{\Delta ^2}\cos ( {2{\bar{\phi}}} )/V_0$ with $\bar{\phi}\equiv\phi_2-\phi_1=\phi_1-\phi_3$. $P_j$ is given in Eq. \eqref{lmeq5}. From $\text{Det}[\mathbf{M}] =0$, we obtain the dispersion relations for the phase fluctuations in the case of an identical density of state and Fermi velocity for the three bands, i.e. $N_j(0)=N(0)$ and $\mathbf{v}_j=\mathbf{v}_f$
\begin{eqnarray}\label{glmeq5}
\Omega _{\text{BAG}}^2 &=& \frac{1}{3}{q^2}v_f^2, \\
\Omega _{\text{L-}}^2 &=& {-\frac{ {\mathcal{D}_1} + {2\mathcal{D}_2}}{2\hbar^2N(0)} + \frac{1}{3}{q^2}v_f^2}, \\
\Omega _{\text{L+}}^2 &=&  { - \frac{{3\mathcal{D}_1}}{2\hbar^2N(0)} +  \frac{1}{3}{q^2}v_f^2} .
\end{eqnarray}
The first mode is the gapless Bogoliubov-Anderson-Goldstone mode, as displayed in the right of Fig. ~\ref{lm_fig1}. The second and third are the Leggett modes $\Omega_\text{L-}$ and $\Omega_\text{L+}$ in the three-band superconductors. Especially, as depicted in the left of Fig. ~\ref{lm_fig1}, the mode $\Omega_\text{L-}$ corresponds to the oscillations of the relative phase $\phi_{23}$ between the gaps of $\Psi_2$ and $\Psi_3$, and becomes gapless at the time-reversal symmetry breaking phase transition depicted in Fig. \ref{lm_fig2}. One may regard $\phi_{23}$ as the order parameter for the time-reversal symmetry: it increases continuously from $0$ at the transition, and therefore, the associated fluctuations become gapless at the transition. In stark contrast to the example in Eqs. \eqref{glmeq1} and \eqref{glmeq2} for conventional symmetry breaking phase transition, there exist stable gapped Leggett modes both before and after time-reversal symmetry breaking transition as shown in Fig. \ref{lm_fig2}, because the relative phase between different condensates is fixed in both the states with and without time-reversal symmetry.

The coupling between superconductors and the gauge field can be introduced into $S_{\theta}$ through the standard replacement $\nabla\theta_l\rightarrow\nabla\theta_l -2\pi \mathbf{A}/\Phi_0$. In this case, it is more convenient to write the phase fluctuations in terms of $\theta_1$, $\theta_{12}\equiv\theta_1-\theta_2$ and
$\theta_{13}\equiv\theta_1-\theta_3$. $\theta_1$ corresponds to the Bogoliubov-Anderson-Goldstone mode, while $\theta_{12}$ and $\theta_{13}$ describe the Leggett modes. The gauge fields couple with $\theta_1$ in the form $(\nabla\theta_1-2\pi \mathbf{A}/\Phi_0)$. After integrating out $\theta_1$, the gapless Bogoliubov-Anderson-Goldstone mode becomes the gapped plasma mode due to the Anderson-Higgs mechanism. \cite{PhysRev.130.439,PhysRevLett.13.508} In contrast, one of the Leggett modes remains gapless at the time-reversal symmetry breaking phase transition since the phase differences $\theta_{12}$ and $\theta_{13}$ do not couple with gauge field $\mathbf{A}$.

In the static region, the gapless Leggett mode manifests itself as a new divergent length scale. \cite{PhysRevB.84.134518, Hu11} When approaching the time-reversal symmetry breaking transition from the state without time-reversal symmetry, this new divergent length is associated with the spatial variation of the amplitude and phase of the superconducting order parameters. On the other hand, this new divergent length corresponds to the spatial variation of the phase of the superconducting order parameters if we approach the time-reversal symmetry breaking transition from the state with time-reversal symmetry.

Kobayashi \emph{et al.} studied the Leggett mode in multiband superconductors with frustrated interband coupling by mapping the multiband tight-binding Hamiltonian with pair-hopping interaction into a frustrated spin Hamiltonian. \cite{KobayashiPRB2013,Kobayashi2013} For three-band superconductors, they also found that the Leggett mode becomes gapless at the time-reversal symmetry breaking transition, consistent with the results in Ref. \cite{Lin2012PRL}. However for four-band superconductors, they revealed the existence of a gapless Leggett mode in a wider phase region, which is not limited to the time-reversal symmetry breaking transition point, because of the degeneracy in the ground states.  The gap value of Leggett modes can be used to characterize the time-reversal symmetry in multiband superconductors with frustrated interband coupling. It was suggested in Refs. \cite{Maiti2013, Marciani2013} to verify the possible time-reversal symmetry broken state $s+i s$ in the doped $\mathrm{Ba_{1-x}K_xFe_2As_2}$ by checking the existence of the gapless Leggett mode. 

The Leggett modes can couple to other neutral modes such as phonons. This coupling may modify the dispersion of the Leggett mode, such as the gap and group velocity. As far as the time-reversal symmetry breaking transition remains continuous, one of the Leggett modes is always gapless at the transition point. However it is also possible that the coupling with other neutral modes results in a first order phase transition, and this case requires a further study.  

\subsection{Experimental observation of the Leggett mode}\label{Sec3D}

In this subsection, we will discuss the possible experimental observation of the Leggett mode. One very useful technique is the Raman spectroscopy. We will first derive the Raman response due to the presence of the Leggett mode, taking the three-band case as an example. The experimental observation of the Leggett mode in $\mathrm{MgB_2}$ will be reviewed. In the second part, we will discuss the thermodynamical signatures of the Leggett mode. The possible observation of the gapless Leggett mode in iron-based superconductors will be discussed.

The Leggett modes can be probed indirectly by electric fields through the coupling to the charge density. Therefore the Leggett modes can be detected by the Raman spectroscopy through the inelastic scattering of photon with the charge density \cite{Abrikosov61,PhysRevB.29.4976,Devereaux95,Lee09}. The interaction between the incident photon and the charge density can be modeled as 
\begin{equation}\label{EOLMeq1}
\tilde \rho (\tau ,q) = \sum\limits_{j = 1}^3 {\sum\limits_{k,\sigma }{{\bar{\gamma}_j }} } (k)\psi _{{{j}}\sigma }^\dag \left( {\tau ,k +\frac{q}{2}} \right){\psi _{{{j}}\sigma}}\left( {\tau ,k -\frac{q}{2}} \right),
\end{equation}
Here $\bar{\gamma}_j(k)$ is the scattering coefficient, which is determined by the polarization of the incident and scattered photon. For the non-resonant electronic Raman scattering, the coefficients $\bar{\gamma}_{j }(k)$ reads
\begin{equation}
\bar{\gamma}_{j }(k)=\sum _{\alpha_R,\ \beta_R }e_{\alpha_R }^I \frac{\partial ^2\epsilon _j}{\partial k_{\alpha_R }\partial k_{\beta_R }}e_{\beta_R }^F,
\end{equation}
where $e_{\alpha_R }^I$ and $e_{\beta_R }^F$ are the polarization vectors of the incoming and outgoing photon respectively. $\alpha_R $, $\beta_R $ denote the coordinates perpendicular to the photon momentum and $\epsilon _j$ is the electron energy. The electronic Raman cross section is proportional to the dynamical structure factor $S_f(\omega ,q\to 0)$, which is related to the retarded correlation function $\chi _{\tilde{\rho } \tilde{\rho}}$ in the following way
\begin{equation}
S_f(\omega , q)=\left[1+n_B(\omega )\right]\left[-\frac{1}{\pi }\text{Im} \chi _{\tilde{\rho } \tilde{\rho }}^R(\omega , q) \right],
\end{equation}
where $n_B(\omega )$ is the Bose distribution function. We need to calculate the correlation function 
\[{{\chi _{\tilde \rho \tilde \rho}}(\tau  - \tau ',q) =  - \left\langle {{\text{T}_\tau }\tilde \rho(\tau ,q)\tilde \rho (\tau ', - q)} \right\rangle },
\] with $\text{T}_\tau$ being time-ordering operator. To compute $\chi _{\tilde\rho \tilde \rho }$ we add a new term in Eq. \eqref{lmeq2}, ${S_\mathcal{J}}(\tau ) =  - \sum\limits_q{{{\tilde \rho }}} (\tau, q)\mathcal{J}(\tau, -q )$ with $\mathcal{J}$ an external field. Then $\chi _{\tilde\rho \tilde \rho }$ can be computed by using the linear response theory with respect to $\mathcal{J}$. Here $S_\mathcal{J}$ in the Nambu space can be written as
\begin{equation}
S_\mathcal{J}= \sum _{j=1}^3\sum _{k,q}\psi _j^{\dagger }\left(\tau , k+\frac{q}{2}\right)\mathcal{G}_{\mathcal{J},
	j}^{-1}(\tau ; k,q)\psi _j\left(\tau , k-\frac{q}{2}\right),
\end{equation}
with ${\cal G}_{\mathcal{J},j}^{ - 1} =  - {\bar{\gamma}_j}(k)\mathcal{J}(\tau , - q){\sigma _3}$. The effective action with incident photons after integrating out the fermionic fields $\psi_j$ becomes
\begin{equation}\label{EOLMeq2}
S = \int d \tau d^3r\sum\limits_{l,j} {{\Psi _l}} {g_{{lj}}}\Psi _j^* - \sum\limits_l {{\rm{Tr}}} \ln \left( {{\cal G}_{\mathcal{J},l}^{ - 1} + {\cal G}_{l}^{ - 1}} \right).
\end{equation}
 We may neglect the fluctuations of the amplitude of the order parameters when the incident wave is weak. The fluctuations for the phase of superconducting order parameters acquire a form ${S } =S_{\theta}+S_{\mathcal{J}}$, with
\begin{equation}\label{EOLMeq3}
\begin{array}{l}
S_{\mathcal{J}}= \frac{1}{2}\sum\limits_{j,q}[\mathcal{J}(q){Z_j}(q)\theta_j^T( - q) \\
+ \mathcal{J}( - q){{\tilde Z}_j}( - q)\theta _j(q)+ \mathcal{J} (q)\mathcal{J}( - q)\Pi _{j,33}^{\gamma \gamma }],
\end{array}
\end{equation}
 and $S_{\theta}$ defined in Eq. (\ref{glmeq3}). Here
\[
{Z_j}(q) = \Psi_j[-\sin \phi _j\Pi _{j,31}^\gamma (q) - \cos\phi_j \Pi _{j,32}^\gamma (q) ],
\]
\[
{{\tilde Z}_j}(q) = \Psi_j[ -\sin \phi _j\Pi _{j,13}^\gamma (q) - \cos\phi _j \Pi _{j,23}^\gamma (q) ],
\]
with the polarization functions
\[
\Pi_{j,ml}^{\gamma \gamma}\equiv {1}/(L^3\beta )\sum _n\int d^3k\Upsilon_{j, ml}\bar{\gamma}_j\left(k+\frac{q}{2}\right)\bar{\gamma}_j\left(k-\frac{q}{2}\right)
\]
\[
\Pi_{j,ml}^{\gamma }\equiv{1}/(L^3\beta )\sum _n\int d^3k\Upsilon_{j, ml}\bar{\gamma}_j\left(k+\frac{q}{2}\right).
\]

\begin{figure}[t]
	\psfig{figure=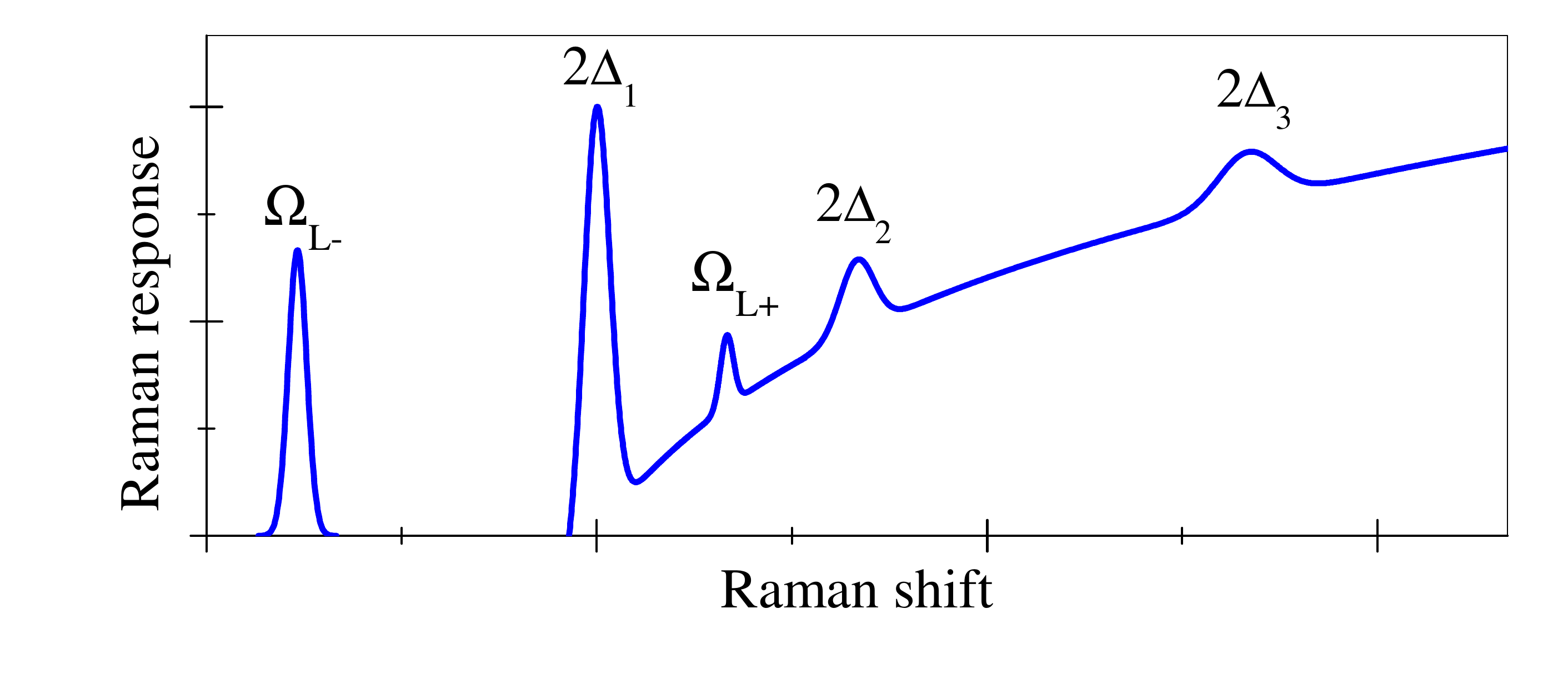,width=\columnwidth} \caption{\label{lm_fig3}
		(color online). Sketch of the Raman response in three-band superconductors with time-reversal symmetry breaking transition. The finite line-width of peaks is caused by damping of the Leggett boson. The background at energy larger than $2\Delta_1$ is due to the quasiparticle excitations. From Ref. \cite{Lin2012PRL}.}
\end{figure}

We then obtain the correlation function after integrating out the phase fluctuations $\theta_j$
\begin{equation}\label{EOLMeq4}
{{\chi _{\tilde \rho \tilde \rho }}(i\Omega ,q = 0)}=\sum\limits_j \left\{{\Pi _{j,33}^{\gamma \gamma }}  - {Z_j} [\mathbf{M} ^{ - 1}]_{jj}{ {{{\tilde Z}_j^T}} }\right\},
\end{equation}
with the matrix $\mathbf{M}$ being given in Eq. \eqref{glmeq4}. The first term accounts for the resonance with quasiparticles at $\hbar\Omega=2\Delta_j$. The second term gives the resonant scattering with the Leggett modes, as displayed in Fig. \ref{lm_fig3}. $\mathbf{M} ^{ - 1}$ becomes singular and gives delta peaks in the spectroscopy when the energy difference between the incident and scattered photons matches the gap of the Leggett modes. The delta-function peaks are rounded in reality by the damping effect arising from the interactions between the Leggett bosons when the oscillations of the Leggett modes become strong, or interaction with other bosonic degrees of freedom or thermal fluctuations, which are neglected in our treatment. The response of a genuinely gapless Leggett mode is hidden into the elastic scatterings. The gapless Leggett mode can be traced out clearly if one can tune the gap of the Leggett mode through changing $\eta$ systematically by electron/hole doping because the interband scattering is renormalized by the density of state as in Eq. (\ref{gseq4}).

\begin{figure*}[t]
	\psfig{figure=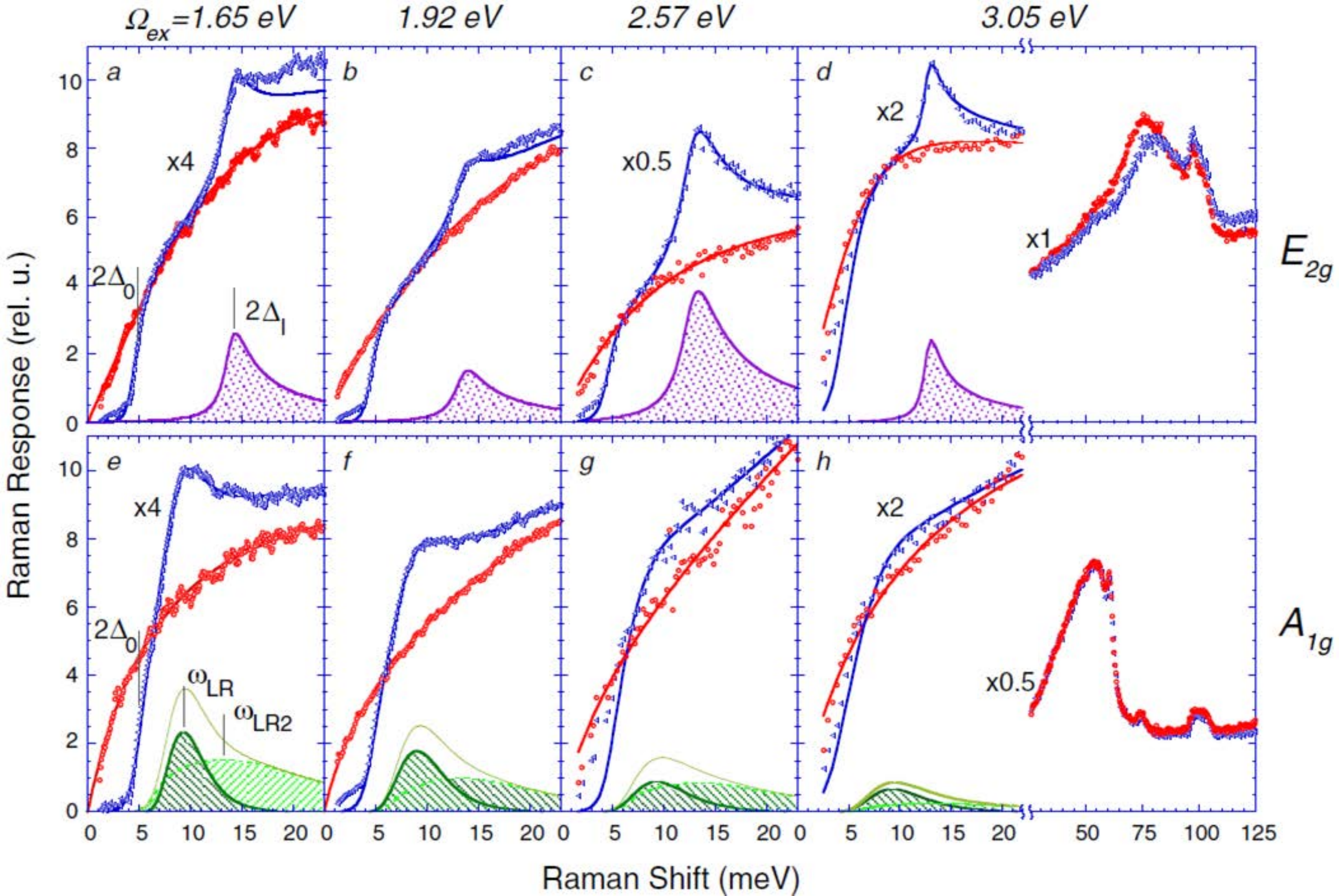,width=17cm} \caption{\label{lm_fig4}
		(color online). The Raman response spectra of an $\mathrm{MgB_2}$ crystal in the normal (red) and superconducting (blue) states for the $E_{2g}$ (top row) and $A_{1g}$ (bottom row) scattering channels. The columns are arranged in the order of increasing excitation energy $\Omega_{\mathrm{ex}}$. Solid lines are fits to the data points. The data in the superconducting state is decomposed into a sum of a gapped normal state continuum with temperature broadened $2\Delta_0=4.6$ meV gap cutoff, the superconducting coherence peak at $2\Delta_1=13.5$ meV (shaded in violet), and the collective modes at $\omega_{\mathrm{LR}}=9.4$ meV and $\omega_{\mathrm{LR2}}=13.2$ meV (shaded in dark and light green). The solid hairline is the sum of both modes. Panels (d) and (h) also show the high energy part of spectra for respective symmetries. The broad $E_{2g}$ band at 79 meV is the boron stretching mode, the only phonon that exhibits renormalization below $T_c$ \cite{PhysRevB.75.020509}. For the $A_{1g}$ channel the spectra are dominated by two-phonon scattering. From Ref. \cite{Blumberg07}.}
\end{figure*}

The Leggett modes have been observed in $\mathrm{MgB_2}$ using polarized Raman scattering measurements in the beautiful experiments by Blumberg \emph{et al.}. \cite{Blumberg07} The main results are summarized in Fig. ~\ref{lm_fig4}. The Raman response in the $E_{2g}$ channel starts to appear at a threshold Raman shift $4.6$ meV, which is assigned as the smaller superconducting energy gap $2\Delta_0$. Another superconducting coherent peak locates at $2\Delta_1=13.5$ meV, which is identified as the larger energy gap. The estimate of $\Delta_0$ and $\Delta_1$ is consistent with those obtained by one-electron spectroscopies. \cite{PhysRevLett.87.177006,souma_origin_2003} Particularly interesting observations are the resonant peaks in the $A_{1g}$ channel. The peak at $\omega_{\mathrm{LR}}=9.4$ meV is identified as a Leggett mode. The measured gap of the Leggett mode is consistent with the theoretical calculations. \cite{Sharapov02} The second resonance at $\omega_{\mathrm{LR2}}=13.2$ meV can be understood with a more elaborate theory by taking four Fermi surfaces of $\mathrm{MgB_2}$ into account. \cite{PhysRevB.82.014507} The observed Leggett mode lies between two superconducting gaps, thus it is short lived and decays into quasiparticle continuum in the band with a smaller superconducting energy gap.

The possible existence of the Leggett mode in $\mathrm{MgB_2}$ with an energy gap about $4$ meV was reported from point-contact and tunneling spectroscopy measurements. \cite{Ponomarev04}

The Leggett mode also manifests itself in several thermodynamic behaviors of $s$-wave superconductors, such as specific heat. For fully gapped superconductors, the quasiparticle contribution to the specific heat at $T\ll T_c$ depends exponentially on temperature $C_e\propto(\Delta/k_B T)^{3/2} \exp (-\Delta/k_B T)$. The contribution of the Leggett modes to the specific heat can be obtained analytically by treating the Leggett bosons as free quantum gas. For the gapped Leggett mode with a gap $\hbar\omega_0$, the specific heat due to the Leggett mode is $C_{L}\propto (\hbar\omega_0/k_B T)\exp(-\hbar\omega_0/k_B T)$ for $k_B T\ll \hbar \omega_0$, and is $C_{L}\propto (k_B T/\hbar\omega_0)^3$ for $k_B T\gg \hbar \omega_0$. For the gapless Leggett mode, the dependence of the specific heat originated from the Leggett mode on $T$ is $T^3$. Thus it is possible to detect the Leggett mode by measuring the electronic specific heat.

It was reported in several experiments a $T^3$ dependence of the electronic specific heat $C_v$ in iron-base superconductors after subtracting the residue electronic contribution (linear in $T$) and phonon contribution (also $T^3$ dependence). \cite{Kim10,Gofryk83,Zeng11} This $T^3$ dependence could also result from a line node in the gap function. This possibility was excluded from the measurements of the dependence of $C_v$ on magnetic fields, which suggests fully gapped order parameters. The authors of Ref. \cite{Gofryk83} suggested that the additional $T^3$ contribution might be originated from some bosonic modes. The existence of the gapless Leggett mode can explain these experimental observations naturally. Such an explanation is quite plausible regarding to the possible time-reversal symmetry breaking transition suggested for the iron-based superconductors. \cite{Maiti2013} The samples used in Refs. \cite{Kim10,Gofryk83,Zeng11} may well be in the vicinity of the time-reversal symmetry breaking transition and more measurements such as the Raman spectroscopy are much anticipated.

\section{Phase soliton}\label{Sec4}

Multiband superconductors with interband Josephson coupling allow for the phase kink or phase soliton excitation due to the degenerate energy minima in the Josephson coupling. For multiband superconductors without time-reversal symmetry, it supports another type of phase soliton between two symmetry-broken domains, similar to the domain walls in ferromagnets. In this section we will review these two types of phase solitons, and discuss the difference between them and their stability. We will also discuss methods to excite phase solitons. In the phase kink region, the time-reversal symmetry is violated locally and under certain conditions, spontaneous magnetic fields appear in the phase kink region. This can be served as experimental signatures of the existence of phase solitons. At the end of the section, the experimental detection of phase solitons will be reviewed.

\subsection{Phase soliton in multiband superconductors with time-reversal symmetry}\label{Sec4A}
 \begin{figure}[t]
 	\psfig{figure=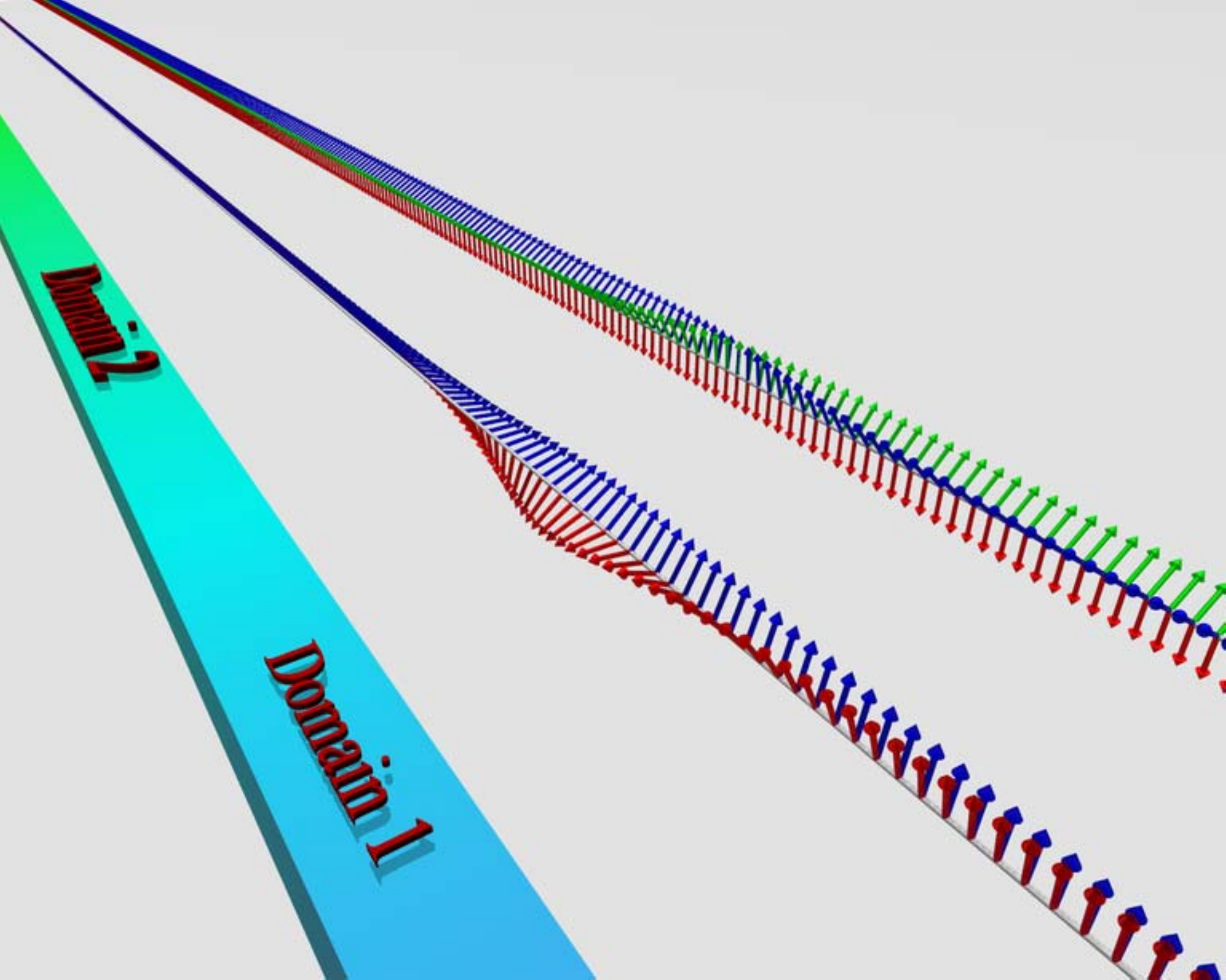,width=\columnwidth}
 	\caption{\label{kink_fig1}(color online) Domain structure in multiband superconductors due to the presence of phase solitons (left); a phase soliton in a two-band superconductor (middle), and that in a three-band superconductor with each domain corresponding to distinct time-reversal symmetry broken states (right). Arrows denote the phase of superconducting order parameter in different bands. From Ref. \cite{LinNJP2012}.}
 \end{figure}

For Josephson coupled multiband superconductors, the Josephson coupling $\gamma_{jl}\cos(\phi_j-\phi_l)$ has multiple degenerate energy minimal at $\phi_j-\phi_l=2n\pi$ for $\gamma_{jl}<0$. Therefore phase kink can be formed between these energy minimum, which corresponds to the homotopy class $\pi_0(S^0)$. The kink solution was first considered by Tanaka in 2001. \cite{Tanaka2001} To illustrate the kink solution or phase soliton in multiband superconductors, let us consider a two-band superconductor in one dimension with a free energy functional given by Eq. \eqref{meq1}. As will be discussed later, the phase soliton solution is only stable in one dimension. In one dimension we can take $\mathbf{A}=0$. Without loss of generality, we consider the $\gamma_{12}<0$ case and $\phi_1=\phi_2=\mathrm{constant}$ in the ground state. We also assume that the amplitude of the order parameters $|\Psi_j|\equiv \Psi_{j0}$ are constant in space, and the validity of this approximation will be clear later. Minimizing Eq. \eqref{meq1} with respect to $\phi_i$, we obtain
\begin{equation}\label{pkeq1}
\frac{{{\hbar ^2}}}{{{m_1}}}\Psi _{10}^2{\nabla ^2}{\phi _1} + \frac{{{\hbar ^2}}}{{{m_2}}}\Psi _{20}^2{\nabla ^2}{\phi _2} = 0,
\end{equation}
\begin{equation}\label{pkeq2}
-\lambda_k^2\partial_x\phi_{12}+\sin(\phi_{12})=0,
\end{equation}
where $\phi_{12}\equiv \phi_1-\phi_2$ and the kink width or soliton size $\lambda_k$ is
\begin{equation}\label{pkeq3}
\lambda_k^2=\frac{\hbar^2}{2|\gamma_{12}|\Psi_{10}\Psi_{20}}\left(\frac{m_1}{\Psi_{10}^2}+\frac{m_2}{\Psi_{20}^2} \right)^{-1}.
\end{equation}
Equation \eqref{pkeq2} is the well known sine-Gordon equation and it supports soliton solutions. One of such soliton solutions is $\phi_{12}=4\tan^{-1}[\exp(x/\lambda_k)]$. From Eq. \eqref{pkeq1} with the boundary condition $\partial_x\phi_1=\partial_x\phi_2=0$ and $\phi_1=\phi_2$ away from the soliton at $x=\pm\infty$, we obtain the profile of $\phi_1$ and $\phi_2$ for the soliton solution
\begin{equation}\label{pkeq4}
\phi_1=4\tan^{-1}[\exp(x/\lambda_k)] \bar{g}, \ \phi_2=4\tan^{-1}[\exp(x/\lambda_k)] (\bar{g}-1),
\end{equation} 
with $\bar{g} = \left(\frac{m_2\Psi_{10}^2}{m_1\Psi_{20}^2} + 1\right)^{-1}$. One typical configuration of the kink solution is schematically shown in the middle of Fig. ~\ref{kink_fig1}. The time reversal symmetry is broken locally in the kink region because $\phi_{12}\neq 0$ or $\phi_{12}\neq\pi$, while it is preserved in the region far away from the kink.

The approximation of constant $\Psi_{10}$ and $\Psi_{20}$ in space is valid when the kink width is much larger than the superconducting coherence length, $\lambda_k\gg\xi_i$. At low temperatures, this approximation is valid for a weak interband coupling $|\lambda_{12}|\ll |\alpha_i|$. However as temperature approaches $T_c$, $\lambda_k$ becomes temperature-independent while $\xi_i\propto 1/\sqrt{(T_c-T)/T_c}$ diverges. The constant $\Psi_{10}$ and $\Psi_{20}$ approximation is no longer valid, and superconductivity at the kink region is suppressed significantly.

The phase soliton in one dimensional wire does not carry magnetic flux. However if we wrap the wire into a ring, then the phase soliton has fractional magnetic flux. \cite{Tanaka2001} The supercurrent for a constant $|\Psi_j|$ in the ring is given by
\begin{equation}\label{pkeq5}
\mathbf{J}_s=\sum_{j}\frac{2e\hbar\Psi_{j0}^2}{m_j}\left(\nabla\phi_j-\frac{2\pi}{\Phi_0}\mathbf{A} \right).
\end{equation}
We integrate along a closed loop in the outer region of the ring where $\mathbf{J}_s=0$ because the magnetic field is fully screened for a ring width much larger than $\lambda$. Moreover the phase $\phi_i$ of superconducting order parameter can only change by $2n_i\pi$ if we move around the ring. The integration yields a total magnetic flux $\Phi$ enclosed by the ring 
\begin{equation}\label{pkeq6}
\Phi  = \frac{{|\Psi _{10}^2|{n_1}/{m_1} + |\Psi _{20}^2|{n_2}/{m_2}}}{{|\Psi _{10}^2|/{m_1} + |\Psi _{20}^2|/{m_2}}}{\Phi _0},
\end{equation}
where $n_j\equiv \oint \nabla\phi_j\cdot d\mathbf{l}/(2\pi)$ is the winding number. Here $\Phi$ is integer quantized only when $n_1=n_2$. The existence of a phase soliton requires that $n_1\neq n_2$, hence the phase soliton in a ring carries fractional quantum flux.

The discussions so far are based on the Ginzburg-Landau approach. Samokhin studied the phase soliton with a microscopic approach and calculated the quasiparticle spectrum in the presence of a phase soliton. \cite{PhysRevB.86.064513} He found the existence of quasiparticle bound states localized near the soliton, with energies being nonuniversal fractions of the bulk superconducting gaps. Such bound states can be measured in tunneling experiments. 

The kink solution or phase soliton also appears in Josephson junctions, where the gauge invariant phase difference is also governed by the sine-Gordon equation Eq. \eqref{pkeq2}. \cite{BaroneBook} In Josephson junctions, the phase difference is coupled with gauge fields, therefore a phase soliton carries $\Phi_0$ flux, and it can be created by applying magnetic fields or driven by currents. The motion of soliton is resonant with the Josephson plasma oscillation, which yields current steps for certain voltages, see Ref. \cite{LinReview2010} for a recent review. In contrast, the phase soliton in a two-band superconductor does not couple with gauge fields and it is neutral. Thus it does not respond to magnetic fields or currents. Nevertheless the phase soliton can be created  by an electric field in nonequilibrium, which will be discussed in Sec. \ref{Sec4D}.

\subsection{Phase soliton in multiband superconductors without time-reversal symmetry}\label{Sec4B}
In three or more bands superconductors with frustrated interband coupling, time reversal symmetry may be broken. In this case, we have two distinct domains with degenerate energy, which is very similar to domains in ferromagnets. In the time-reversal symmetry broken state, these two domains have order parameter $\hat{\Psi}=(\Psi_1, \Psi_2, \Psi_3, \cdots)$ or  $\hat{\Psi}^*=(\Psi_1^*, \Psi_2^*, \Psi_3^*, \cdots)$ with $\hat{\Psi}\neq \hat{\Psi}^* \exp(i\theta)$, i.e. one cannot obtain one domain from the other by global rotation of phase. Therefore there can be stable kink solution between two domains $\hat{\Psi}$ and $\hat{\Psi}^*$ in one dimension, \cite{tanaka_chiral_2010,LinNJP2012,PhysRevLett.107.197001} which is quite different from the kink solution discussed in multiband superconductors with time-reversal symmetry in Sec. \ref{Sec4A}. Note that in multiband superconductors without time-reversal symmetry, it still supports kink solution with one domain $\hat{\Psi}$ or $\hat{\Psi}^*$ similar to that in multiband superconductors with time-reversal symmetry. 

To illustrate the idea, we consider a minimal model with three identical bands $\alpha_{j}=\alpha$, $\gamma_{ij}=\gamma>0$, and $m_i=m$. In this case, time-reversal symmetry is violated and there are two degenerate ground states with $\hat\Psi=\Delta (1, e^{i2\pi/3}, e^{i4\pi/3})$ and $\hat\Psi^*=\Delta (1, e^{-i2\pi/3}, e^{-i4\pi/3})$, which are sketched in Fig. \ref{kink_fig1} (right). The phase kink is described by \cite{LinNJP2012}
\begin{equation}\label{pkeq7a}
\partial_x \phi_1=0,
\end{equation}
\begin{equation}\label{pkeq7b}
\partial_x (\phi_{12}+\phi_{13})=0,
\end{equation}
\begin{equation}\label{pkeq7c}
\frac{1}{2 m \gamma }\partial_x ^2\phi _{12}+\sin \phi _{12}+\sin \left(2\phi _{12}\right)=0,
\end{equation}
for constant amplitudes of order parameters valid at $\gamma\ll|\alpha|$. The same double sine-Gordon equation in Eq. \eqref{pkeq7c} was also derived by Yanagisawa \emph{et al.}. \cite{yanagisawa_vortices_2012}. The potential corresponding to Eq.~(\ref{pkeq7c}) is $V_p=\cos\phi_{12}+\cos(2\phi_{12})/2$, which has many degenerate energy minima at $\phi_{12,m}=\pm 2\pi/3+2n\pi$. A phase kink can be constructed between any pair of the energy minima with qualitatively the same physical properties and stability. Using the Bogomolny inequality \cite{Manton04}, we find a phase kink solution analytically
\begin{equation}\label{pkeq8}
\phi _{12}=2 \arctan  \left[\sqrt{3}\text{tanh}\left(-\frac{\sqrt{3 m\gamma }}{2}x\right)\right],
\end{equation}
with an energy
\begin{equation}\label{pkeq9}
\mathcal{E}_k=\frac{4}{3}\sqrt{m\gamma} \left(3 \sqrt{3}-\pi \right).
\end{equation}

\subsection{Stability of phase soliton}\label{Sec4C}
 \begin{figure*}[t]
 	\psfig{figure=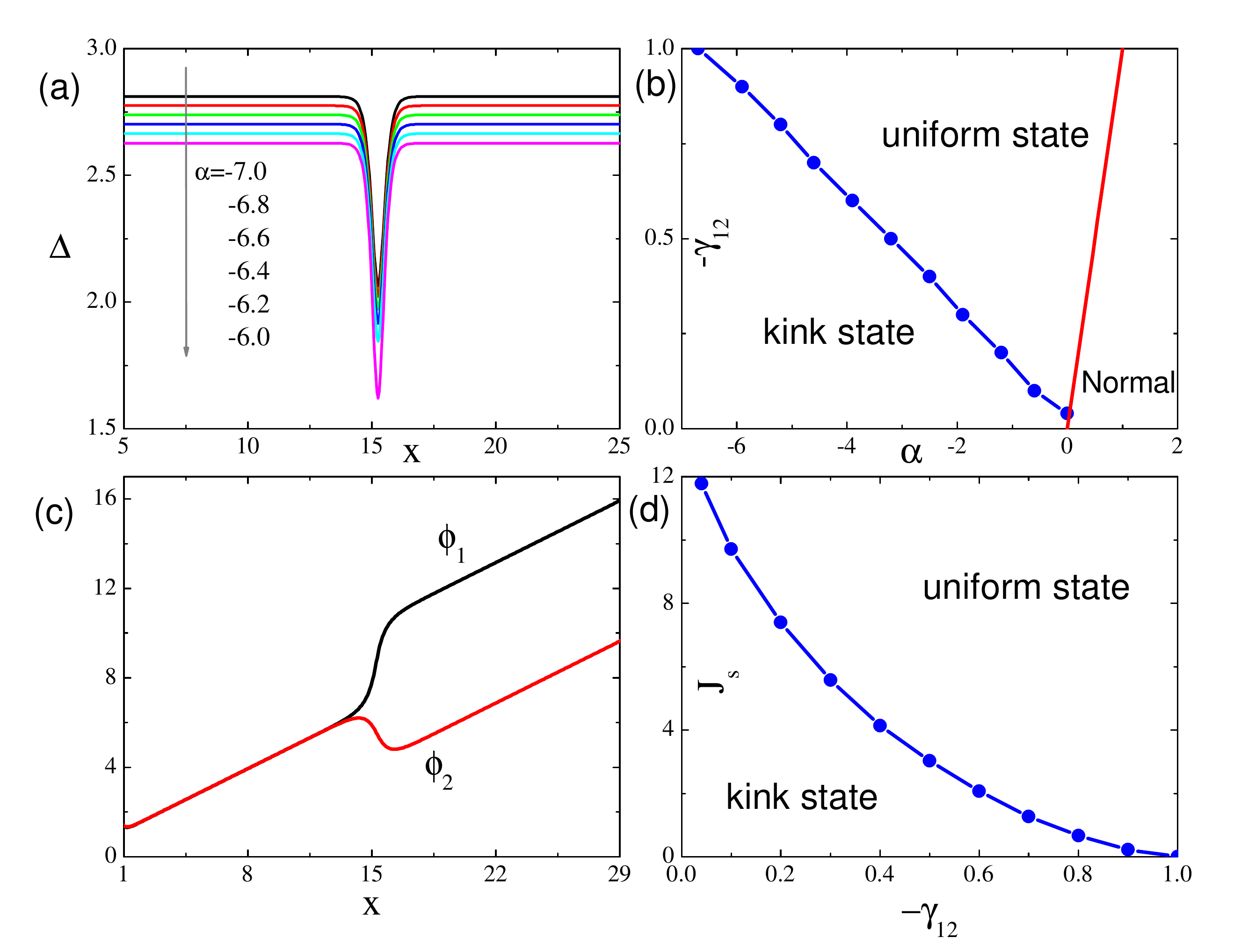,width=14cm}
 	\caption{\label{kink_fig2}(color online) (a) Suppression of superconductivity in the phase soliton when temperature denoted by $\alpha$ is increased in a two-band superconductor with $\gamma_{12}=-0.9$. (b) Phase diagram for the stability of phase kink.  (c) Structure of the phase kink in the presence of supercurrent. Here $\alpha_j=-7$, $\gamma_{12}=-0.5$ and the supercurrent $J_s=2.94$. (d) Stability of the phase kink upon current injection, with $\alpha_j=-7$, $\beta_j=1$ and $m_j=2$. From Ref. \cite{LinNJP2012}.}
 \end{figure*}
 In this subsection, we study the stability of the phase soliton in Eq. (\ref{pkeq2}) for multiband superconductors with time-reversal symmetry by accounting for the suppression of the amplitude of order parameters. The presence of the phase soliton suppresses the amplitudes of the order parameters, which depends on the ratio of the width of the phase kink $\lambda_k$ to coherence length $\xi_i$. The amplitudes of order parameters are greatly depressed when $T$ is tuned to $T_c$ because $\xi$ increases while $\lambda_k$ almost does not change. Thus at a threshold $T$, the phase soliton becomes unstable and the system evolves into a uniform state with $\nabla \phi_i=0$. The dynamics of the instability transition can be detected in experiments by measuring the voltage in the phase soliton because the change of the phases of superconducting order parameter results in a voltage according to the ac Josephson relation. This process can be regarded as a new type of phase slip, which differs from that in single-band superconductors driven by quantum or thermal fluctuations. \cite{TinkhamBook}

The above discussions are borne out by numerical calculations of the time-dependent Ginzburg-Landau in Eqs. \eqref{meq2} and \eqref{meq3} for a two-band superconductor with identical bands $\alpha_j=\alpha$, $\beta_j=\beta$, $m_j=m$ in one dimension. We initially put a phase soliton at the center of a superconducting wire. We then increases temperature by changing $\alpha$ and obtain a stable configuration of superconducting order parameters. As displayed in Fig. \ref{kink_fig2}(a), superconductivity is greatly suppressed in the phase kink region and it becomes weaker upon increasing $T$. The phase soliton becomes unstable at a critical $\alpha_c$ [symbols in Fig.~\ref{kink_fig2}(b)]. Then the system transits into the uniform state and a voltage pulse is generated during this process. As shown in Fig. ~\ref{kink_fig2}(b), the phase soliton is stable in a small temperature ($\alpha$) window for a strong interband coupling $|\gamma_{12}|$ (The sign of $\gamma_{12}$ does not matter here). Therefore, the phase solitons in multiband superconductors with time-reversal symmetry are more stable for weak interband couplings.

A multiband superconducting wire with a phase soliton can be regarded as a Josephson junction because of the weakened superconductivity near the soliton region. In the ground state, the phase differences between two domains separated by the phase soliton is nonzero according to Eq. \eqref{pkeq4}. The wire with a phase soliton thus realize a $\phi$-junction \cite{Buzdin08}, or $\pi$-junction \cite{Bulaevskii77} if the two bands are identical. We then investigate the effect of an external current on the stability of the phase soliton. The external current is introduced by twisting the phase of superconducting order parameter at the two ends of the wire. The phase kink is deformed in the presence of current as depicted in Fig. \ref{kink_fig2}(c). At a threshold current, the deformation renders the phase kink unstable and this threshold current can be regarded as the critical current for the Josephson junction. The dependence of the critical current on $\gamma_{12}$ is present in Fig.~\ref{kink_fig2}(d) and it decreases with $|\gamma_{12}|$. At the instability current when the system evolves from the kink state to the uniform state, a phase slip occurs associated with a voltage pulse similar to the case with increasing temperature.

The phase kink in Eq. \eqref{pkeq8} between two time-reversal symmetry broken states $\hat{\Psi}$ and $\hat{\Psi}^*$ is different from that in superconductors with time-reversal symmetry. In the former case, to remove the kink, one needs to change $\hat{\Psi}^*$ into $\hat{\Psi}$ or vice versa, which requires to overcome a huge energy barrier proportional to the volume of domains. Thus the phase soliton in this case is topologically protected as a result of breaking $Z_2$ symmetry. In the latter case, the domains separated by the phase soliton are essentially the same except for a common phase factor as depicted in Fig. \ref{kink_fig1} (middle). One can remove the phase soliton by rotating the phase of a domain without costing energy in the domain. Energy costs only happen in the kink region and do not depend on the size of domains. 

\subsection{Creation of phase soliton}\label{Sec4D}
The phase solitons are stable topological excitations, which generally are not present in the ground state. In this subsection, we discuss possible ways to create the phase solitons. Since the phase soliton does not couple directly with gauge fields, one cannot create them by applying magnetic fields. First let us consider the dynamical excitation of phase kink in two-band superconductors with time-reversal symmetry by electric fields in nonequilibrium region, following the arguments by Gurevich and Vinokur \cite{Gurevich03}. They considered a two-band superconducting wire attached to a normal electrode, from which a current is applied. For a large enough current, electric field penetrates into the superconductor. However the electric field cannot exist uniformly in the superconductor, otherwise Cooper pairs are accelerated indefinitely and superconductivity would be destroyed. The electric field is localized in a finite region, where phase slips occur according to the ac Josephson effect. For multiband superconductors with different relaxation time in different bands, the rate of phase slip for different bands is different, therefore the phase difference between different bands increases linearly with time. As a consequence, the phase solitons are nucleated at the edge and then are pushed towards the center of the wire.

We adopt the time-dependent Ginzburg-Landau theory in Eqs. \eqref{meq2} and \eqref{meq3} to describe the nonequilibrium dynamics of superconducting order parameters. In one dimension, we can put $\mathbf{A}=0$. Assuming that the amplitude of the order parameter $|\Psi_j|=\Psi_{j0}$ is constant in space, we obtain two equations for the phase $\phi_j$ in a two-band superconductor,
\begin{equation}\label{eqck3}
\frac{{{\hbar ^2 \Psi _{10}^2}}}{{2{m_1}{D_1}}}\left({\partial _t}{\phi _1} + \frac{{2e}}{\hbar }\varphi \right) = \frac{{{\hbar ^2}\Psi _{10}^2}}{{2{m_1}}}{\nabla ^2}{\phi _1} - {\gamma_{12}}{\Psi _{10}}{\Psi _{20}}\sin ({\phi _2} - {\phi _1})
\end{equation}
\begin{equation}\label{eqck4}
\frac{{{\hbar ^2\Psi _{20}^2}}}{{2{m_2}{D_2}}}\left({\partial _t}{\phi _2} + \frac{{2e}}{\hbar }\varphi \right) = \frac{{{\hbar ^2}\Psi _{20}^2}}{{2{m_2}}}{\nabla ^2}{\phi _2} + {\gamma_{12}}{\Psi _{10}}{\Psi _{20}}\sin ({\phi _2} - {\phi _1})
\end{equation}
Multiplying Eqs. \eqref{eqck3} and \eqref{eqck4} by proper factors and subtracting Eq. \eqref{eqck4} from Eq. \eqref{eqck3}, we obtain an equation for $\phi_{12}\equiv \phi_1-\phi_2$
\begin{equation}\label{eqck5}
{\tau _{12}}{\partial _t}{\phi _{12}} = \lambda _{12}^2{\nabla ^2}{\phi _{12}} - {\alpha _{12}}\sigma\nabla\cdot \mathbf{E} + \mathrm{sign}[{\gamma_{12}}]\sin ({\phi _{12}}),
\end{equation}
with
\begin{equation}\label{eqck6}
{\tau _{12}} = \frac{{{\hbar ^2}{\Psi _{10}}{\Psi _{20}}}}{{(\Psi _{20}^2{m_1}{D_1} + \Psi _{10}^2{m_2}{D_2}){|\gamma_{12}|}}},
\end{equation}
\begin{equation}\label{eqck7}
\lambda _{12}^2 = \frac{{{\hbar ^2}{\Psi _{10}}{\Psi _{20}}}}{{2({m_2}\Psi _{10}^2 + {m_1}\Psi _{20}^2)|\gamma_{12} |}},
\end{equation}
\begin{equation}\label{eqck8}
{\alpha _{12}} = \frac{{\hbar \Psi _{10}\Psi _{20}(D_1-D_2)}}{{4e |\gamma_{12}|({m_2}\Psi _{10}^2 + {m_1}\Psi _{20}^2)}}\left(\frac{{\Psi _{20}^2 D_1}}{{{m_2}}} + \frac{{\Psi _{10}^2}D_2}{{{m_1}}}\right)^{-1}.
\end{equation}
We have used the expression for supercurrent during the derivation
\begin{equation}\label{eqck9}
{\mathbf{J}_s} = \frac{{2e\hbar }}{{{m_1}}}\Psi _{10}^2\nabla ({\phi _1} - {\phi _2}) + \left(\frac{{2e\hbar }}{{{m_1}}}\Psi _{10}^2 + \frac{{2e\hbar }}{{{m_2}}}\Psi _{20}^2\right)\nabla {\phi _2},
\end{equation}
and $\nabla\cdot \mathbf{J}_s=-\sigma \nabla\cdot \mathbf{E}$ because of the current conservation in one dimension,
\begin{equation}\label{eqck10}
\mathbf{J}_s+\sigma \mathbf{E}=\mathbf{J}_{\mathrm{ext}},
\end{equation}
with $\mathbf{J}_{\mathrm{ext}}$ being the bias current.

\begin{figure}[t]
 	\psfig{figure=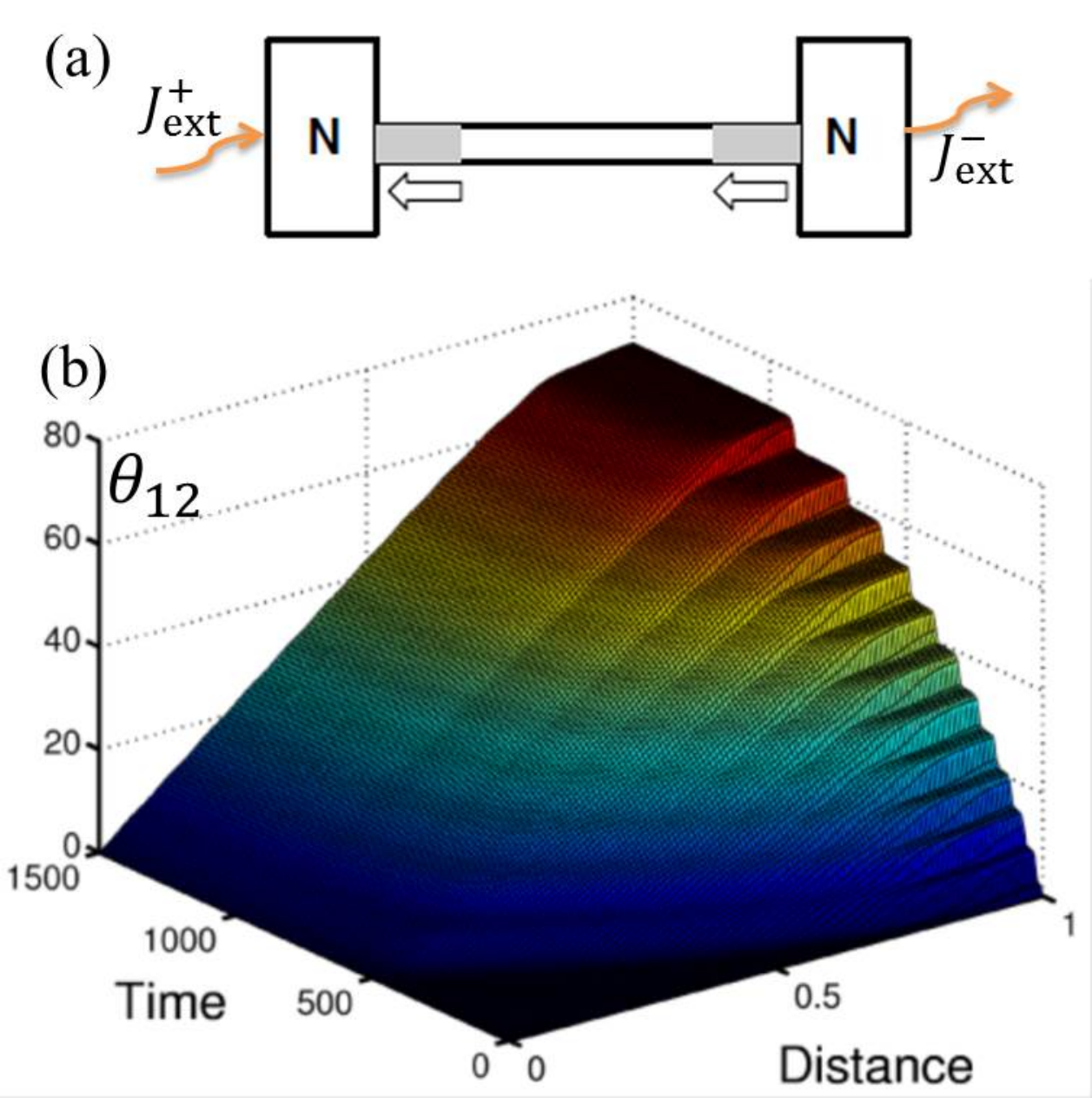,width=\columnwidth}
 	\caption{\label{kink_creation}(color online) (a) Schematic view of a two-band superconducting wire attached to two normal electrodes (N) through which an external current $J_{\mathrm{ext}}$ is injected. (b) Dynamics of formation of phase solitons in the wire of length $L$ after $J_{\mathrm{ext}}(t)$ was instantaneously turned on from 0 to $1.025J_t$ at $t = 0$. Only the right half $L/2\leq x\leq L$ is shown. Time is in unit of $\tau_{12}$ and distance from the center of wire $x=L/2$ is normalized as $(2x-L)/L$. Here $\lambda_e=L/20$ and $\lambda_{12}=0.1\lambda_e$.  From Ref. \cite{Gurevich03}.}
 \end{figure}

Adding Eq. \eqref{eqck4} to Eq. \eqref{eqck3} and using Eq. \eqref{eqck9} and Eq. \eqref{eqck10}, we obtain an equation for $\mathbf{E}$
\begin{equation}\label{eqck11}
-{\alpha _e}{\partial _t}\nabla {\phi _{12}} + {\tau _e}({\partial _t}{\mathbf{J}_{\mathrm{ext}}}/\sigma  - {\partial _t}\mathbf{E}) - \mathbf{E} =  - \lambda _e^2\nabla({\nabla }\cdot\mathbf{E}),
\end{equation}
with
\begin{equation}\label{eqck12}
{\tau _e} = \frac{\sigma}{4e^2} \left(\frac{{\Psi _{10}^2 }}{{{m_1}}} + \frac{{\Psi _{20}^2}}{{{m_2}}}\right)^{-1},
\end{equation}
\begin{equation}\label{eqck13}
\lambda _e^2 = \frac{ \sigma}{{4e^2}} \left(\frac{{{\Psi _{10}^2}}}{{{m_1}{D_1}}} + \frac{{{\Psi _{20}^2}}}{{{m_2}{D_2}}}\right)^{-1},
\end{equation}
\begin{equation}\label{eqck14}
\alpha_e=2|\gamma_{12}|\Psi_{10}\Psi_{20}\alpha_{12}.
\end{equation}
Equations \eqref{eqck5} and \eqref{eqck11} together with boundary conditions describe the dynamics of a two-band superconducting wire subject to a bias current. These equations were first derived in Ref. \cite{Gurevich03}. For a weak current, the electric field is screened by superconductors in a length scale of $\lambda_e$. The system is in a phase-locked static state except for the penetration of phase kink near edge. For currents above a threshold value $J_t$, the phase solitons start to enter into the wire because of the interband breakdown. For $\lambda_e\ll \lambda_{12}\ll L$ with $L$ being the length of the wire, $J_t=2\lambda_{12}/\alpha_{12}$. For $\lambda_{12}\ll \lambda_{e}\ll L$, $J_t=\lambda_{e}/\alpha_{12}$. \cite{Gurevich03} Here $J_t$ is much smaller than the pair breaking current. Gurevich and Vinokur solved Eqs. \eqref{eqck5} and \eqref{eqck11}  numerically for $J_{\mathrm{ext}}>J_t$, for a setup sketched in Fig. ~\ref{kink_creation} (a). The phase solitons are created continuously at the edges of wire and then propagate into the wire, as shown in Fig. ~\ref{kink_creation} (b). Since the phase difference $\phi_{12}$ is coupled to the electric field, one may also create phase solitons by shining a microwave to a two-band superconductor. Later Gurevich and Vinokur showed that it is also possible to create phase solitons in a two-band superconductor with bias current in equilibrium. \cite{Gurevich06} 

Vakaryuk \emph{et al.} proposed to stabilize the phase soliton utilizing the proximity effect. \cite{PhysRevLett.109.227003} They considered a two-band superconductor with $s\pm$ pairing symmetry in proximity to a $s$-wave single-band superconductor. The proximity to the $s$-wave superconductor tends to align the phases of superconducting order parameter in the $s\pm$ superconductor, while the $s\pm$ pairing symmetry favors a $\pi$ phase shift in the phase of superconducting order parameters. For a strong proximity effect, the formation energy of phase soliton can be reduced even to a negative value, thus renders the phase soliton thermodynamically stable. The reduction of the formation energy of phase soliton does not occur for the $s++$ pairing symmetry. Thus in this way, one may be able to nail down the pairing symmetry of a two-band superconductor by exploiting the proximity effect to a $s$-wave superconductor. The authors proposed to measure the magnetization in a ring, which is made of a two-band superconductor with the $s\pm$ pairing symmetry in proximity to a patch of $s$-wave superconductor, to observe the phase soliton, because the phase soliton carries magnetic flux in a ring geometry.

To create a phase kink between the time-reversal broken pair states $\hat{\Psi}$ and $\hat{\Psi}^*$ in a multiband superconducting wire, one may repeat cooling process for one part of the wire from normal state while keep the rest part in superconducting state. \cite{Hu11} In certain circumstances, the cooled part may reach a state that is different from the other part of the wire, provided the cooling process is fast, thus form a kink between the two domains with $\hat{\Psi}$ and $\hat{\Psi}^*$. As demonstrated in Ref. \cite{PhysRevLett.112.017003}  the phase kinks can also be created by homogenous fast cooling in bulk superconductors according to the Kibble-Zurek mechanism \cite{Kibble1976,Zurek1985}. These created kinks are stabilized by random pinning centers or the preexisting vortices.

\subsection{Experimental signatures and observations of phase soliton}\label{Sec4E}

 \begin{figure}[b]
 	\psfig{figure=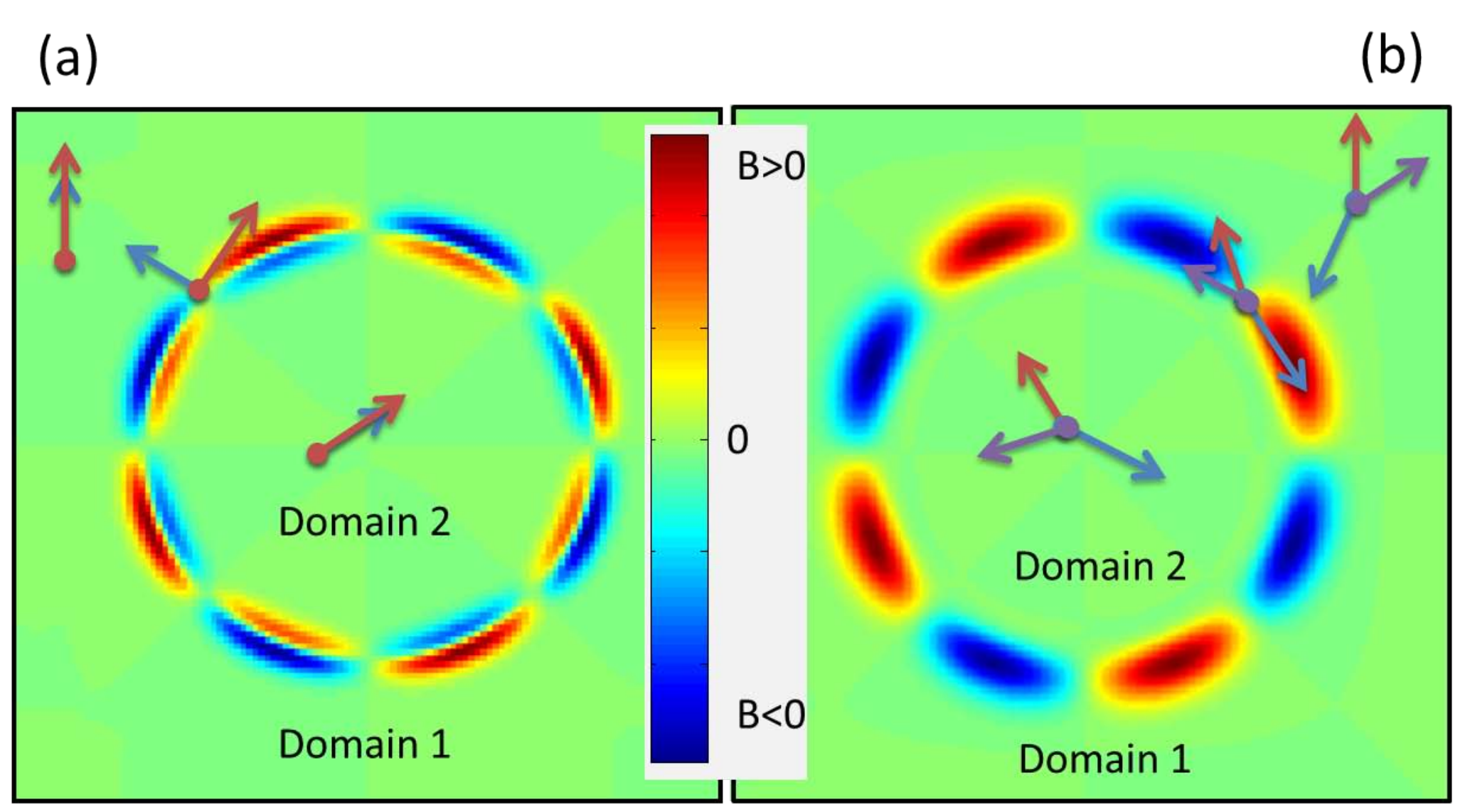,width=\columnwidth}
 	\caption{\label{kink_fig4}(color online) Numerical results of the magnetic field distribution: (a) a circular domain wall in a two-band superconductor; (b) a circular domain wall between two time-reversal symmetry broken states in a three-band superconductor. For (a), $\alpha_j=-20$, $\gamma_{12}=-1$, $\beta_j=1$, $m_1=1$ and $m_2=3$ in the numerical calculations; for (b),  $\alpha_j=0$, $\beta_j=m_j=p_j=1$, $\gamma_{12}=1$, $\gamma_{13}=1.2$ and $\gamma_{23}=1.5$. From Ref. \cite{LinNJP2012}.}
 \end{figure}
 
In this subsection, we discuss the experimental signatures for the phase solitons. In the phase solitons, the time-reversal symmetry is broken locally and we expect spontaneous magnetic fields under proper conditions. Here we will demonstrate the existence of such magnetic fields due to the presence of phase solitons both in nonequilibrium and equilibrium. 

To study the generation of magnetic fields, one has to go beyond one dimension. We first prepare a closed domain wall (phase kink) in two dimensions as initial conditions and solve the time-dependent Ginzburg-Landau equations numerically. During the time evolution in simulations the domain wall organizes itself into a circular shape regardless of its initial shape in order to minimize the domain wall energy, see Figs. \ref{kink_fig4} (a) and (b). There are spontaneous magnetic fields with alternating directions at the domain walls. As displayed in Fig. \ref{kink_fig4} (a) for the phase kinks in superconductors with time-reversal symmetry [see Eq. (\ref{pkeq4})], the induced magnetic field changes polarization in both radial and azimuthal directions. For the kinks between two time-reversal symmetry broken states [see Eq. (\ref{pkeq8})], the magnetic field changes polarization only in the azimuthal direction as shown in Fig. \ref{kink_fig4} (b). The circular domain wall then shrinks and finally disappears, which results in a uniform state. Therefore the domain walls or phase kinks in dimensions higher than one are intrinsically unstable. The life time of the circular domain wall may be long when the size of domain enclosed by the domain wall is large, which allows for a possible experimental detection by measuring the induced magnetic fields.

\begin{figure}[t]
 	\psfig{figure=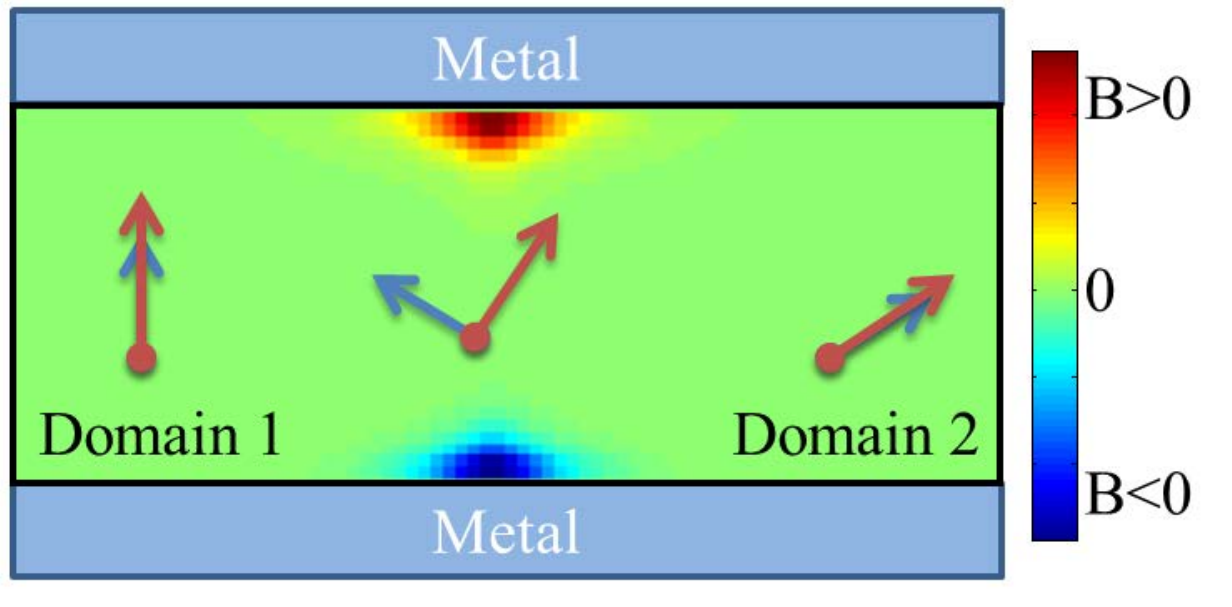,width=\columnwidth}
 	\caption{\label{kink_fig3}(color online) Numerical results of the magnetic field distribution for a two-band superconducting strip with phase kink at the middle in contact with normal metals. Here $\alpha_j=-20$, $\gamma_{12}=-1$, $\beta_j=1$, $m_1=1$ and $m_2=3$, with proximity lengths $p_{jj}=2$ and $p_{j\neq l}=\infty$.}
 \end{figure}

We then study the spontaneous magnetic fields produced by the phase kink in equilibrium. We consider a superconducting strip with a phase soliton at its center in proximity to a normal metal, as sketched in Fig. \ref{kink_fig3}. The time-reversal symmetry is violated at the phase soliton, therefore the spatial variation of amplitude is coupled with that of phase of superconducting order parameters, which can be checked by expanding the interband coupling term $\gamma_{ij}\Delta_i\Delta_j\cos (\phi_i-\phi_j)$ to quadratic order in variation of superconducting order parameters. The amplitudes of the superconducting order parameters are modified by the proximity effect, and this modification produces supercurrent hence generates magnetic fields. We solve the time-dependent Ginzburg-Landau equations with the boundary condition Eq. \eqref{gseqg7}. As shown in Fig. \ref{kink_fig3}, magnetic fields are generated at the proximity region when the phase soliton is present. The magnetic fields change sign at the opposite interface and the total magnetic flux over the sample is zero.

We remark that the spontaneous magnetic fields are about  $H\sim 10^{-5}H_{c2}$ for typical parameters in simulations, which are strong enough to be measured experimentally by scanning SQUID, Hall, or magnetic force microscopy, etc.

Similar dipolar magnetic fields associated with the phase kinks were also observed in numerical simulations recently in Ref. \cite{PhysRevLett.112.017003} in superconducting constrictions or bulk superconductors without time-reversal symmetry, which could be used to detect the possible $s+is$ pairing symmetry in $\mathrm{Ba_{1-x}K_xFe_2As_2}$. \cite{Maiti2013}
 
There is no experimental observation of the phase kink in bulk multiband superconductors to date. The possible existence of the phase soliton has been inferred from measurements in artificial two-band superconductors by Bluhm \emph{et al.} in 2006. \cite{Bluhm06} They fabricated superconducting aluminum rings of various sizes, deposited under conditions likely to generate a layered structure. They were able to control the number of layers and the coupling between layers by varying the annulus width of the ring. Thus the ring can behave effectively as a single-band superconductor or two-band superconductor with a tunable interband Josephson coupling. They then measured the current $I$ in the ring as a function of external magnetic flux $\Phi_a$ enclosed in the ring and temperatures. For a narrow annulus width with one superconducting layer, the measured $I$-$\Phi_a$ curves can be described satisfactorily with a theory for single-band superconductors. For intermediate coupled artificial two-band superconductors, they found metastable states with different winding numbers for different condensates, which was inferred from comparing the measured $I$-$\Phi_a$ curves to a theory based on the two-band Ginzburg-Landau theory. These observations suggest the possible existence of the phase soliton in these artificial two-band superconductors. In the strong coupling regime the measured $I$-$\Phi_a$ signal again can be fitted by a theory for single-band superconductors, because the phases of superconducting order parameter for different layers are locked together which prevents the formation of phase solitons.

\section{Vortex}\label{Sec5}

Vortices are well known topological excitations in superconductors, which arises due to the macroscopic quantum nature of superconducting state. As superconductivity is described by a complex wave function, the single valueness of this wave function requires that the superconducting phase changes by multiple integers of $2\pi$ around a closed loop. When the phase change by $2\pi$ inside superconductors, a vortex excitation appears. For single component $s$-wave superconductors, a vortex has a normal core with size of the superconducting coherence length $\xi$ and magnetic field region of size of the London penetration depth $\lambda$. The total magnetic flux associated with a vortex is quantized to $\Phi_0=hc/2e$. The interaction between normal cores is attractive while the interaction due to the magnetic region is repulsive, thus the net interaction between vortices is determined by the ratio $\lambda$ to $\xi$. \cite{PhysRevB.3.3821,Jacobs79} For type II superconductors when $\lambda/\xi>1/\sqrt{2}$, vortices repel each other and they are stable; while for type I superconductors with $\lambda/\xi<1/\sqrt{2}$, there is attraction between vortices and vortices become unstable upon formation of normal domains. At the special point $\lambda/\xi=1/\sqrt{2}$ vortices do not interact with each other. \cite{Bogomolnyi76} It is possible for the superconducting phase change by $2n\pi$ along a closed loop with an integer $n>1$. These vortices with larger winding number $n$ are called giant vortices carrying $n\Phi_0$ quantum flux. In bulk superconductors, the energy of giant vortices is proportional to $n^2$ thus they are not energetically favorable. However in mesoscopic superconductors when the size of superconductors is comparable to $\xi$, giant vortices may be stabilized by geometric confinement. \cite{PhysRevLett.81.2783,PhysRevB.65.104515,PhysRevLett.107.097202,PhysRevLett.93.257002} Vortices are crucial to determine physical properties, such as transport and electromagnetic response, of a superconductor. Many efforts have been taken to understand the statics and dynamics of vortices in single component superconductors in the past decades, see Refs. \cite{Blatter94, Brandt95} for a review.

Due to the multiband nature, vortices in multiband superconductors possess unique properties that are not shared by single-band superconductors. In this section, we shall explore these novel properties. First we will present the concept of fractional vortex and study their interaction. Then we will review several theoretical proposal to stabilize fractional vortices. We then proceed to discuss vortices with attraction at large separation and repulsion at short separation. The consequences of the nonmonotonic inter-vortex interaction will be reviewed. 

\subsection{Fractional vortex}\label{Sec5A}
\begin{figure}[t]
 	\psfig{figure=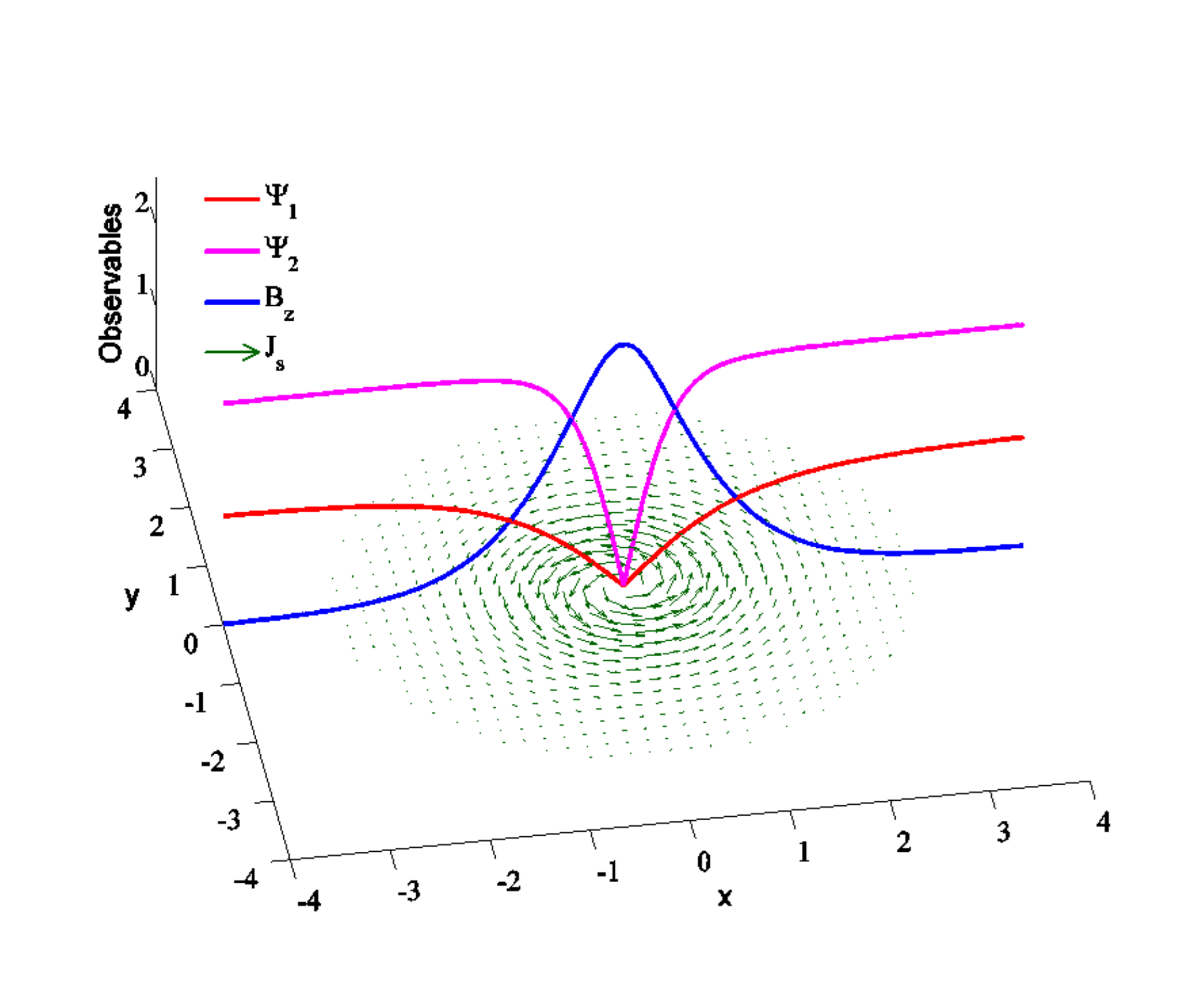,width=\columnwidth}
 	\caption{\label{vortex2B_profile}(color online) Schematic view of superconducting order parameters, magnetic field and supercurrent profiles for a (composite) vortex in two-band superconductors.}
 \end{figure}

Multiband superconductors have multiple complex gap functions, thus it is possible that the phase associated with these gap functions changes by different integer multiple of $2\pi$ along a closed loop. In this case, the quantum flux associated with this vortex is not an integer multiple of $\Phi_0$ and this vortex is called a fractional vortex. \cite{Babaev02, Babaev07}  Let us consider a two-band superconductor with the free energy functional in Eq. \eqref{meq1}. We integrate the supercurrent Eq. \eqref{pkeq5} along a close loop far from the vortex core where $\mathbf{J}_s=0$. Then the total flux is given by Eq. \eqref{pkeq6}. The flux is integer quantized $\Phi=n\Phi_0$ only when $n_1=n_2=n$. In other cases, we have a fractional quantized vortex. As vortex energy increases with the winding number, the most important fractional vortices are those with $n_1=0$, $n_2=1$ or $n_1=1$, $n_2=0$. The fractional vortex in bulk superconductors however is thermodynamically unstable because its energy diverges logarithmically with system size, see Eq. \eqref{vieq6} below. Therefore these two fractional vortices alway lock together to form a composite vortex with $n_1=n_2=1$ in equilibrium. A schematic view of order parameters, supercurrent and magnetic field for a composite vortex is displayed in Fig. ~\ref{vortex2B_profile}. Nevertheless under certain circumstances, composite vortices with $n_1=1$, $n_2=1$ can dissociate into fractional vortices as will be discussed in the next subsections.

Let us compare a fractional vortex to a phase soliton in a superconducting ring discussed in Sec. \ref{Sec4A}. They are the same in terms of the phase winding number and the associated magnetic flux. However they have entirely different topological nature. The phase soliton belongs to the homotopy class $\pi_0(S^0)$ and is stable only in one dimension; while the fractional vortex belongs to the homotopy class $\pi_1(S^1)$ and can be stable in two or three dimensions. 
 
There is no experimental observation of the fractional vortex in multiband superconductors at the time of writing.

\subsection{Interaction between fractional vortices}\label{Sec5B}
In this subsection we formulate the pairwise interaction between fractional vortices in the London limit when $\xi_i\ll \lambda$, taking a two-band superconductor as an example. In this case, the normal cores of vortices do not play a role in determining the vortex interaction. There are four sources of interactions. Vortices as magnetic objects, vortices with the same polarization repel each other in a short range due to the exchange of massive photons inside superconductors. They can also repel in a long-range due to the exchange of massless photon outside superconductors, which is especially important for thin films. Fractional vortices in different condensates attract because of the coupling to the same gauge field. They attract also as a consequence of the interband coupling.  

We consider the London free energy functional for a two-band superconductor with a Josephson-like interband coupling. In contrast to the Ginzburg-Landau free energy functional valid near $T_c$, the London free energy functional is valid in the whole temperature region. The free energy density is
\begin{align}\label{vieq1}
\mathcal{F}_L=\frac{1}{8\pi }\sum _{j=1}^2\left[\frac{1}{\lambda _{j}^2}\left(\mathbf{A}-\frac{\Phi _0}{2\pi }\nabla \phi _{j}\right)^2+(\nabla \times \mathbf{A})^2\right]\nonumber\\
+2\bar{\gamma}_{12}\cos \left(\phi _1-\phi _2\right),
\end{align}
where $\lambda_{j}=\sqrt{(m_{j} c^2)/(4\pi \bar{n}_{j} e^2)}$ is the London penetration depth for each condensate with superfluid density $\bar{n}_{j}$. The effective penetration depth for the two-band system is $\lambda^{-2}=\sum_{j=1}^2 \lambda_{j}^{-2}$. We can split $\mathcal{F}_L$ into two parts $\mathcal{F}_L=\mathcal{F}_m+\mathcal{F}_c$, \cite{Silaev11} with the magnetic coupling 
\begin{equation}\label{vieq2}
\mathcal{F}_m=\frac{1}{8\pi }\left[\mathbf{B}^2+\lambda ^2(\nabla \times \mathbf{B})^2\right].
\end{equation}
$\mathcal{F}_m$ accounts for the magnetic coupling between vortices and it is the same as that for single-band superconductors because there is only one gauge field $\mathbf{A}$ in superconductors. The coupling of the superconducting phases in different bands is described by $\mathcal{F}_c$
\begin{equation}\label{vieq3}
\mathcal{F}_c=\frac{\Phi _1\Phi _2}{32\pi ^3\lambda ^2}\left[\nabla \left(\phi _1-\phi _2\right)\right]^2+2\bar{\gamma}_{12} \cos \left(\phi _1-\phi _2\right).
\end{equation}
Assuming straight vortex lines, the magnetic field profile for fractional vortices can be obtained by minimizing $\mathcal{F}_L$ with respect to $\mathbf{A}$, which yields the London equation
\begin{equation}\label{vieq4}
\lambda ^2\nabla \times \nabla \times \mathbf{B}+\mathbf{B}=\sum _{j=1}^2\sum _{l}\Phi _{j}\delta \left(\mathbf{r}-\mathbf{r}_{j,l}\right).
\end{equation}
Here $\mathbf{r}_{j, l}=(x_{j, l}, y_{j, l})$ is the vortex coordinates for the fractional vortex in the $j$-th condensate, and 
\begin{equation}\label{vieq5}
\Phi _{j}=\lambda ^2\Phi _0/\lambda _{j}^2, 
\end{equation}
is its fractional quantum flux. For a fractional vortex where $\phi_2$ does not change and $\phi_1$ changes by $2\pi$ around $\mathbf{r}_{0}$, the self-energy per unit length is
\begin{equation}\label{vieq6}
\mathcal{E}_{fv}=\left(\frac{\Phi _1}{4\pi  \lambda }\right)^2\ln \left(\frac{\lambda }{\xi _1}\right)+\frac{\Phi _1\Phi _2}{16\pi ^2\lambda ^2}\ln \left(\frac{L}{\xi _1}\right)+|2\bar{\gamma}_{12}|\int dr^2[1-\cos(\phi_1)].
\end{equation}
The first term at the right-hand side is due to $\mathcal{F}_{m}$ and the rest terms are contributed from $\mathcal{F}_{c}$. Here $L$ is the linear system size. Because of the neutral mode described by the term proportional to $[\nabla(\phi_1-\phi_2)]^2$ in Eq. (\ref{vieq3}), $\mathcal{E}_{fv}$ diverges in the thermodynamic limit. The energy of the fractional vortex also diverges linearly in $L$ due to the Josephson coupling in Eq. (\ref{vieq3}). For these reasons, a fractional vortex is thermodynamically unstable in bulk multiband superconductors \cite{Babaev02}. For a composite vortex with $\phi_1=\phi_2$ or  $\phi_1=\phi_2+\pi$, we have $\mathcal{F}_c=0$ and its self-energy is finite.

The equilibrium configuration of $\phi_{j}$ is obtained by minimizing Eq. \eqref{vieq3} with respect to $\phi_{j}$
\begin{equation}\label{vieq7}
\frac{\Phi _1\Phi _2}{16\pi ^3\lambda ^2}\nabla ^2\left(\phi _1-\phi _2\right)+2\bar{\gamma}_{12}  \sin \left(\phi _1-\phi _2\right)=0,
\end{equation}
together with the boundary condition accounting for vortex cores
\begin{equation}\label{vieq8}
\nabla\times(\nabla \phi_{j})=2\pi\sum _{j=1}^2\sum _{l}\delta(\mathbf{r}-\mathbf{r}_{j, l}).
\end{equation}
Due to the presence of nonlinear term $\sin \left(\phi _1-\phi _2\right)$, the interaction between vortices is nonlinear and is of many-body interaction. In the presence of a strong magnetic field, such a nonlinear term can be neglected for the following reason. The term $\nabla^2(\phi_1-\phi_2)$ is of the order $1/\bar{a}^2$ with $\bar{a}$ being the average distance between vortices in the same condensate. For a strong field when $\bar{a}\ll\lambda_J\equiv\sqrt{{\Phi _1\Phi _2}/({32\pi ^3\lambda ^2|\bar{\gamma}_{12}|}})$, the sine term is much smaller than the first term in Eq. \eqref{vieq7} and can be neglected. In this case, $\mathcal{F}_c$ is equivalent to the energy for the $XY$ model. For $\rm{MgB_2}$, $|\bar{\gamma}_{12}|\approx 75\ \rm{J/m^3}$ \cite{Gurevich03b}, and we can neglect the sine term for fields stronger than $4$ T at temperature $T=0$ K. For $\rm{V_3Si}$ \cite{Kogan09} and $\rm{FeSe}_{1-x}$ \cite{Khasanov10}, because of the much weaker interband coupling the required field is smaller.

Neglecting the Josephson coupling in $\mathcal{F}_c$, the pairwise interaction between the fractional vortices can be derived analytically because both $\mathcal{F}_m$ and $\mathcal{F}_c$ are quadratic in $\mathbf{B}$ and $\phi_{j}$. Here $\mathcal{F}_m$ describes the short-range interband and intraband magnetic repulsion between vortices with the same polarization. The term proportional to $(\nabla\phi_{j})^2$ in $\mathcal{F}_c$ accounts for the long-range repulsion between vortices in the same band. The term proportional to $-\nabla\phi_1\nabla\phi_2$ yields a long-range attraction between vortices in different bands. Putting all together, we obtain the repulsion between fractional vortices with flux $\Phi_l$ in the same band separated by a distance $\mathbf{r}_{l, ij}\equiv\mathbf{r}_{l,i}-\mathbf{r}_{l,j}$
\begin{equation}\label{vieq9}
V_{\rm{intra}}(r_{l, ij})=\frac{\Phi _{l}^2}{8\pi ^2\lambda ^2}K_0\left(\frac{r_{l, ij}}{\lambda }\right)-\frac{\Phi _1\Phi _2}{8\pi ^2\lambda ^2}\ln \left(r_{l, ij} \right),
\end{equation}
where $K_0(r)$ is the modified Bessel function. The attraction between two fractional vortices in different bands with a separation $\mathbf{r}_{12, ij}\equiv\mathbf{r}_{1,i}-\mathbf{r}_{2,j}$ is
\begin{equation}\label{vieq10}
V_{\rm{inter}}({r}_{12, ij})=\frac{\Phi _1\Phi _2}{8\pi ^2\lambda ^2}\left[K_0\left(\frac{{r}_{12, ij}}{\lambda }\right)+\ln \left({r}_{12, ij}\right)\right].
\end{equation}
Equations (\ref{vieq9}) and (\ref{vieq10}) are valid away from vortex cores.

\subsection{Dissociation of composite vortex lattice in the flux flow region}\label{Sec5C}
In this subsection, we discuss the dissociation of a composite vortex lattice in a two-band superconductor in the flux flow region due to the difference in coherence length and superfluid density for different bands. \cite{szlin13PRL} The higher the superfluid density $\bar{n}_j$, the larger the magnetic flux $\Phi_j$ for the fraction vortex and hence a stronger Lorentz force in the presence of a current. On the other hand, the shorter the coherence length $\xi_j$, the bigger the viscosity $\eta_j$ and hence a smaller velocity at a given force. As a consequence, the fractional vortex lattice in certain band tends to move faster. For a small external current, the disparity of vortex motion in different bands can be compensated by the attraction between vortices in different bands. However for a large current, such attraction becomes insufficient to balance the difference in the vortex viscosity and driving force for different bands. As a result, fractional vortex lattices in different bands decouple from each other and they move with different velocities. We will also discuss the consequences of the dissociation of composite vortex lattice, such as the appearance of the Shapiro steps and increase of the flux flow resistivity. We remark that the dissociation transition is possible only in the vortex lattice region. For a single composite vortex, because the attraction between two fractional vortices diverges logarithmically with separation, see Eq. \eqref{vieq10}, it is impossible to completely decouple these two fractional vortices by applying a current.

We first introduce an equation of motion for fractional vortices. The vortex viscosity results from dissipation, which is caused by motion of normal cores. Employing the Bardeen-Stephen model \cite{Bardeen65}, the viscosity of a fractional vortex in each band is given by 
\begin{equation}\label{dcveq1}
\eta_{j}=\Phi_0^2/(2\pi\rho_j c^2\xi_{j}^2),
\end{equation}
where $\rho_j$ is the electric resistivity in the $j$-th band. We assume that the vortex structure does not change in the flux flow region and assume an overdamped dynamics for vortices. The equation of motion for fractional vortices is
\begin{align}\label{dcveq2}
\nonumber \eta_j\partial_t r_{j, i}=\frac{1}{8\pi ^2\lambda ^3}\sum _l\left[\Phi _{l }^2 K_1\left(\frac{r_{j ,il}}{\lambda }\right)+\frac{\Phi _{1 }\Phi _{2 }\lambda}{r_{j, il}}\right]\\
+\frac{\Phi _{1 }\Phi _{2 }}{8\pi ^2\lambda ^3}\sum _l\left[K_1\left(\frac{r_{12 ,il}}{\lambda }\right)-\frac{\lambda }{r_{12, il}}\right]+ \frac{J_{\mathrm{ext}} \Phi_{j}}{c},
\end{align}
with $J_{\mathrm{ext}}$ the external current. The first term at the right-hand side of Eq. \eqref{dcveq2} is the repulsion between vortices in the same bands and the second term is the attraction between vortices in different bands. The last term is the Lorentz force. The effect of random pinning centers are not important in the flux flow region, because they are quickly averaged out by vortex motion, resulting in an improved lattice order. \cite{Koshelev94,Besseling03} The interaction between fractional vortices in the same bands cancels out in the vortex lattice phase. We assume a square vortex lattice with a lattice constant $\bar{a}$. By accounting for the dominant wavevector $\mathbf{G}=(\pm 2\pi/\bar{a}, 0)$ for the vortex lattice moving along the $x$ direction, the equation of motion for the center of mass of fractional vortex lattice in each band $R_{j}$ can be written as
\begin{align}\label{dcveq3}
{\eta _2'}\partial _t\left(R_2-R_1\right)=-\left(1+{\eta _2'}\right)\sin \left(R_2-R_1\right)+\left({\Phi _2'}-{\eta _2'}\right)J_{\mathrm{ext}},
\end{align}
\begin{equation}\label{dcveq4}
\partial _tR_1+{\eta _2'}\partial _tR_2=\left(1+{\Phi _2'}\right)J_{\mathrm{ext}}.
\end{equation}
We have introduced dimensionless unit: current is in unit of $c F_d/\Phi_1$; time is in unit of $\eta_1 a/ (2\pi F_d)$; length is in unit of $a/(2\pi)$. Here $\Phi_2'\equiv\Phi_2/\Phi_1$, $\eta_2'\equiv\eta_2/\eta_1$, and 
\begin{equation}\label{dcveq4Fd}
F_d=\frac{\Phi _1\Phi _2\bar{a}}{64\pi ^6\lambda ^4},
\end{equation}
is the maximal attraction between two fractional vortices in different bands. Equations \eqref{dcveq3} and \eqref{dcveq4} are similar to those for the vortex motion in superconducting bilayers. \cite{Clem74}

\begin{figure}[t]
 	\psfig{figure=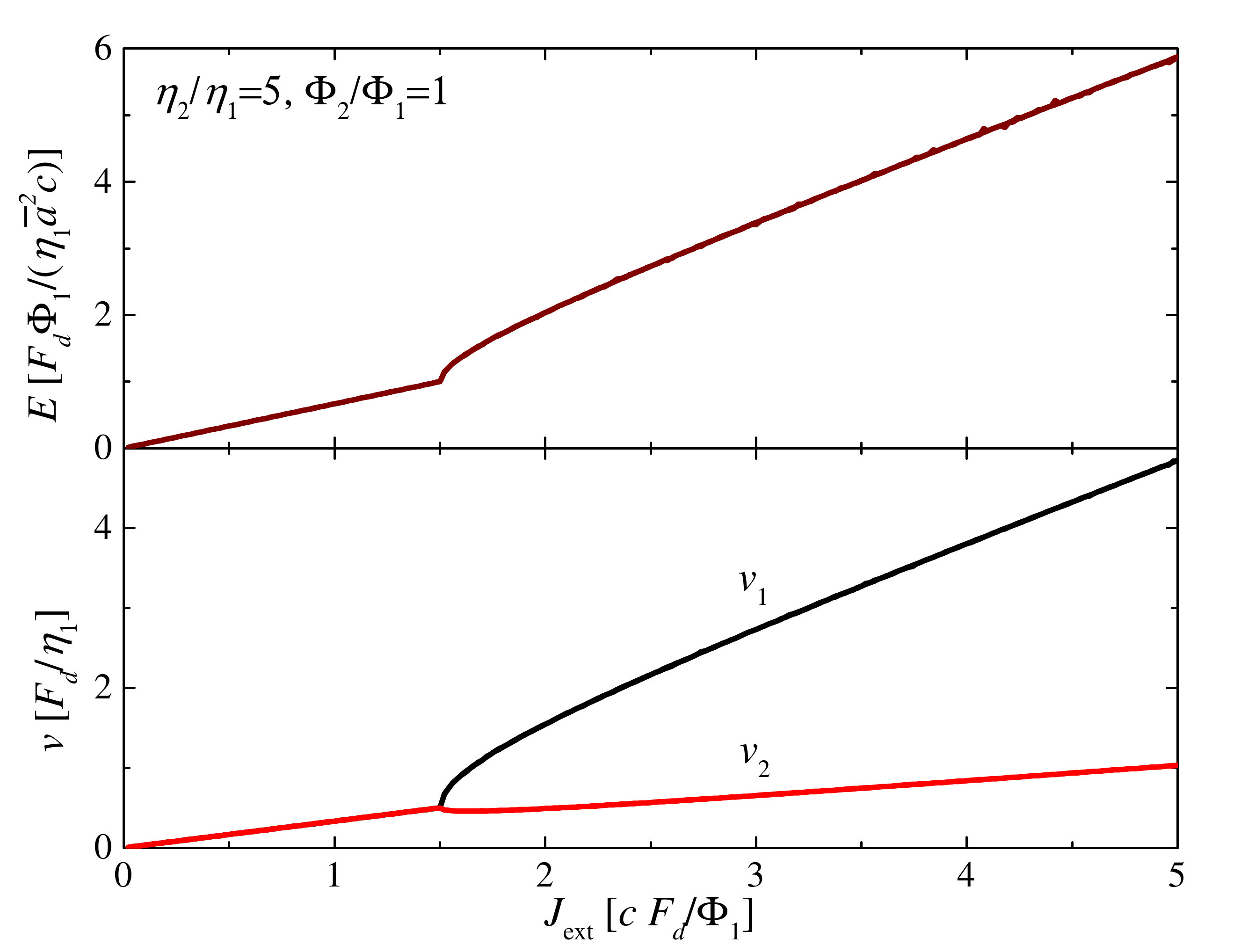,width=\columnwidth}
 	\caption{\label{cvd_fig2}(color online) Dependence of electric field $E$ and velocity $v_1$, $v_2$ on the current $J_{\mathrm{ext}}$, obtained from Eqs. \eqref{dcveq3}, ~\eqref{dcveq4} and \eqref{dcveq8}. From Ref. \cite{szlin13PRL}.}
 \end{figure}

The two fractional vortex lattices in different bands move at the same velocity
\begin{equation}\label{dcveq5}
v_1=v_2=\left(1+{\eta _2'}\right)^{-1}\left(1+{\Phi _2'}\right)J_{\mathrm{ext}}
\end{equation}
for a small current. The composite vortex starts to deform in this region with a separation between the center of mass of these two fractional vortex lattices
\begin{equation}\label{dcveq6}
a_s=\sin^{-1}\left[\left(1+{\eta _2'}\right)^{-1}\left({\Phi _2'}-{\eta _2'}\right)J_{\mathrm{ext}}\right].
\end{equation}
These two lattices decouple at a threshold current
\begin{equation}\label{dcveq7}
J_d=\left|\left(1+{\eta _2'}\right)\left({\Phi _2'}-{\eta _2'}\right)^{-1}\right|,
\end{equation}
and they move with different velocities
\begin{align}\label{dcveq8}
v_{j}=\left(1+{\eta _2'}\right)^{-1}\times\nonumber\\
\left[\left(1+{\Phi _2'}\right)J_{\mathrm{ext}}-\frac{\eta _1}{\eta _{j}}\sqrt{\left({\Phi _2'}-{\eta _2'}\right)^2J_{\mathrm{ext}}^2-\left(1+{\eta _2'}\right)^2}\right].
\end{align}
In this region, the composite vortex lattice dissociates into two fractional vortex lattices moving at different velocities. For a large current $J_{\mathrm{ext}}\gg J_d$, each lattice move independent with a velocity $v_1=J_{\mathrm{ext}}$ and $v_2=J_{\mathrm{ext}} /\eta_2'$. We plot $v_{j}$ as a function of $J_{\mathrm{ext}}$ in Fig. \ref{cvd_fig2}.

The corresponding I-V characteristics can be obtained from the power balance condition
\begin{equation}\label{dcveq9}
\eta_1 v_1^2+\eta_2 v_2^2= J_{\mathrm{ext}} E \bar{a}^2,
\end{equation}
with $E$ the electric field. The I-V curve is present in Fig. \ref{cvd_fig2}, where the flux flow resistivity $dE/dJ_{\mathrm{ext}}$ increases in the decoupled phase. Observation of such enhanced resistivity in experiments might be challenging because the dissociation current $J_d$ is usually large, where the Larkin-Ovchinnikov instability of vortex lattice may be important \cite{Larkin76}. The flux flow resistivity also increases in the Larkin-Ovchinnikov instability region. Nevertheless, the dissociation transition can be confirmed unambiguously by measurement of the Shapiro steps in the decoupled phase, as discussed below.   

 \begin{figure}[b]
 	\psfig{figure=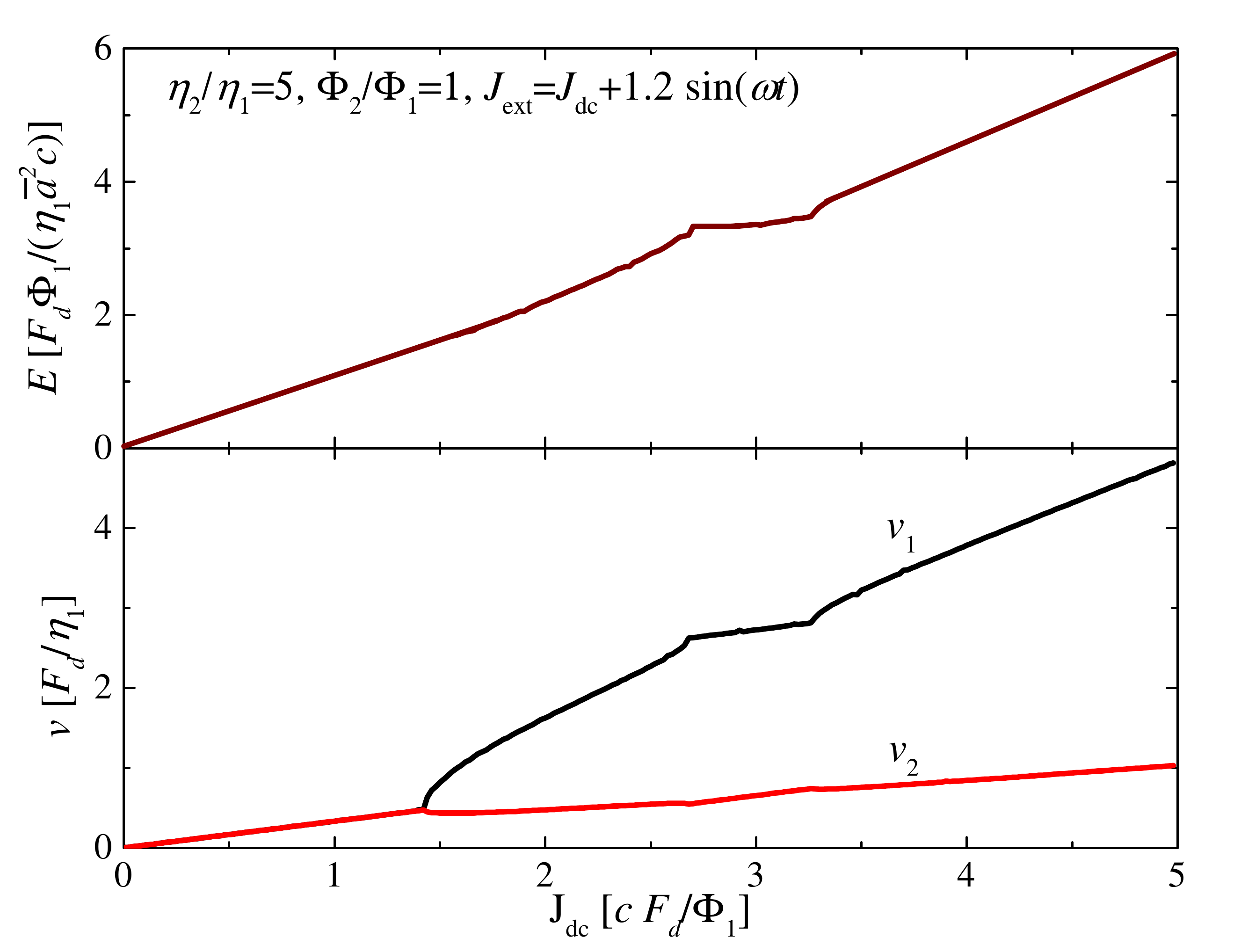,width=\columnwidth}
 	\caption{\label{cvd_fig3}(color online) The same as Fig. \ref{cvd_fig2}, but with an ac current in addition to a dc current. From Ref. \cite{szlin13PRL}.}
 \end{figure}

Equation (\ref{dcveq3}) also describes the phase dynamics in an overdamped Josephson junction. When an ac current is added in superposition to the dc current, there will be Shapiro steps when the resonance condition is satisfied. In the decoupled phase, if one takes one lattice as a reference, the other lattice experiences a periodic potential induced by the reference lattice. The oscillation of the moving lattice induced by the ac current may be in resonance with the oscillation due to the periodic potential of the reference lattice if the period of the ac current matches with the period of the potential. This results in the Shapiro steps in I-V curves. With a current $J_{\mathrm{ext}}=J_{\rm{dc}}+{\rm{Re}} \left[J_{\rm{ac}}\exp[i (\omega t+\theta)]\right]$, the center of mass of each lattice is $R_{j}={v}_{j}t+{\rm{Re}} \left[\tilde{A}_{j}\exp[i({v}_2-{v}_1)t]\right]$  in the region $|v_1-v_2|=\omega\gg 1$. From Eqs. \eqref{dcveq3} and \eqref{dcveq4}, we obtain 
\[
\tilde{A}_{j}=\frac{{\eta _{j}}[1-i{\Phi _{j}}J_{\text{ac}}\exp(i\theta)/{\Phi _1}]}{{\left(v_2-v_1\right){\eta _1}}}.
\] 
The dc current is 
\[
J_{\text{dc}}=\left({\Phi _2'}-{\eta _2'}\right)^{-1}{\rm{Re}}\left[{\eta _2'}\left(v_2-v_1\right)+\left(1+{\eta _2'}\right){\left(\tilde{A}_2-\tilde{A}_1\right)}/{2}\right].
\]
$\theta$ adjusts correspondingly when one changes $J_{\rm{dc}}$, because $v_1-v_2$ is locked with the driving frequency $\omega$, and a Shapiro step is traced out. The height of the Shapiro step is 
\begin{align}\label{dcveq10}
J_{sp}=\frac{\left(1+{\eta _2'}\right)\left({\eta _2'}{\Phi _2'}-1\right)}{({v_2-v_1})\left({\Phi _2'}-{\eta _2'}\right)}.
\end{align}
We solve Eqs.  \eqref{dcveq3} and \eqref{dcveq4} numerically with $J_{\mathrm{ext}}=J_{\rm{dc}}+1.2\sin(\omega t)$ and the results are shown in Fig. \ref{cvd_fig3}. The Shapiro steps appear when $\omega=v_1-v_2$. Generally the Shapiro steps also occur at $n \omega=(v_1-v_2)$ with a much smaller height. Here we only observe the prominent step with $n=1$. The Shapiro steps at $n\omega=v_{j}$ (with the reduced units) can also be induced by the periodic passing of vortex lattice through defects. \cite{Fiory71,Schmid73} These steps can be separated from those induced by relative motion of two fractional vortex lattices in Eq. \eqref{dcveq10}, because their resonance condition is different.

Generally the decoupling of composite vortex lattice is present in all multiband superconductors and it depends on the two parameters $\eta_2/\eta_1$ and $\Phi_2/\Phi_1$. The dissociation current $J_d$ is high for a small disparity in $\lambda_{j}$ and $\xi_{j}$. Here we estimate $J_d$ for $\rm{MgB_2}$. We use $\lambda_1=47.8$ nm, $\lambda_2=33.6$ nm, $\xi_1=13$ nm, and $\xi_2=51$ nm at $T=0$ K \cite{PhysRevLett.102.117001},  $\rho_{j}=10^{-9}\ \rm{\Omega\cdot m}$ \cite{Xi08} and $\bar{a}=40$ nm corresponding to magnetic fields at $B\approx 1$ T. We then estimate $J_d=5\times 10^9\ \rm{A/m^2}$, which is much smaller than the depairing current. The velocity of vortex lattice at the dissociation transition is $v_1=v_2\approx 3\ \rm{m/s}$, which is smaller than the typical Larkin-Ovchinnikov instability velocity for vortex lattice. \cite{Doettinger94} In samples with pinning centers, one first needs to overcome the pinning potential to observe the dissociation transition. The effective dissociation current thus is the sum of the depinning current and the dissociation current $J_d$ for clean systems. The dissociation transition may also be observed in the proposed liquid hydrogen two-component superconductor with a small $J_d$ because of the large mass difference between the proton and electron.

The dissociation of composite vortex lattice discussed here is similar to the decoupling of vortex motion in multilayer superconductors \cite{Giaever65} and in cuprate superconductors \cite{Busch92,Wan93,Safar94}. For the dissociation transition in multiband superconductors, vortices have fractional quantum flux after dissociation and the dissociation takes place in the band (momentum) space. While for multilayer superconductors, vortices in different layers carry quantized flux $\Phi_0$ and the decoupling occurs in the real space. This decoupling in multilayer superconductors has been discussed theoretically \cite{Cladis68,Clem74,Uprety95} and observed experimentally \cite{Giaever65,Busch92,Wan93,Safar94} decades ago. The Shapiro steps were also measured in multilayer superconductors in the decoupled phase \cite{Gilabert94}. These observations corroborate the possible experimental observation of the predicted dissociation of composite vortex lattice in multiband superconductors. 

\subsection{Stabilization of fractional vortex by pinning arrays}\label{Sec5D}

 \begin{figure}[t]
 	\psfig{figure=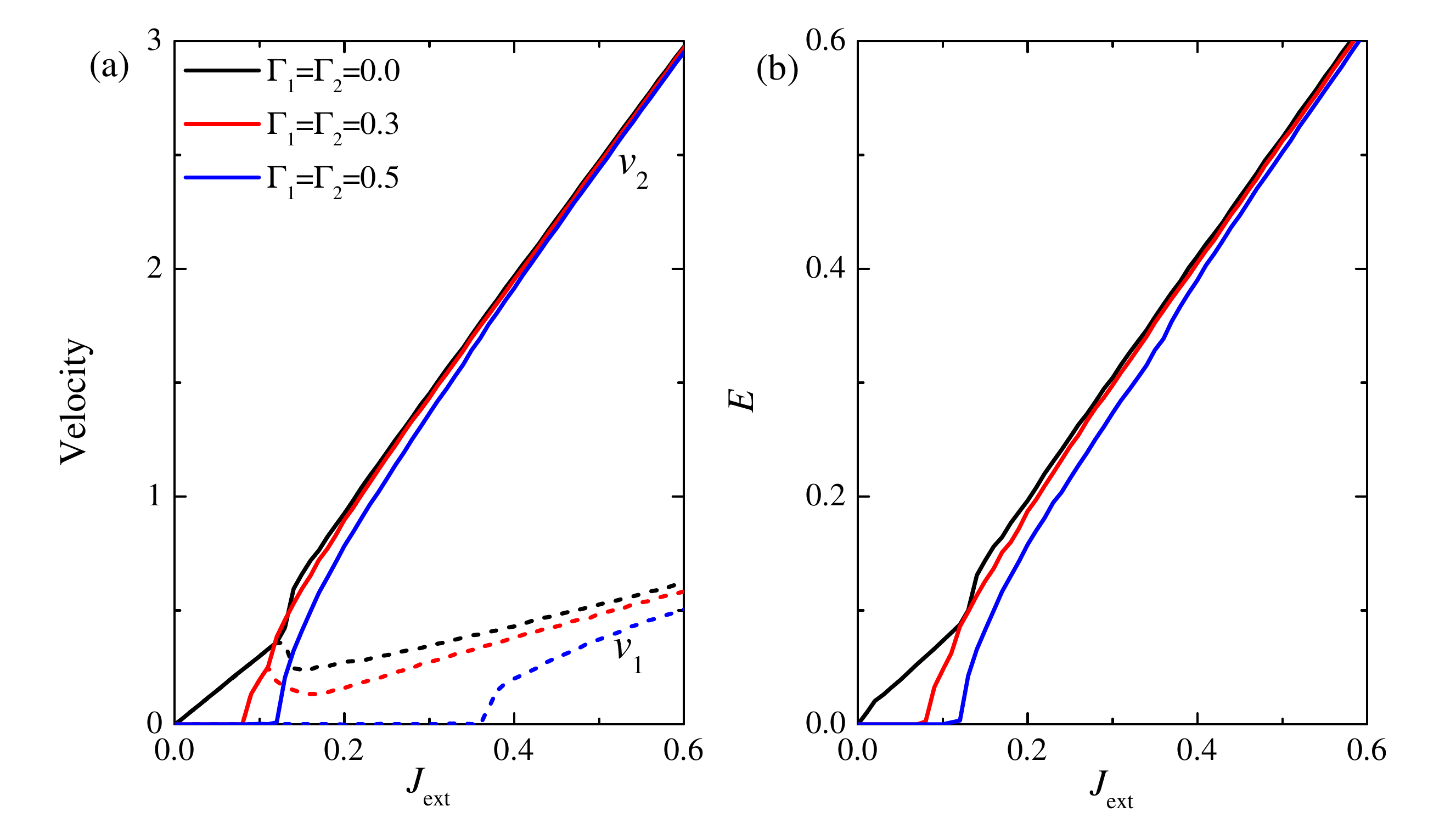,width=\columnwidth}
 	\caption{\label{pa_fig1}(color online) (a) Dependence of the velocity of fractional vortices on current and (b) I-V characteristics for different pinning strengths $\Gamma_j$. For a weak pinning strength, the two vortex lattices depin simultaneously at the depinning current and then move together with the same velocity. For a large current, they split into two moving fractional vortex lattices with distinct velocities. In the case with a stronger pinning, they depin at different currents.  Here the density of pinning centers are the same as that of composite vortex $n_p=n_v$ and $\Phi_2/\Phi_1=5.0$.  We use a square pinning array with a lattice constant $a_{p,x}=a_{p,y}=5.0$. From Ref. \cite{LinPRB13}.}
 \end{figure}
 \begin{figure}[b]
 	\psfig{figure=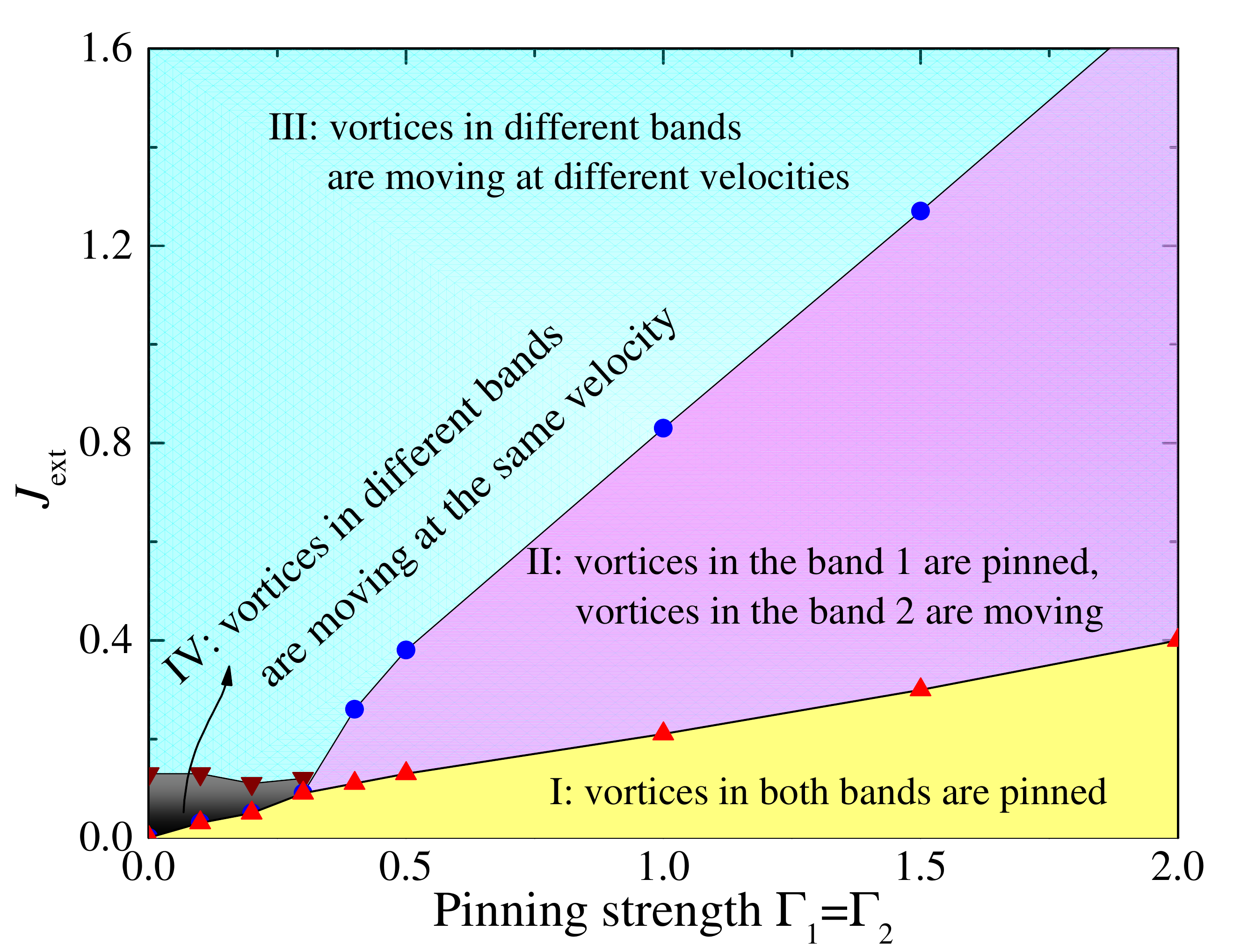,width=\columnwidth}
 	\caption{\label{pa_fig2}(color online) (a) Dynamic phase diagram for two-band superconductors with a pinning array subject to a dc current. Four different regions can be identified. Here $n_p=n_v$ and $\Phi_2/\Phi_1=5.0$. We use a square pinning array with a lattice constant $a_{p,x}=a_{p,y}=5.0$. From Ref. \cite{LinPRB13}.}
 \end{figure}
In this subsection, we discuss the possibility to stabilize fractional vortices by pinning arrays. \cite{LinPRB13} When the external current is turned off suddenly in the decoupled phase where two fractional vortex lattices move with different velocities, it is possible that the fractional vortices get trapped by pinning centers if the density of pinning centers is higher than the vortex density. We will construct a dynamic phase diagram for vortices in a two-band superconductor with pinning arrays. The presence of pinning arrays also yields novel self-induced Shapiro steps in the decoupled phase. 

We model the interaction between the pinning site at $\mathbf{r}_{p}$ and the fractional vortex in the $j$-th band at $\mathbf{r}_{j}$ as 
\begin{equation}\label{dcveq11}
	U_{j,p} \left(\mathbf{r}_{j}-\mathbf{r}_{p}\right)=-\Gamma_{j} \exp \left[-{\left(\mathbf{r}_{j}-\mathbf{r}_{p}\right)^2}/{l _{j}^2}\right],
\end{equation}
where $l_j$ is the pinning range and $\Gamma_{j}$ characterizes the pinning strength. The equation of motion for the fractional vortices is
\begin{align}\label{dcveq12}
	\eta_j\partial_t \mathbf{r}_{j, i}=-\nabla_{{r}_{j, i}}{ (V_{\rm{intra}}+V_{\rm{inter}}+U_{j,p})}+ {\mathbf{J}_{\mathrm{ext}} \Phi_{j}}/{c},
\end{align}
where $V_{\rm{intra}}$ and $V_{\rm{inter}}$ are given in Eqs. (\ref{vieq9}) and (\ref{vieq10}) respectively. Equation \eqref{dcveq12} is solved numerically using the second order Runge-Kutta method. We use dimensionless units for force: ${\Phi _1 \Phi _2}/({8 \pi ^2 \lambda ^3})$; length: $\lambda$; time: $8 \pi ^2 \eta _1 \lambda ^4/(\Phi_1\Phi_2)$; and current: $c\Phi_2/(8 \pi ^2 \lambda ^3)$, and we set $\eta_1=\eta_2$ and $l_j=1$.

\begin{figure}[t]
	\psfig{figure=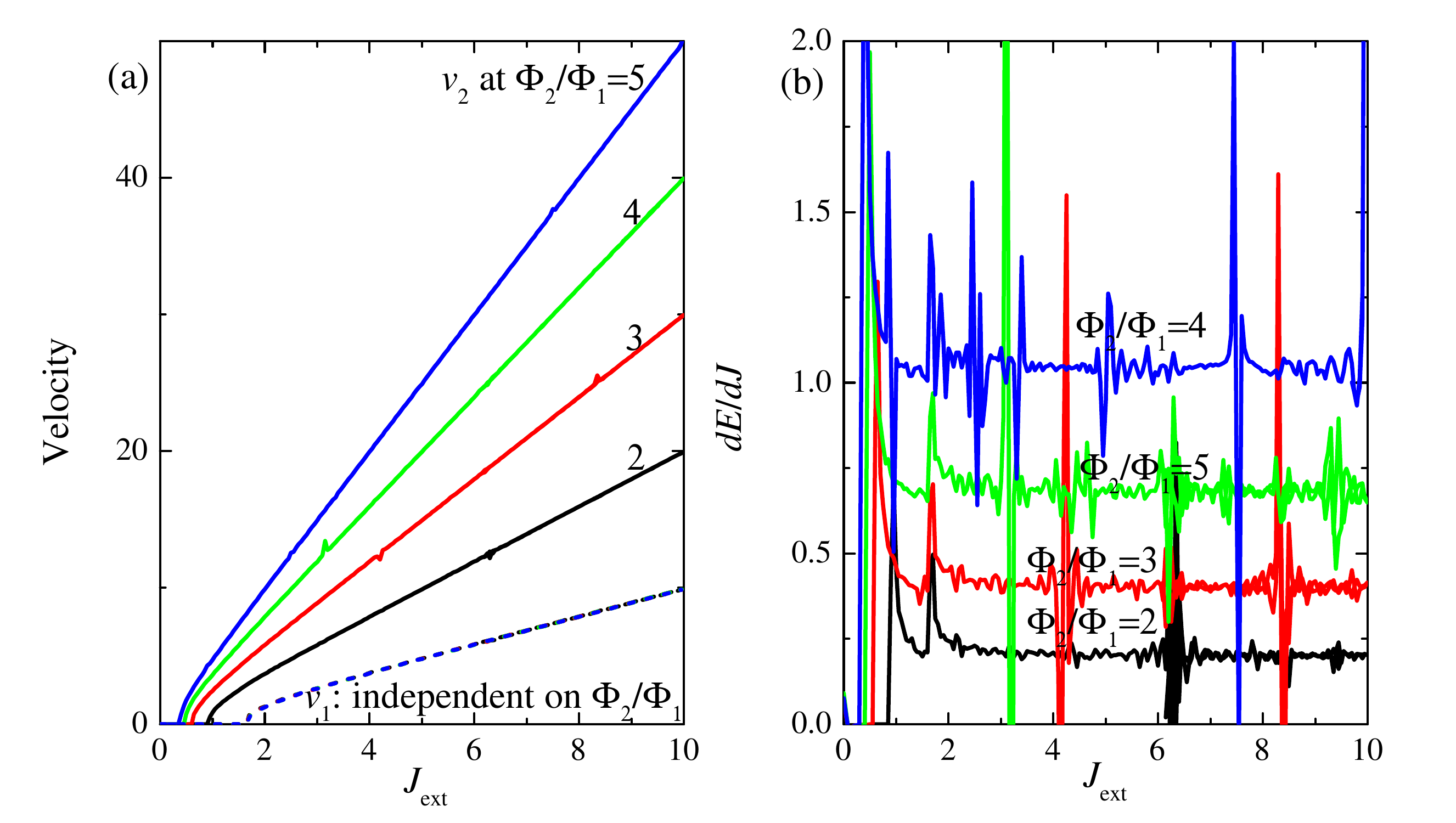,width=\columnwidth}
	\caption{\label{pa_fig3}(color online) (a) Current-velocity curves and (b) the corresponding differential resistivity $dE/dJ_{\mathrm{ext}}$ for different $\Phi_2/\Phi_1$. The spikes in the velocity are due to the resonance among the two moving fractional vortex lattices and the pinning array. The first two peaks from the left in (b) are due to the depinning transition of the two composite vortex lattices. The other peaks in (b) are due to the Shapiro steps. Here $n_p=n_v$ and $\Gamma_1=\Gamma_2=2.0$.  We use a square pinning array with a lattice constant $a_{p,x}=a_{p,y}=5.0$. From Ref. \cite{LinPRB13}.}
\end{figure}

The attraction between two fractional vortex lattices in different bands is a periodic function of space with a period equal to the lattice constant. The maximal attraction is $F_d$ given in Eq. \eqref{dcveq4Fd}. When the maximal pinning force $F_{j, p}=-\nabla U_{j, p}$ is much smaller than $F_d$, i.e. $F_{j,p}\ll F_d$, the two fractional vortex lattices depin simultaneously at a current $J_p=(F_{1,p}+F_{2, p})c/\Phi_0$, as shown in red curve in Fig. \ref{pa_fig1} (a). They travel with the same velocity after depinning until the current is large enough to decouple them as in the case of clean systems. In this region, the fractional vortices form a composite vortex with deformation. All these can be seen in the I-V curve, as depicted in Fig. \ref{pa_fig1} (b). In the other limit when $F_{j,p}\gg F_d$, the two fractional vortex lattices depin at different currents $J_{j,p}=F_{j, p}c/\Phi_j$, and they move at different velocities once depinned. The dynamic phase diagram is constructed in Fig. \ref{pa_fig2}. The depinning current increases with the pinning strength $\Gamma_j$. Meanwhile the region of flux flow of the composite vortex lattice shrinks and finally disappears. Then the two fractional vortex lattices depin at different currents. At a high current $J_{\mathrm{ext}}\gg 1$, the two fractional vortex lattices travel with different velocities $v_j\approx J_{\mathrm{ext}}\Phi_j/(c\eta_j)$. All these dynamical phase transitions manifest themselves in the I-V curves.

\begin{figure}[t]
	\psfig{figure=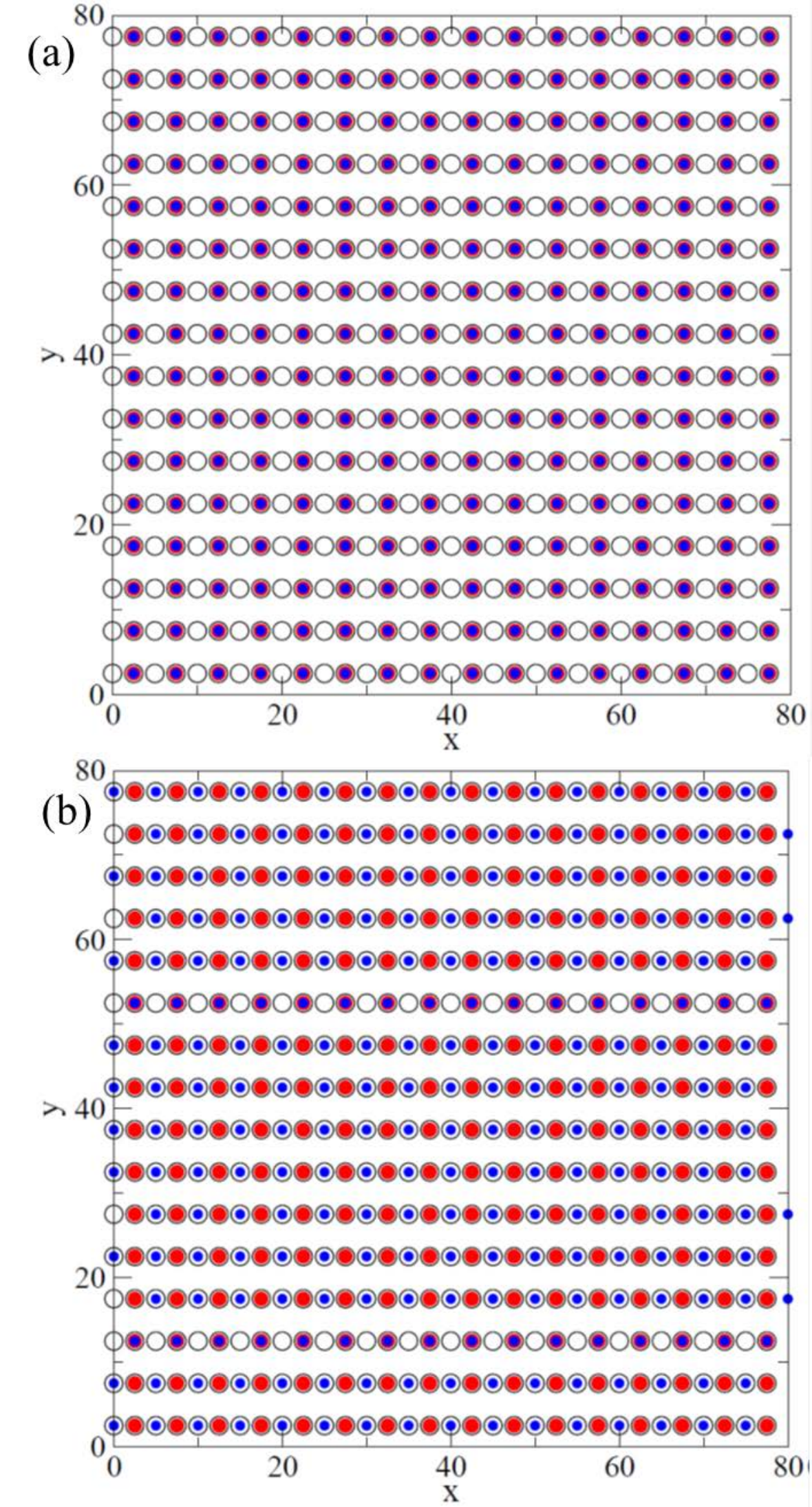,width=8.0cm}
	\caption{\label{pa_fig4}(color online) Snapshot of the fractional vortex configuration (a) in the ground state and (b) after a current quench. Open circles are pinning sites; blue and red circles represent the fractional vortices in different bands. Some fractional vortices are trapped at different pinning sites after the current quench. Here $n_p=2n_v$, $\Gamma_1=\Gamma_2=2.0$, $l_1^2=l_2^2=0.5$, $J_{\mathrm{ext}}=3.2$ and $\Phi_2/\Phi_1=5.0$. We use a rectangular pinning array with a lattice constant $a_{p,y}=2a_{p,x}=5.0$. From Ref. \cite{LinPRB13}.}
\end{figure}

We then study the self-induced Shapiro steps in the decoupled phase where two fractional vortex lattices in bands $1$ and $2$ are moving with different velocities. The self-induced Shapiro steps are distinct from the conventional Shapiro steps \cite{Shapiro1963} because they appear without the application of an external ac drive. The appearance of the self-induced Shapiro step is also an experimentally observable signature to detect the dissociation of composite vortex lattice in the presence of pinning arrays. The basic idea for the self-induced Shapiro steps is as follows. The slower moving fractional vortex lattice in band $1$ feels a periodic potential induced by the pinning array. It also experiences a periodic force due to the fast moving lattice in band $2$. The Shapiro steps are induced when $j_1 \mathcal{T}_1=j_2 \mathcal{T}_2$, where $\mathcal{T}_1=\bar{a}/v_1$ is the period of the ac force due to the pinning array acting on the vortex in band $1$, and $\mathcal{T}_2=\bar{a}/(v_2-v_1)$ is the period of the ac force due to the fast moving lattice in band $2$. For simplicity, we have assumed that both the pinning array and the vortex lattice have the same lattice constant $\bar{a}$. In terms of the velocities of the fractional vortex lattices the Shapiro steps occur when $(j_1+j_2)v_1=j_2 v_2$. We observe spikes in velocity curves as a consequence of the resonance as shown in Fig. \ref{pa_fig3} (a) in the numerical simulations. The resonance can be seen more clearly in the differential resistivity $dE/dJ_{\mathrm{ext}}$ plotted in Fig. \ref{pa_fig3} (b) whenever $v_j$ satisfies the resonance condition.

We proceed to consider the case that the pinning density is twice of the vortex density $n_p=2n_v$ to optimize the trapping of fractional vortices by pinning arrays. In the ground state as depicted in Fig.~\ref{pa_fig4}(a), the composite vortices reside in a checkerboard pattern with every other pinning site occupied. The composite vortex lattice is dissociated into two fractional vortex lattices by applying a large current. The current is then turned off and most of the fractional vortices in bands $1$ and $2$ get trapped at different pinning sites. This results in a metastable state where every pinning center is occupied by a fractional vortex, as depicted in Fig.~\ref{pa_fig4}(b). The life time of the fractional vortex can be long for a strong pining potential. The commensuration between vortex configuration and the pinning sites increases the life time further by reducing the fluctuations from vortex-vortex interaction. The trapped fractional vortices could be observed in experiments with various imaging techniques such as a SQUID.

\begin{figure}[t]
	\psfig{figure=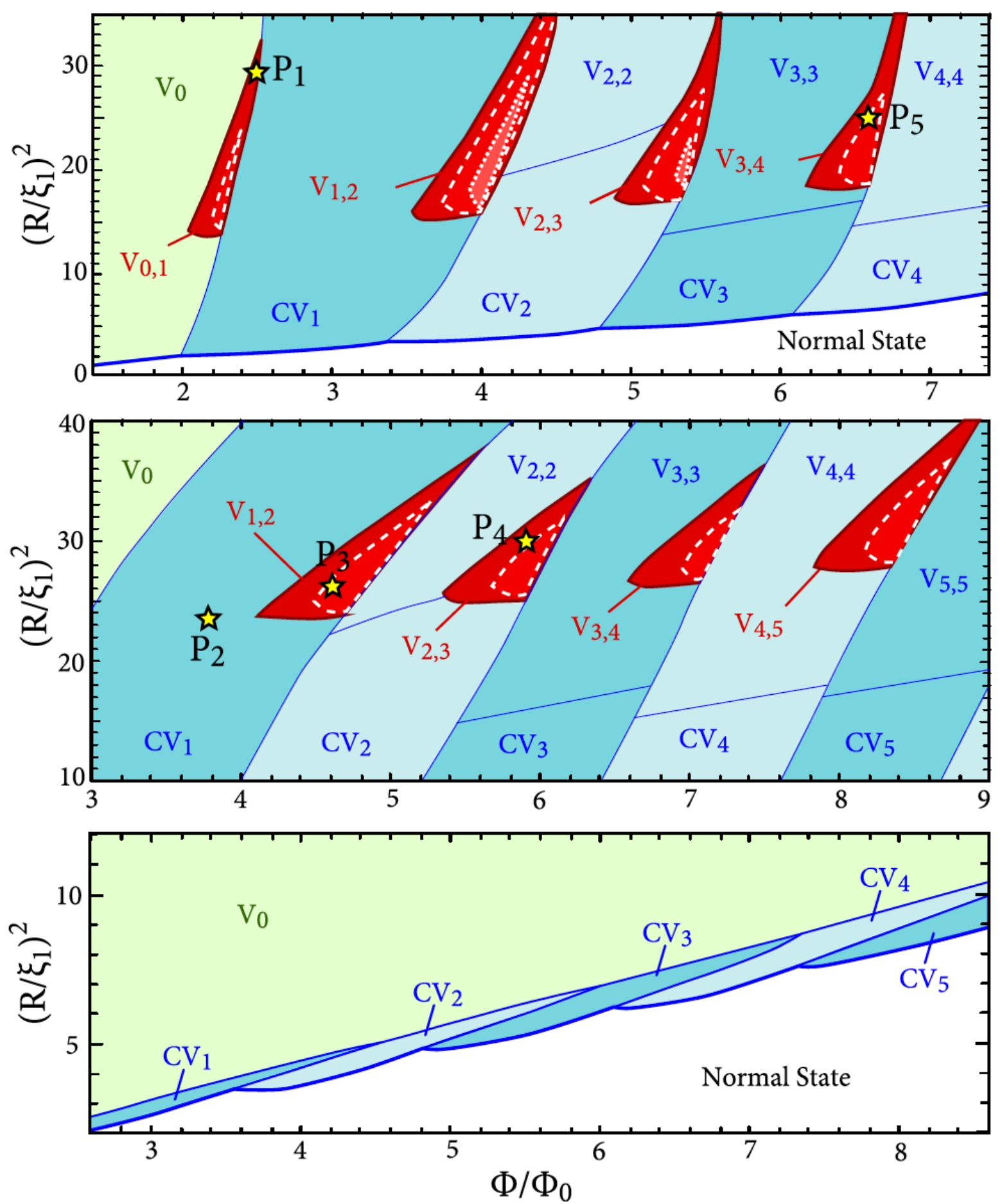,width=\columnwidth}
	\caption{\label{eqfv_fig1}(color online) (a) Phase diagram of a two-band superconducting cylinder with a radius $R$ obtained by numerical minimization of the Ginzburg-Landau free energy functional in Eq. \eqref{meq1} for $\kappa'\equiv \Phi_0 m_j \beta_j^{1/2}/(2\pi)^{3/2}\hbar^2 = 10$ (top), 3 (middle) and 0.5 (bottom) [$x_{20}\equiv (R/\xi_1)^2(\alpha_2/|\alpha_1|+\beta_2/\beta_1)= 12$, 20 and 0.5, respectively] with $\gamma'\equiv -\gamma_{12} /(\hbar^2 /2m_j R^2)$ =0.01. Here $m_1=m_2$ and $\beta_1=\beta_2$. It is divided into domains of superconducting states with no vortex (V$_0$), with a central giant vortex of winding number $n$ in each condensate (CV$_n$), and with $n_j$ separated vortices in the $j$-th condensate (V$_{n_1,n_2}$). Dashed and dotted lines delineate the domains where fractional flux vortices exist as stable phases for $\gamma'$ =0.05 and 0.1, respectively. The stars on the plots denote points at which current distributions are shown in Fig. 3 of Ref. \cite{Chibotaru10}. From Ref. \cite{Chibotaru10}.}
\end{figure}

\subsection{Other mechanisms to stabilize fractional vortex}\label{Sec5E}

It is also possible to stabilize a fractional vortex in equilibrium. Silaev considered vortices near the surface of a two-band superconductor using the London free energy functional in Eq. \eqref{vieq1} by neglecting the interband Josephson coupling. \cite{Silaev11b} He found that fractional vortices are stable near the surface of the superconductor due to the cancellation of the unscreened supercurrent by the image antivortices. He also studied the penetration of fractional vortices through the boundary when the external magnetic field is increased. For superconductors with different coherence lengths, the fractional vortices with a larger normal core first enter into the superconductors. When the external field is increased further, the fractional vortices with a smaller normal core then enter into the superconductors, and they merge with the fractional vortices with a larger normal core to form composite vortices. These composite vortices then proliferate into the bulk superconductor. This two-step penetration process is manifested as two jumps in the magnetization curve as a function of external magnetic fields.

\begin{figure*}[t]
	\psfig{figure=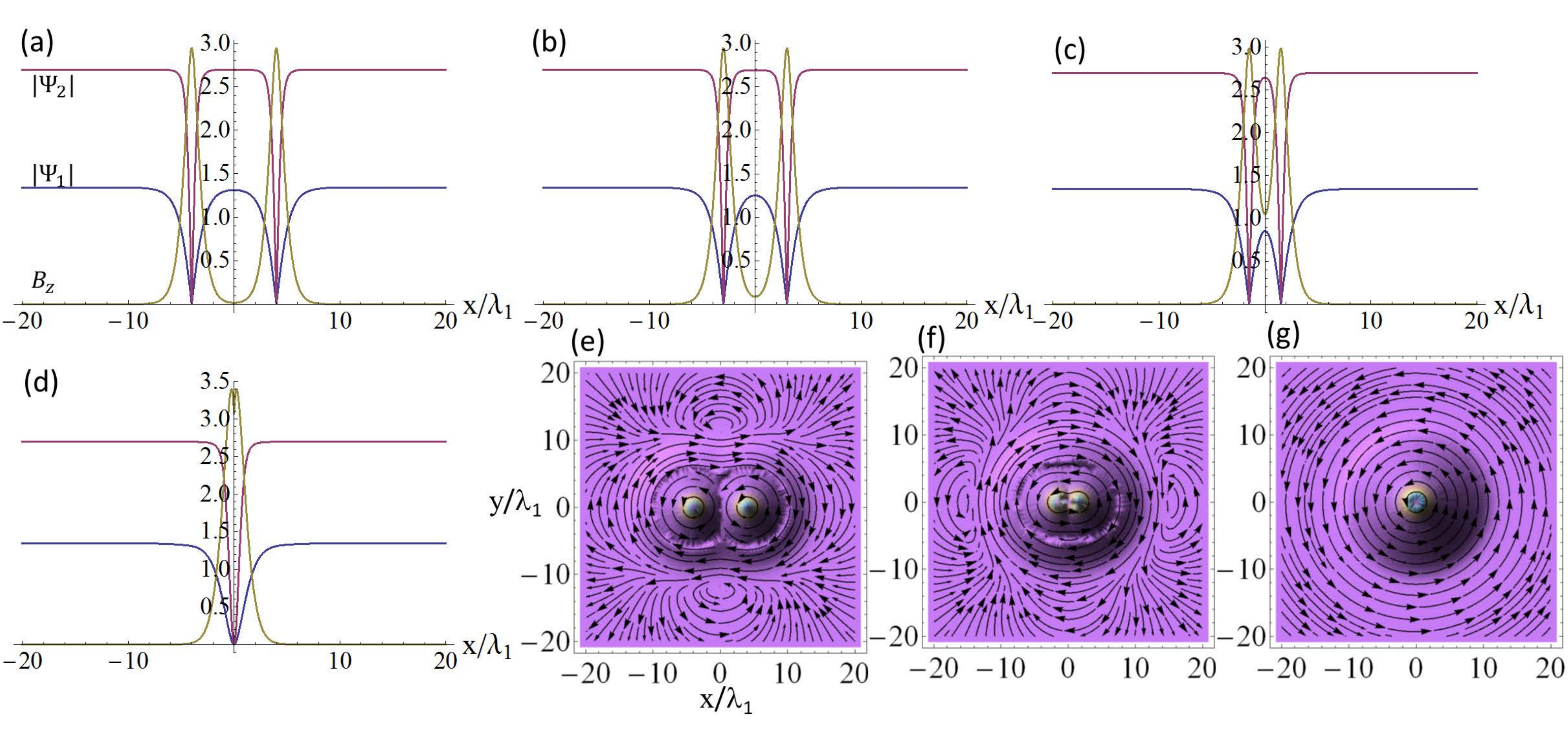,width=18cm} \caption{\label{nmv_fig0} (color online).
		Distribution of the amplitude of order parameters $|\Psi_j|$ (purple and blue lines) and magnetic field $B_z$ (yellow line) at $y=0$ with a vortex separation  (a) $d=8\lambda_1$, (b) $d=6\lambda_1$, (c) $d=3\lambda_1$, and (d) $d=0\lambda_1$. Distribution of the supercurrent at vortex separation (e) $d=8\lambda_1$, (f) $d=3\lambda_1$, and (g) $d=0$. From Ref. \cite{LinPRB11}.}
\end{figure*}
 
Fractional vortices can also be stabilized in a mesoscopic sized two-band superconductor, where the logarithmic divergence of the fractional vortex self-energy is cutoff by the system size. \cite{chibotaru_thermodynamically_2007,Chibotaru10,Geurts10,PhysRevB.84.144504,Pina12,PhysRevB.89.024512} In Ref. \cite{Chibotaru10}, the authors minimized numerically the two-band Ginzburg-Landau free energy functional in Eq. \eqref{meq1} and found a stable fractional vortex configuration, as shown by the red region in Fig. ~\ref{eqfv_fig1}. For superconducting condensates with different coherence length and superfluid density, they respond to the geometry confinement in a different way. Meanwhile, both superconducting condensates are coupled with the same gauge field and they may also couple via the Josephson coupling, which favors integer quantized vortices because the phases of superconducting order parameter tend to lock with each other. The competition of these two effects gives rise to a plethora of vortex states in mesoscopic superconductors, as plotted in Fig. \ref{eqfv_fig1}. 

Sm\o rgrav \emph{el al.} considered a two-band superconductor in magnetic fields with a strong disparity in phase stiffness for different superconducting condensates by Monte Carlo simulations. \cite{Smorgrav05} At zero temperature, the fractional vortices in different condensates form a triangular lattice of composite vortex. Upon heating they found that the sublattice of fractional vortex with lower phase stiffness first melts due to the proliferation of vortex loop driven by thermal fluctuations. This phase transition belongs to the three dimensional $XY$ universality class. In this temperature region, the fractional vortices in different bands are decoupled. When temperature is increased further, the remaining fractional vortex lattice with higher phase stiffness melts via first order phase transition, and the system enters into the vortex liquid phase.

\subsection{Vortex with nonmonotonic interaction}\label{Sec5F}
In multiband superconductors, it is possible that vortices repel at short distant and attract at large separation as first pointed out by Babaev and Speight. \cite{Babaev05} Here we review briefly the nonmonotonic interaction between vortices and its consequences. Another review is available in Ref. \cite{babaev_type-1.5_2012}. Let us consider a two-band superconductor with $\xi_2\ll \lambda \ll \xi_1$. As depicted in Fig. \ref{nmv_fig0} when two isolated vortices with quantum flux $\Phi_0$ approach each other, the normal core of vortex in the first band $|\Psi_1|$ first overlaps, which induces an attraction between vortices. As they get closer, the electromagnetic interaction becomes dominant because of the overlapping of magnetic fields and vortices repel each other, which results in nonmonotonic interaction between vortices. As the separation is reduced further, the normal core of vortex in the second band starts to overlap. They finally merge into a giant vortex with vorticity equal to two.

 \begin{figure}[b]
 	\psfig{figure=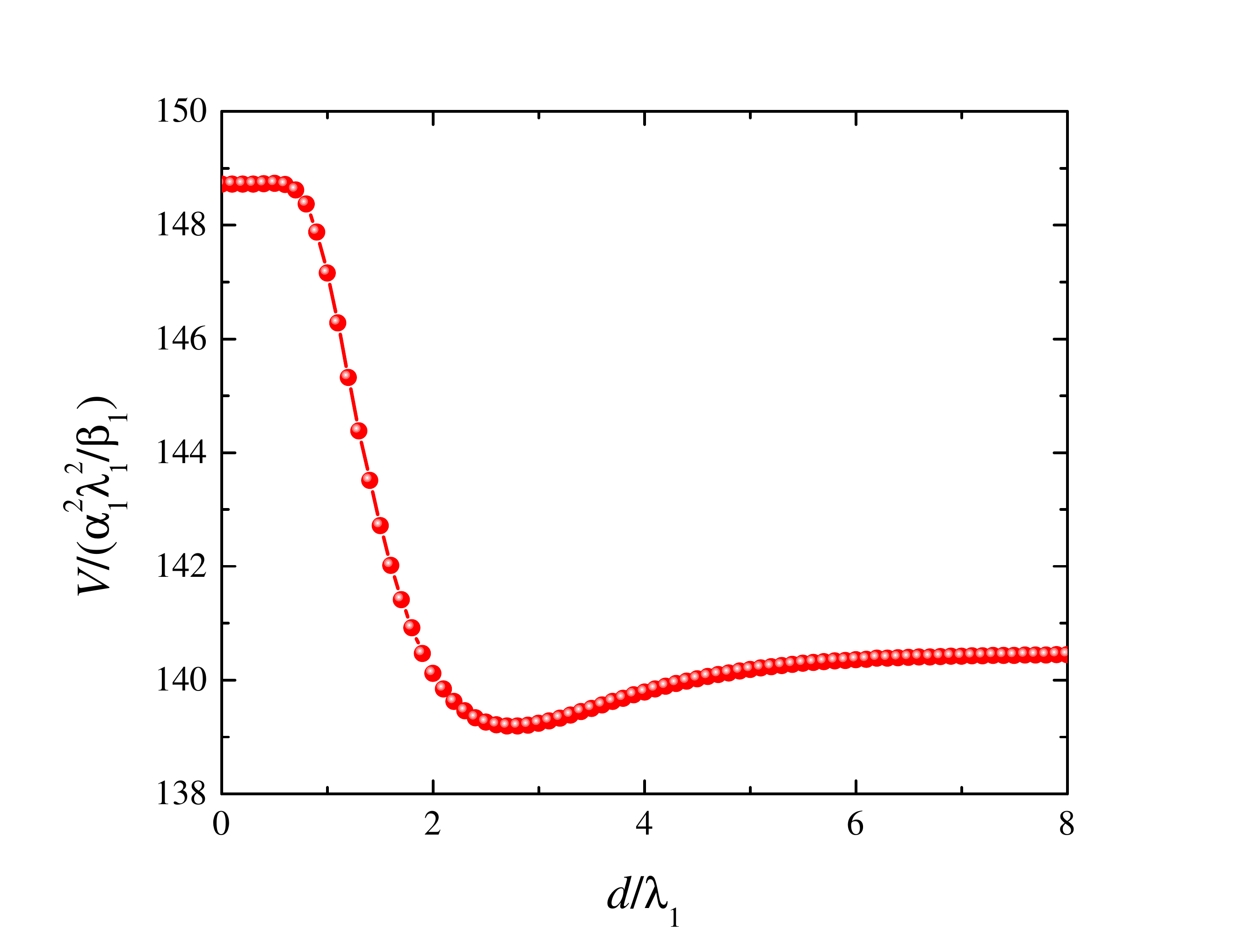,width=\columnwidth}
 	\caption{\label{nmv_fig1}(color online) Dependence of inter-vortex interaction potential $V$ per unit length on the separation $d$ between two vortices. From Ref. \cite{LinPRB11}.}
 \end{figure}

The calculation of interaction between vortices as a function of separation poses a challenge to theory since vortices are extended objects. In the London limit, the normal core becomes a point-like object and one can fix the vortex at a desired position $\mathbf{r}_i$ using the boundary condition $\nabla\times\nabla\phi(\mathbf{r})=2\pi \delta (\mathbf{r}-\mathbf{r}_i)$. For general cases interested here, one has to introduce constraints to fix two vortices at a desired separation.  One may impose pinning potential to vortices by fixing the amplitude and/or phase of superconducting order parameter in a certain region near the vortex cores. \cite{Misko03} However, this method may introduce artifacts when two vortices are close to each other since the order parameters change dramatically near the vortex core. Furthermore, it is sometimes insufficient to pin vortices by imposing the local constraints because the interaction becomes strong when vortices are close. In Ref. \cite{LinPRB11}, we implemented and generalized the variational method for single-band superconductors \cite{Jacobs79} to calculate the inter-vortex interaction in a two-band superconductor. The vortex separation is fixed by choosing proper trial functions for $\Psi_1$, $\Psi_2$ and $\mathbf{A}$. By varying the separation $d$ continuously, we obtained nonmonotonic interaction profile between two vortices as shown in Fig. ~\ref{nmv_fig1}, where there exists a local minimum. The profiles of $\Psi_j$, magnetic field $B_z$ and supercurrent as a function of separation obtained by variational calculations are displayed in Fig. \ref{nmv_fig0}. 

We also introduced two vortices in a square disk with size $L$ through the boundary condition \cite{Doria89}
\begin{equation}\label{eqs3n8}
\mathbf{A}(r+L_\mu)=\mathbf{A}(r)+\nabla\chi_\mu, \ \ \Psi_j(r+L_\mu)=\Psi_j(r)\exp(i2\pi\chi_\mu/\Phi_0),
\end{equation}
with $\mu=x, y$ and $\chi_x=H_a L y$ and $\chi_y=0$. Here $H_a$ is the applied magnetic field and should obey the vortex quantization condition via $\oint d\mathbf{l}\cdot\mathbf{A}=2 n\Phi_0$, which yields $H_a=2n\Phi_0/L^2$. By minimizing the two-band Ginzburg-Landau free energy in Eq. \eqref{meq1} numerically, we found a bound solution with vortex separation independent of the disk size $L$ for a large $L$, which indicates unambiguously an energy minimum in the inter-vortex interaction profile.  The vortex separation is consistent with the results in Fig. ~\ref{nmv_fig1} obtained by variational calculations.

\begin{figure*}[t]
 	\psfig{figure=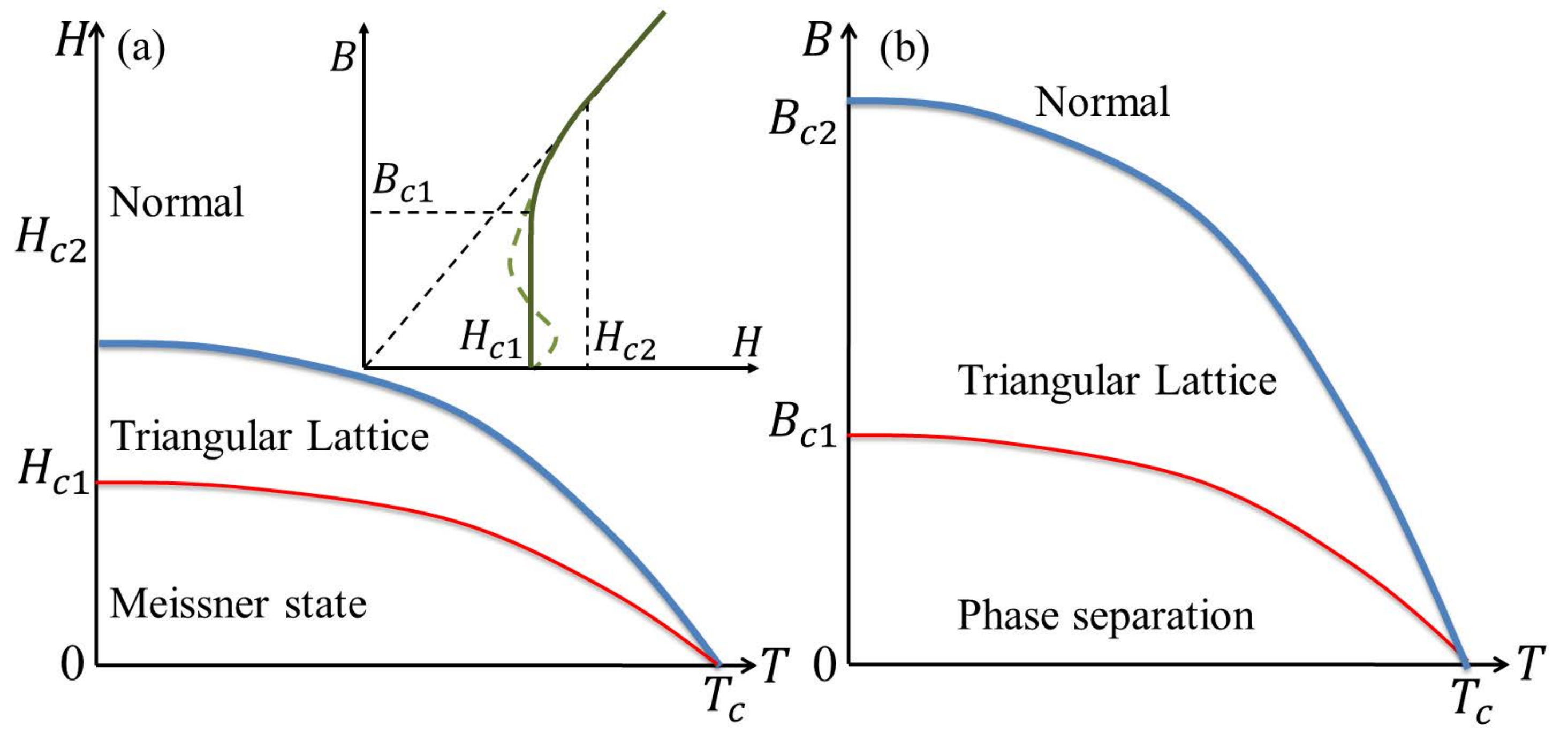,width=14cm}
 	\caption{\label{nmv_fig2}(color online) Mean-field $H$-$T$ (a) and $B$-$T$ (b) phase diagrams of superconductors with competing inter-vortex interaction. The	red (thin) line in (a) indicates the first-order phase transition, while that in (b) indicates the upper boundary for the phase separation ( the lower boundary is the horizontal axis), and the blue (thick) lines are for the second-order phase transition. Inset is for the dependence of magnetic induction on the applied field, and the dashed line represents the hysteresis associated with the first-order phase transition.}
 \end{figure*}

The presence of nonmonotonic inter-vortex interaction modifies drastically the magnetic response of a superconductor. For superconductors with nonmonotonic interaction between vortices, upon increasing external magnetic fields $\mathbf{H}$, clusters of vortex penetrate into superconductors associated with a discontinuous jump in magnetic induction from zero to $B_{c1}$, which practically may be seen as a hysteresis loop in magnetization curve as depicted in the inset of Fig. \ref{nmv_fig2} (a). Thus the transition from the pure Meissner state to vortex cluster phase is of the first order phase transition. It was pointed out that at a low density of vortices, vortex clusters coexist with Meissner state. \cite{Babaev05} As the vortex cluster has positive surface energy, these clusters are of circular shape. \cite{LinPRB11}  The interaction between the vortex clusters is long-range and repulsive induced by magnetic interaction outside the superconductor, and vortex clusters are distributed evenly in clean superconductors. \cite{PhysRevB.86.180506} The vortex density then increases gradually with the external magnetic field until $H_{c2}$ at which superconductivity is destroyed completely [see inset of Fig. \ref{nmv_fig2}(a)]. Here $H_{c2}$ is the same as that of type II superconductors, which is solely determined by the condensate with the shortest coherence length. The mean-field $H$-$T$ and the corresponding $B$-$T$ phase diagram for superconductors with nonmonotonic vortex interaction are depicted in Fig. \ref{nmv_fig2}. Here the $B$ is the magnetic induction $\mathbf{B}=\mathbf{H}+4\pi \mathbf{M}$ with $\mathbf{M}$ the magnetization.

The long-range repulsion between vortices due to the dipolar interaction at large separation, the short-range attraction at intermediate separation and strong repulsion at short separation yield complex vortex configurations in an intermediate vortex density. A low-density clump phase, an intermediate density stripe phase, an anticlump phase, and a high-density uniform phase have been observed in simulations. \cite{OlsonReichhardt2010,PhysRevB.83.020503,zhao_vortex_2012,PhysRevB.83.014501,PhysRevB.90.020509,drocco_static_2013,zhao_analysis_2012,varney_hierarchical_2013,xu_simulation_2014} As vortices approach each other, the nonlinear effect becomes important such that the interaction between vortices may no longer be pairwise, i.e. many-body interaction such as three-body interaction becomes important. \cite{PhysRevB.84.134515,edstrom_three_2013} Such a non-pairwise interaction could result in even more complex vortex distributions, such as vortex glassy phases. \cite{PhysRevE.88.042305}

Recently stripe and gossamer phases of vortex were observed in $\mathrm{MgB_2}$ using imaging methods. \cite{PhysRevLett.102.117001,PhysRevB.81.020506,PhysRevB.85.094511} Coexistence of Meissner state and vortex state was observed in $\mathrm{Sr_2RuO_4}$ using muon-spin rotation measurements. \cite{PhysRevB.89.094504} These observations were interpreted in favor of the existence nonmonotonic interaction between vortices. However caution should be taken because at low vortex densities the pinning of vortices by defects unavoidable in superconductors may be dominant over the vortex interaction, and produces similar vortex patterns as observed. 

The theoretical discussions so far are based on the multiband Ginzburg-Landau free energy functional. When apply to real superconductors, the Ginzburg-Landau theory is valid only for temperature very close to $T_c$. As discussed in Sec. \ref{Sec2B}, multiband superconductors with interband coupling behave as single-band superconductors near $T_c$. One may extend the applicable region of the Ginzburg-Landau theory by expanding to higher order in $(T-T_c)/T_c$. \cite{Shanenko11, Vagov12, Komendova11} Nevertheless qualitative features about the vortex interaction may be extracted from the Ginzburg-Landau theory even at low temperatures. To describe the vortex interaction at low temperatures in a rigorous way, a microscopic theory beyond the Ginzburg-Landau theory is required. Such a microscopic theory was developed in Ref.  \cite{Silaev11} using the two-band Eilenberger formalism, where the authors demonstrated the existence of nonmonotonic inter-vortex interaction for appropriate parameters.

Finally we would like to remark that the nonmonotonic inter-vortex interaction can also be found in single-band superconductors with $\lambda/\xi$ close to $1/\sqrt{2}$, such as high purity Nb crystal. The nonmonotonic interaction arises from the BCS correction to the Ginzburg-Landau theory. Due to the competing interaction, vortex clusters coexisting with the Meissner phase are stabilized, which was observed in Nb crystals. For details, please refer to Ref. \cite{Brandt2011} and references therein.  

The physics of vortex with nonmonotonic interaction in multiband superconductors is still under active research. The nonmonotonic interaction between vortices was also found in three-band superconductors with frustrated interband couplings. \cite{PhysRevB.84.134518,takahashi_unconventional_2013,takahashi_ht_2014} The effect of the interband Josephson coupling \cite{PhysRevLett.105.067003, Geurts10,PhysRevB.83.174509} and the condition for the nonmonotonic interaction \cite{PhysRevB.83.214523} were studied. The applicability of the Ginzburg-Landau theory to investigate the nonmonotonic inter-vortex interaction was discussed in Refs. \cite{Kogan2011,PhysRevB.86.016501,PhysRevB.86.016502}. The comparison between vortex with nonmonotonic interaction in single-band and multiband superconductors was made in Refs. \cite{Brandt2011,babaev_type-1.5_2013}.

\section{Discussions}\label{Sec6}

It is possible that the vortices, phase solitons and Leggett modes interact with each other. For instance, in three-band superconductors the presence of a vortex distorts the amplitude and phase of the superconducting order parameters and excites the Leggett mode. This distortion propagates in superconductors and can be felt by another vortex. In this way a mutual interaction is established between vortices. \cite{PhysRevB.84.134518} One particularly interesting situation is when the Leggett mode becomes gapless. In this case the interaction between vortices due to the exchange of the Leggett excitation becomes long-range. On the other hand, the motion of vortex excites superconducting amplitude-phase mixed mode or the Leggett mode, thus provides additional viscosity to the vortex motion. \cite{PhysRevB.88.220504} The phase kink in two dimensions forms a circular shape to minimize the energy because the energy cost to excite phase kink is positive. For the same reason, the circle shrinks in time and finally the kink disappears. Therefore the kink solution in two dimensions is unstable, in accordance with the Derrick's theorem \cite{Derrick1964}. The interaction between the kink and vortex can stabilize the kink solution. Near the kink region, superconductivity is suppressed and therefore vortices are pinned in the kink region. The repulsion between vortices prevent kink from collapsing thus stabilize the kink-vortices composite structure. \cite{PhysRevLett.107.197001}

There are also many interesting physics arising from the multiband nature of superconductors, which is not discussed in the previous sections. Here we mention them briefly, and the list here is rather partial and biased. For details readers may consult the original papers.  Flux flow and pinning of the vortex was studied in Ref. \cite{PhysRevB.70.100502}. Anomalous flux flow resistivity in $\mathrm{MgB}_2$ was observed in Ref. \cite{PhysRevB.68.060501}. Field dependence of the vortex core size in a multiband superconductor was measured in Ref. \cite{PhysRevLett.95.197001}.  Skyrmions in multiband superconductors were discussed in Refs. \cite{PhysRevB.89.104508,PhysRevB.87.014507,PhysRevB.90.064509}. Hidden criticality inside the superconducting state in multiband superconductors was studied in Ref. \cite{Komendova2012}. Entropy-induced and flow-induced superfluid states were proposed in Ref. \cite{PhysRevLett.113.055301}. Thermal fluctuations in multiband superconductors were investigated in Refs. \cite{PhysRevB.72.064523,PhysRevB.84.214515,Koshelev2014}. Phase slip was studied in Ref. \cite{fenchenko_phase_2012}. Magnetic field delocalization and flux inversion in fractional vortices was investigated in Ref. \cite{PhysRevLett.103.237002}. The unusual dependence of superfluid density and specific heat was calculated in Ref. \cite{Kogan09}. For more discussions on the thermodynamical properties in multiband superconductors, please refer to Ref. \cite{zehetmayer_review_2013} for a review.

Much of the novel physics for multiband superconductors discussed in this review can be realized in Josephson junctions.  In junctions, the superconducting electrodes can be regarded as distinct superconducting condensates separated in real space and coupled by the Josephson interaction. Josephson junctions with electrodes made of single-band $s$-wave superconductors can be regarded as artificial two-band superconductors. The sign of the Josephson coupling can be tuned by using different blocking layers. For instance one can achieve a $\pi$ phase shift between superconducting electrodes by using a ferromagnetic blocking layer \cite{Bulaevskii77}, which corresponds to the $s\pm$ pairing symmetry in two-band superconductors. One can also use a two-band superconductor as one electron and a single-band superconductor as the other electrode to realize an artificial three-band superconductor. Frustration can be introduced when the two-band superconductors have $s\pm$ pairing symmetry.  Time-reversal symmetry breaking in these junctions was discussed in Refs. \cite{tanaka_phase_2001,Ng09,PhysRevB.86.214502,LinPRB12} and phase solitons with fractional quantum magnetic flux was discussed in Ref. \cite{LinPRB12}. Such configurations were also proposed to detect the $s\pm$ pairing symmetry. \cite{PhysRevB.86.174515,koshelev_proximity_2011,PhysRevLett.103.207001,PhysRevLett.102.227007,PhysRevB.80.020503,PhysRevB.83.212501,Chen10} The main difference between Josephson junctions and multiband superconductors is that the superconducting phase differences in junctions are coupled to gauge fields while the phase differences between bands in multiband superconductors are not.  Josephson junctions using multiband superconductors as electrodes are interesting systems and novel phenomena emerge,  \cite{Ota09,PhysRevB.81.014502,PhysRevB.82.140509,PhysRevB.81.214511,huang_josephson_2014,yerin_frustration_2014} which deserve a separate investigation. Interested readers may refer to review papers \cite{Brinkman03,Xi09,Seidel11} on this topic.  

One may obtain qualitative features of the phase solitons and vortices in multiband superconductors using a simplified model. However to apply to real materials, one has to consider a realistic model derived for a specific material. Multiband superconducting materials with a weak interband coupling can facilitate the experimental observations of the Leggett mode, the phase solitons and fractional vortex. The experimentally observed Leggett mode is unstable while the phase soliton and fractional vortex have never been observed yet in any bulk multiband superconductor at the time of writing. The interband couplings for the prominent multiband superconductors $\mathrm{MgB_2}$ and iron-based superconductors are not weak. It is a big challenge and opportunity to find multiband superconductors with weak interband couplings.

\begin{acknowledgments}
The author is grateful to Xiao Hu, Lev N. Bulaevskii and Charles Reichhardt for fruitful collaborations over the past few years. He also thanks Egor Babaev, Alex E. Koshelev, Vladimir Kogan, Alex Gurevich, Milorad Milo\u{s}evi\'{c}, Valentin Stanev and Yukihiro Ota for helpful discussions. He apologizes to the many whose work we were not able to treat in the depth it deserved in this review. This work was carried out under the auspices of the NNSA of the US DoE at LANL under Contract No. DE-AC52-06NA25396, and was supported by the US Department of Energy, Office of Basic Energy Sciences, Division of Materials Sciences and Engineering. 
\end{acknowledgments}

%

\end{document}